# Nature of the Structural Symmetries Associated with Hybrid Improper Ferroelectricity in $Ca_3X_2O_7$

S. Liu[1], H. Zhang[1], S. Ghose[2,*], M. Balasubramanian[3], Zhenxian Liu[4], S. G. Wang[5], Y-S. Chen[5], B. Gao[6,7], J. Kim[6,7], S.-W. Cheong[6,7], and T. A. Tyson[1,7,*]
[1]Department of Physics, New Jersey Institute of Technology, Newark, NJ 07102
[2]National Synchrotron Light Source II, Brookhaven National Laboratory, Upton, NY 11973
[3]Advanced Photon Source, Argonne National Laboratory, Argonne, IL 60439
[4]Institute of Materials Science, Department of Civil and Environmental Engineering, The George Washington University, Washington, DC 20052
[5]Center for Advanced Radiation Sources, University of Chicago, Argonne, IL 60439
[6]Department of Physics and Astronomy, Rutgers University, Piscataway, NJ 08854
[7]Rutgers Center for Emergent Materials, Rutgers University, Piscataway, NJ 08854

*Corresponding Authors: T. A Tyson, e-mail: tyson@njit.edu
S. Ghose, e-mail: sghose@bnl.gov

## Abstract

In hybrid improper ferroelectric systems, polarization arises from the onset of successive nonpolar lattice modes. In this work, measurements and modeling were performed to determine the spatial symmetries of the phases involved in the transitions to these modes. Structural and optical measurements reveal that the tilt and rotation distortions of the $MnO_6$ or $TiO_6$ polyhedra relative to the high symmetry phases driving ferroelectricity in the hybrid improper $Ca_3X_2O_7$ system (X=Mn and Ti) condense at different temperatures. The tilt angle vanishes abruptly at $T_T$ ~ 400 K for $Ca_3Mn_2O_7$ (and continuously for X=Ti) and the rotation mode amplitude is suppressed at much higher temperatures $T_R$ ~1060 K. Moreover, Raman measurements in $Ca_3Mn_2O_7$ under isotropic pressure reveal that the polyhedral tilts can be suppressed by very low pressures (between 1.4 and2.3 GPa) indicating their softness. These results indicate that the $Ca_3Mn_2O_7$ system provides a new platform for strain engineering of ferroelectric properties in film based systems with substrate induced strain.

## I. Introduction

A broad range of mechanisms are known to produce ferroelectricity in solid materials including intrinsically atomic effects such as the lone pairs (e.g. $6s^2$ in $BiFeO_3$ [1]) or ion size mismatch leading to off-center lattice distortions in the classical $BaTiO_3$ system [2]. From a microscopic perspective, these materials fall into two main classes having net electric polarization due to atomic displacement, such as the $ABO_3$ system, where an off-center B atom displacement is proposed. Alternatively, order-disorder transitions have also been proposed in these systems. In general, the microscopic mechanisms driving ferroelectric behavior are thought to cause a gamma point instability in the ordering of a polar mode of the parent structure. There is a direct connection between the polarization amplitude and the structural polar mode.

Distinct from these mechanisms above are a class of improper ferroelectrics in which the polarization is a secondary order parameter. Recently, the layered perovskites (n=2 Ruddelsen-Popper structure with double stacked $MnO_6$ polyhedra, $A_3B_2O_7$) was predicted to support ferroelectric induced by the introduction of a rotation pattern due to two independent nonpolar rotations of the $BO_6$ polyhedra [3] relative to the high symmetry I4/mmm (#139) phase at.high temperature. The structure is characterized by an $X_2^+$ rotation of the $MnO_6$ polyhedra in the high symmetry plane and a $X_3^-$ tilt relaive to the high symmmerty (long) axis. The polarized state of this system is determined by the combined rotation and tilt structural modes. In these hybrid improper ferroelectrics, DFT calculations reveal that the symmetry of the phase exhibiting ferroelectricity has space group $A2_1am$ (#36, $Cmc2_1$ in standard setting). The rotations couple directly with the magnetic order of the Mn sites (having G type antiferromagnetic order) at low temerature [4].

The nature of the transition from the high-temperature I4/mmm phase to the low-temperature $A2_1am$ phase is not well understood. Deterination of whether the rotation and tilt structural changes onset collectivley or separately, the nature of the transitions (first or second order), and the appropriate long range and local symmetries present is essential to the development of accurate theoretical models.

Limited structural powder diffraction work suggests that the $Ca_3Mn_2O_7$ system undergoes a transition into an intermediate Acaa phase (#68, Ccca standard setting) [5]. In polycrystalline samples, the proximity of the Acaa and $A2_1am$ phase in energy was found to produce a broad mixed phase region. Additional symmetry possibilities of the intermediate state include the Acam, Amam and F2mm space groups. While clear switchable ferroelectric polarization has been observed in Ti and Sn-based systems (B site), the manganese perovskite system has not been seen to exhibit this critical behavior needed for device applications. A complex domain structure with the stacking of a and b domains along the c axis (long axis) has been found to suppress polarization in $Ca_3Mn_2O_7$ [6]. The primary difference between $Ca_3Mn_2O_7$ (with the coupling of magnetic order and structure but no finite electric polarization) and the $Ca_3Ti_2O_7$ (with large electric polarization, 8 $\mu C/cm^2$) [7] is the stability of $A2_1am$ low-temperature phase in the Ti-based system over a broad range of temperatures and its switchable room tempearature electric polarization.

Systematic studies of the $Ca_3Mn_{2-x}Ti_xO_7$ mixed system have been initiated to understand the electric polarization domain structure and magnetic order in this system [5]. Recent Raman measurements up to 350 K on single crystals of $Ca_3Mn_2O_7$ indicate a significant softening of a mode near 107 $cm^{-1}$ which was assigned to be an antiphase tiling mode bases on phonon calculations utilizing classical interaction potentials [8]. Shifts in the modes corresponding to tilting and rotation were examined, and it is argued that reduction of the frequency of the tilting mode with temperature indicates that this is a transition from $Cmc2_1$ ($A2_1am$) to an assumed Ccca (Acaa) intermediate phase is driven by a tilt rotation. An intermediate phase region with the coexistence of competing soft modes with distinct phases was found to be consistent with these measurements. However, no direct structural information was presented.

Neutron diffraction and laboratory-based x-ray diffraction structural measurements on the B site (Ti or Mn) doped system $Ca_3Mn_{0.9}Ti_{0.1}O_7$ found antiphase tilting of the $MnO_6$ polyhedra decreases in amplitude while the rotation remains unchanged with increasing temperature and pressure [9]. The tilting was found to decrease smoothly (from ~8° to ~6.5°), going from ~10 K to ~375 K, with a kink near the magnetic ordering temperature $T_N$. In the analysis in that work, the single-crystal neutron scattering measurements reveal a polar Aba2 space group at 450 K. It was indicated that $Ca_3Ti_2O_7$, by contrast, would have enhancement of both antiphase tilt and in-plane distortion with increasing temperature.

To fully understand the nature of the transitions in this material, structural measurements on the endmember $Ca_3Mn_2O_7$ and $Ca_{2.4}Sr_{0.6}Ti_2O_7$ (representing CaTiO3) were conducted over a broad temperature range. Samples derived from single crystals were used in all measurements. Heat capacity (2 K to 380 K) and differential scanning calorimetry (DSC, 300 K to 500 K) measurements were conducted. Local structural pair distribution function (PDF) measurements were conducted between 100, and 1100 K at the NSLS II XPD (28-ID-2) Beamline and APS beamline 6-ID-D. X-ray absorption (XAFS and XANES) measurements were conducted with respect to the Mn site for the temperature range 300 K and 1100 K (at beamline 8-ID at NSLS II), and with respect the Ca site (at APS beamline 20 BM- for temperatures near the ~400 K transition). High-resolution synchrotron single crystal diffraction measurements were conducted between 100 K and 480 K (at APS beamline 15-ID-D). High-pressure Raman measurements were conducted at NSLS2 beamline 22-IR-1 to follow the changes in symmetry with pressure and determine the pressure ranges for the phase changes. Experimental and computation details including phonon mode information, Raman spectra, heat capacity and TGA measurements, XAFS refinements and XAS results and single crystal diffraction structural data (at 104 K, 300 K and 480 K) are provided in the Supplementary Document [10].

Calorimetry measurements (differential scanning calorimetry (DSC) and heat capacity) reveal a second order transition near 110 K corresponding to the antiferromagnetic ordering temperature and a first-order transition near 367 K on cooling and near 405 K on warming in $Ca_3Mn_2O_7$. This first order transition is found in structural measurements corresponding to a loss of tilt angle (vanishing of $X_3^-$ tilt amplitude) above this transition on warming and yields an nonpolar space group (I4/mmm or C2/m with split in-plane oxygen sites). The transiton temperature is termed $T_T$. Distinct domains with the same space group are found. These measurements are supported by Raman observations. Local structural measurements reveal enhanced Ca-Mn bond on warming above $T_T$ consistent with loss of electric polarization. No evidence of phase coexistence is found near $T_T$. In this temperature region, thermal measurements reveal no clear phase transitions in polycrystalline samples. PDF and XAFS measurements reveal a transition near 1060 K consistent with the loss of the $X_2^+$ rotation amplitude (termed $T_R$).

In this work, parallel structural and spectroscopic measurements, conducted on the corresponding Ti-based system $Ca_{3-x}Sr_xTi_2O_{7\,()}$ ($Ca_{2.4}Sr_{0.6}Ti_2O_7$), reveal the same trend. However the loss of the $X_3^-$ tilt amplitude is found to be a continuous transition. DFT calculations on $Ca_3Mn_2O_7$ indicate that the relevant self force constant for the O sites are significantly softer for a local tilt distortion compared to local twist or rotational local distortion, consitent with the measuremets. The phonon densitiy of states of both systems are qualitatively similar. The essentail overview of the general $Ca_3X_2O_7$ system is that near $T_T$ (~400 K) the tilt mode vanishes, while the rotational mode amplitude decrease continuously with increase temperature. High-pressure raman measurements indicate loss of tilt angle for very low pressures (between 1.4 and 2.3 GPa) revealin the softness of the tilt mode. Preparing thin film samples with substrate strain values near this critical cross-over value in strain (near ferro/paraelectric transition will enable switching of polarization (on/off) using a piezoelectric substrate. The results also suggest that enhancement of the A-O bond strength will lift the temperate $T_T$ at which the polarization vanishes. Work on the X=Mn system is presented in the main text and the X=Ti system results are included in the supplementary document.

## II. Qualitative Structure, Predicted Phonon Spectra and Raman Measurements

Figure 1(a) shows the crystal structure of $Ca_3Mn_2O_7$ with respect to the $Cmc2_1$ space group (#36, standard setting) with the long-axis along a. Note that there are three distinct types of O sites in this n=2 Ruddlesden-Popper perovskite system. There are apical O sites bonded only to one Mn site and weakly bonded to the Ca sites, apical interior sites shared by two Mn sites along the a-axis (a-axis, long axis for this space group setting) and planar O sites shared between two Mn site in the b-c plane. Fig. 1(b) shows the $MnO_6$ polyhedral rotation about the long axis ($X_2^+$, black curved arrow) and the tilts ($X_3^-$, blue arrows) of the polyhedra normal to the long axis.

Calculation of the force constants by DFT methods enables computation of the phonon density of states. Fig. 1(c) shows the phonon density of states (DOS) projected onto the Mn, Ca and O sites exclusively. What can be seen is that although the Ca atoms are lower in mass than the Mn atoms, the main peak in the phonon DOS at low energy is primarially from the Ca site. The Mn contributions are broad and extend up to ~500 $cm^{-1}$ while the O DOS covers the full energy range. Hence Ca atom motions dominate the low-frequency phonon modes while oxygen contributes at all frequencies. The similarity of the force constants indicates similar phonon DOS for $Ca_3Mn_2O_7$ and $Ca_3Ti_2O_7$. Indeed theoretical work on $Ca_3Ti_2O_7$ yield qualitatively similar phonon DOS [11] to that presented in Fig. 1(c).

Raman measurements were conducted on single crystal, and polycrystalline samples for temperatures between 100 and 500 K, covering the high end of any possible low-temperature transitions. In Fig. 2(a) we show spectra for the incident laser beam normal to the long axis (x-axis). Note the abrupt change in frequency in the peak near 615 $cm^{-1}$ on cooling below ~370 K (breathing mode of $MnO_6$ polyhedra, see below). Near the same temperature, the peak near 460 $cm^{-1}$ also vanishes. We note that a weak low-frequency peak near 146 $cm^{-1}$ in the spectrum is found to vanish with increasing temperature (see arrow in Fig. 2(a)). The area of this peak vs. temperature is given in Fig. 2(b) indicates the abrupt drop-off with temperature. For the same measurement geometry, the peak position of the main peak near 615 $cm^{-1}$ is shown on warming and cooling in Fig 2(c) and in Fig. S4(b). The abrupt transition is characteristic of first-order behavior. Measurement of a polycrystalline sample over the same temperature range reveals no step change, consistent with earlier measurements [4] on the same type of material suggesting mixed phase behavior. Fig. S1 (supplementary document) shows the calculated displacement modes from DFT for phonons near 128 $cm^{-1}$ and 141 $cm^{-1}$ (Fig. S1(a), Fig. S1(b) and Table S1.). The lowest energy mode (128 $cm^{-1}$) corresponds to the shear motion of the Ca ion near the apical O atoms.

While the calculated mode at 141 $cm^{-1}$ corresponds to the combined planar motion of the Ca sites and tilt rotations of the $MnO_6$ polyhedra. For both modes, all Ca atomic motions are in the yz plane containing the electric polarization vector. The main peak at 612 $cm^{-1}$ (Fig. S1(c)) is characterized by a breathing mode of the $MnO_6$ polyhedra. The full set of calculated phonon-frequencies at the gamma point are given in Table S1. The corresponding Raman spectra in a offset configuration for the laser beam normal to and along the long axis are given in Fig. S2. The measurements for the beam along the symmetry axis show suppression of the peak near 120 $cm^{-1}$ with increasing temperature (Fig S3, and additional surface plots in Fig S4).

## III. Calorimetric Measurements

Heat capacity and DSC measurements on crystals of $Ca_3Mn_2O_7$ were used to identify the nature of the observed transitions. Low-temperature heat capacity measurements (Fig. S5) on both warming and cooling reveal a smooth transition near 110 K corresponding to the magnetic ordering temperature ($T_N$). The absence of hysteresis and the step-like shape indicates that the transition is second order. DSC measurements on crystals of at higher temperature (Fig. 3) reveal an abrupt structural transition at 405 K on warming and at 367 K on cooling the sample, indicating that the transition is first order in nature. No transition features are seen in the corresponding polycrystalline sample measurements (derived from solid state reaction, Fig.S6(a)).

DSC measurements for $Ca_{2.4}Sr_{0.6}Ti_2O_7$ indicate a very weak feature with the characteristic step change observed for continuous or glass-like transitions (Fig. S6(b)). Raman measurements on single crystal $Ca_{2.4}Sr_{0.6}Ti_2O_7$ (stabilized by Sr doping) between 100 and 500 K (Figs. S7 and S8) indicate low-frequency phonons near 93 and 112 $cm^{-1}$ soften smoothly with temperature indicating a continuous transition. Similar behavior is seen in polycrystalline samples of $Ca_3Ti_2O_7$ with low-frequency phonons near 121 $cm^{-1}$ (Figs. S9 and S10). The combined results indicate that the transition at $T_T$ is first order for the X=Mn system and continuous for the X=Ti system.

## IV. Local Structural Measurements (X-Ray Absorption Spectroscopy)

Detailed structural measurements on $Ca_3Mn_2O_7$ were conducted to determine the changes in local and long-range structure with temperature. XAFS measurements relative to the Mn site were used to determine the changes in local structure between 300 and 1100 K (Fig. 4). Fits were made for Mn-O, Mn-Ca and Mn-Mn bond (covering the range up to ~ 4 Å, Fig S11). At high temperatures (Fig. 5(a) and 5(b)) the Mn-O and Mn-Ca bonds exhibit changes associated with ordering of the structure (near $T_R$). Examination of the structure-function over the full temperature range reveals stiff Mn-O bonds and stiff Mn-Mn correlation with weak changes in the peak amplitudes with temperature (See Fig. S11) as predicted by the DFT simulations (Table I). The largest changes with temperature occur in the Ca-Mn correlation peak between 300 and 600 K. The Mn lattice sites are seen to form a very rigid network with major changes corresponding to the angular motion of rigid $MnO_6$ polyhedra units and the Ca ions. In the temperature region between 300 and 700 K (Fig. S12), an enhancement of the Mn-O peak width is seen near 400 K on warming.

Examination of the Ca site (Ca-O, Ca-Mn and Ca-Ca bonds up to ~ 4 Å) near the vicinity of the transition at 400 K reveal an enhancement of Ca-Mn correlation (peak width of Ca-Mn distribution) on heating through $T_R$ indicating an increase on local symmetry at this transition. This can be seen in Figs. S13 and Table S2, where the Gaussian width of the Ca-Mn are found to become narrow on going above the $T_R$.

To follow the local structure trends in the X=Ti system, Ti K-edge XANES spectra of crystal derived $Ca_{2.4}Sr_{0.6}Ti_2O_7$ samples were measured for temperatures between 300 and 540 K (Fig. S14(a)). It was found that the main pre-edge feature (feature B in Fig S14(b)) increases in intensity with increasing temperature. Simulated XANES spectra for perovskite systems (Fig. S14(b)) showed that the reduction of tiling amplitude relative to the long axis (called a-axis here) increases the amplitude of peak B while reduction of twisting about z-axis reduces the peak amplitude (dotted line). Hence the observed continuous increase the B feature amplitude for temperatures from 300 to 540 K is related to a continuous loss of tilt angle with increasing temperature above 300 K in $Ca_{2.4}Sr_{0.6}Ti_2O_7$. Hence, like the $Ca_3Mn_2O_7$ system, the $Ca_{2.4}Sr_{0.6}Ti_2O_7$ (representing X=Ti) system shows a reduction in tilt angle with increasing temperature.

## V. Nanoscale Structural Measurements (X-Ray Pair Distribution Function Analysis)

More detailed structural information on $Ca_3Mn_2O_7$ and $Ca_3Ti_2O_7$ type systems was obtained on an intermediate length scale by PDF measurements between 100 K and 1100 K (in real space taken over the range 1.3 to 20 Å (Fig. S15), single crystal based materials). To determine the space group in $Ca_3Mn_2O_7$ on this intermediate structural level on crossing the transition near 400 K ($T_T$), fits to unit cells with possible space groups suggested previously [3] were examined including the groups $Cmc2_1$ (#36), Amm2 (#38), Amam (#63), Acam (#64), and Acaa (#68). The Rw parameter indicating the fit quality was compared over the temperature range 200 to 450 K for each space group (see Fig. S16(a)). ($R_W = \left\{\frac{\sum_{i=1}^{N} w(r_i)[G_{Obs}(r_i)-G_{Calc}(r_i)]^2}{\sum_{i=1}^{N} w(r_i)[G_{Obs}(r_i)]^2}\right\}$ was scaled by the number of independent parameters minus the number of free fittng parametres [12]). Additional, comparisons of the space group $Cmc2_1$ (#36) with the nonpolar space group I4/mmm (#139) and C2/m (#12, see below), both with cells having ½ the volume of the $Cmc2_1$ uint cell, were also conducted (Fig. S16(b)). The results indicate that below ($T_T$), $Ca_3Mn_2O_7$ crystals possess $Cmc2_1$ space group symmetry on an intermediate structural range. While above it, the space group is closer to the C2/m space group on this intermediate length scale (1.3 to 20 Å used in fits). Examining the a/c ratio (Fig. S17) reveals an abrupt transition near 400 K on warming and near 370 K on cooling consistent with the DSC measurements (Fig. 3). The high temperature PDF measurement over the range 300 to 1100 K (Fig. S18) reveals an abrupt change in $R_W$ near 1060 K and smooth increase in the b/c ratio at this transition but no change along the c axis consistent with rotation of the $MnO_6$ polyhedra ($T_R$).

In Fig. 6. we examine the change in the structure of $Ca_3Mn_2O_7$ on this length scale in more detail (with respect to the cell $Cmc2_1$ cell in Fig 1). In Fig. 6(a), the volume on cooling (3 K steps) reveals two distinct regions of thermal expansion behavior. The volumetric thermal expansion coefficient ($\alpha_V$) takes on the value 3.5(1) x $10^{-5}$ $K^{-1}$ above ~370 K and the value 2.63(4) x $10^{-5}$ $K^{-1}$ below it. Near 110 K evidence is seen for a negative thermal expansion corresponding to the onset of the magnetic ordering and spin lattice coupling. The inset of the figure reveal hysteresis behavior covering the region between ~350 and ~400 K. (Consistent with earlier published work on polycrystalline samples [4], our PDF data on polycrystalline samples (Fig. S19, inset) reveals hysteresis behavior covering a much broader region of temperature (~250 to ~400 K).) As indicated in Fig. 6(b), examination of the PDF refinement xyz data indicates that the Mn x postion (long axis) show abrupt change both at $T_N$ and at $T_T$. However, the Ca x atomic position show changes only at $T_T$. With the enhanced symmetry indicated by the XAFS results on the Ca local structure, this indicates a loss of polarization at $T_T$. The exact nature of the structural change was qualitatively determined by looking at the atomic displacements of the 400 K PDF structure relative to the structure at 300 K. In Fig 6(c), we see that the major displacements are for axial O atoms leading to reduction of the $MnO_6$ tilt amplitude ($X_3^-$ amplitude).

## VI. Single Crystal Diffraction Measurements (Space Group above $T_T$)

Accurate structural parameters of $Ca_3Mn_2O_7$ were derived from detailed synchrotron single crystal diffraction measurements on ~20 microns crystals between 300 and 480 K on warming. The low-temperature structure (104 K) was also determined. A high count rate Pilatus 1M detector ($10^7$ cps maximum/pixel) was utilized to obtain large signals for structure factors corresponding to scattering from both the light O atoms and heavy atoms (Mn/Ca). (Use of this detector with the high flux at a synchrotron beamline was found to enable the structural solution of atomic positions of B atoms in rare earth systems such as $HoAl_3 (BO_3)_4$ yielding accurate B-O distances (see Ref. [13]).) For each temperature data set, a systematic space group search was conducted [14]. Fits of full diffractions data sets to the $Cmc2_1$ structure were found to be stable for the temperature range 300 (fit parameter R1 = 5.2%) to 380 K (R1 = 5.6 %). At 400 K the best structural refinements were obtained for the nonpolar spacegroups I4/mmm (R1 = 5.6 %) and C2/m (R1 = 7.1%). Fits to the polar Aba2 space group at 400 K yielded R1 = 10% and large residual charge density. Measurements up to 480 K were taken, and the fits to I4/mmm and C2/m space groups yielded R1= 4.2 % and R1 = 4.8%, respectively.

The full structural data from the refinements of $Ca_3Mn_2O_7$ at 480 K are given in Tables S3 and S4 for I4/mmm and Table S5 for the C2/m space group (both with cells ½ the $Cmc2_1$ cell). The room temperature structural data ($Cmc2_1$) are given in Tables S6 and S7. Tables S4 and S7 and Fig. S22 can enable comparison of the bond distribution difference between the 300 K and 480 K structures. The room temperature results reproduce the previously published work [15]. Low temperature data are given in Table S8 (104 K).

Using either of the high-temperature single crystal structure solutions (I4/mmm or C2/m), it is found that the tilt angle in $Ca_3Mn_2O_7$ vanishes above $T_T$. Quantifying the result of Fig. 4(c), the $X_3^-$ tilt amplitude vanshes and hence, the electric polarization determined by the existence of both $X_3^-$ and $X_2^+$ distortion is prediceted to vansih abruptly near 400 K on heating. The single crystal results are consistent with the abrupt in-plane shift at the Ca site seen in the PDF measurement. Fig. 7 shows the 480 K structure (I4/mmm space group) while the the corresponding figure for C2/m space group is given in Fig. S20. Note that in both structural solutions the rotation amplitue ($X_2^+$ ) is nonzero at 480 K. The in-plane O sites have double occupancy indicating the presence of 50/50 domains (by volume) with positive and negative in plane rotations ($X_2^+$ mode) of the $MnO_6$ polyhedra but of the same magnitude (see structural results in Tables S3 and S5). It should be noted that the abrupt change in structure and loss of tilt seen in the single crystal measurements are consistent with first order behavior seen in the DSC and Raman measurements as well as with the abrupt changes in structure found in the PDF refinements. The electric polarization in $Ca_3Mn_2O_7$ is expected to vanish abruptly near $T_T$ while that in $Ca_3Ti_2O_7$ is expeceted to vanish continusly after passing throug $T_T$.

## VII. High-Pressure Raman Measurements

To understand the softness of the tilting configuration in the $Ca_3Mn_2O_7$ system quantitatively, high-pressure Raman scattering measurements were conducted. In Fig.8, data are shown for the pressure range 1 to 13.4 GPa for the same crystal orientation as in Fig. 3. The feature near 140 cm$^{-1}$, labeled A (Fig. 8(a)), is seen to vanish between 1.4 and 2.3 GPa. In temperature-dependent ambient pressure Raman measurements (Fig. 2), this feature was associated with the loss of tilt and transition into a nonpolar phase as indicated by the combined structural and optical measurements. We find that at very low pressures, the system goes from the $Cmc2_1$ space group into a nonpolar space group (C2/m). Between 4.7 and 5.8 GPa (inset, Fig. 8(a)) the peak near 380 cm$^{-1}$ splits into two components and an additional feature near 190 cm$^{-1}$ onsets and grows larger as pressure rises (indicating symmetry reduction). Fig. 8(b) and the corresponding inset again show the disappearance of the same feature between 1.4 and 2.3 GPa (near 460 cm$^{-1}$) which vanishes at $T_T$ in the ambient pressure data (Fig. 3).

## VIII. Discussion of Combined Results

Development of an intuitive approach for the origin of the softness of the tilt mode is possible via simulations. The DFT derived self-force constants [16] (on site terms in force constant matrix which indicate the force on the isolated atom with respect unit displacements) are presented in Table I for each unique site the directions are labeled by the unit cell in Fig. 1 (for both $Ca_3Mn_2O_7$ and $Ca_3Ti_2O_7$). The force constants of the Mn sites are seen to be approximately the same for displacements along a, b, and c axis for the Mn sites. They are also significantly larger than those of any other atomic sites. Combined with typically strong Mn-O bonds, the results predict rigid $MnO_6$ polyhedra as exemplified by the weak temperature dependence of the Mn-O XAFS peak (Fig. 4).

For the apical O sites, the force constants in the yz plane are smaller than those for motion transverse to this plane ($k_y,k_z \sim 1/3\ k_x$ (long axisThe same holds for the apical interior O atoms. On the other hand, the planar (O3, O4) atoms have $k_y,k_z \sim 2\ k_x$. This predicts that local tilting (Fig. 1(b)) of the $MnO_6$ polyhedra corresponding mainly to y/z displacement of the apical O and x displacement of the planar O atoms will be softer than rotation of the $MnO_6$ polyhedra about the x-axis (long axis). In the case of the rotation, the planar O atoms have $k_y,k_z \sim 2\ k_x$ making this type of distortion relatively stiffer. The same trend was found for self-force constants of the $Ca_3Ti_2O_7$ system. (The force constants are rhobust respect to cell size and U/J parameters. See Table I and supplementary document Tables S9 to S11.)

Based on these arguments, it is predicted that <u>local tilting</u> of single $MnO_6$ polyhedron in the lattice, as opposed to coherent long-range tilting of all $MnO_6$ polyhedra simulated in previous work [3], is a lower energy distortion than <u>local rotation</u> in both $Ca_3Mn_2O_7$ and $Ca_3Ti_2O_7$**.** Also, in contrast with the current work, Ref. [9] suggests a difference in behavior between $Ca_3Mn_2O_7$ and $Ca_3Ti_2O_7$ with enhancement of both antiphase tilt and in-plane rotation occurring with increasing temperature in $Ca_3Ti_2O_7$, this is not observed in this work.

In addition to the possibility of local tilt or local rotation structural phases being the cause of the observed order of the transitions, we note that the relative stability of the $X_3^-$ and $X_2^+$ distortions is reversed by compressive biaxial strain of ~1.5 % [3] compared to the strain free or tensile strained systems under DFT+U. This strain value is at the level of accuracy for the determination of lattice parameters found in standard DFT simulations. It would be useful to explore more accurate methods than DFT+U to determine the stable structural phases.

## IX. Summary

We note that the $Ca_3Mn_2O_7$ and $Ca_3Ti_2O_7$ exhibit a transition near 400 K with verified loss of inversion center in the case of $Ca_3Mn_2O_7$ (for temperatures below ~400 K). The characteristic feature in these systems is the low energy local tilting of the Mn/Ti-$O_6$ polyhedra groups. To sustain the electric polarization at higher temperatures, it is necessary to strengthen the bonds on the Ca site (A site) in these systems. We also note that the low isotropic pressure needed to suppress the polarization state (1-2 GPa) will be significantly reduced if uniaxial pressure is applied. This may make it possible to switch the electrical polarization on and off by depositing this material as a thin film on a piezoelectric substrate for films strained just below this critical strain value. The measurements also suggest that simulations beyond standard DFT+U may be needed to determine the ordering of the energetics of the distortions.

## X. Acknowledgments

This work is supported by NSF Grant No. DMR-1809931. Work at Rutgers University is supported by DOE Grant No. DE-GF02-07ER46382. This research used beamlines 8-ID, 22-IR-1 and 28-ID-2 of the National Synchrotron Light Source II, a U.S. Department of Energy (DOE) Office of Science User Facility operated for the DOE Office of Science by Brookhaven National Laboratory under Contract No. DE-SC0012704. Single crystal x-ray diffraction measurements were performed at NSF's ChemMatCARS Sector 15, which is principally supported by the National Science Foundation/Department of Energy under Grant NSF/CHE-1346572. Use of the Advanced Photon Source was supported by the U.S. Department of Energy, Office of Science, Office of Basic Energy Sciences, under Contract No. DE-AC02-06CH11357. The Physical Properties Measurements System was acquired under NSF MRI Grant DMR-0923032 (ARRA award). This research used resources of the National Energy Research Scientific Computing Center, a DOE Office of Science User Facility supported by the Office of Science of the U.S. Department of Energy under Contract No. DE-AC02-05CH11231. We are indebted to Prof. E. A. Nowadnick at NJIT for critical discussions and advice on simulations and central experiments.

**Table I. $Ca_3Mn_2O_7$ Self-Force Constants***

| Ion Site | $k_x$ (long axis) (eV/Å$^2$) | $k_y$ (eV/Å$^2$) | $k_z$ (eV/Å$^2$) |
|---|---|---|---|
| **$Ca_3Mn_2O_7$** (1x2x2) cell | | | |
| Mn | 26 | 28 | 28 |
| Ca1 | 8.2 | 8.2 | 8.0 |
| Ca2 | 11 | 7.9 | 7.2 |
| O1 (apical interior) | 21 | 9.0 | 5.9 |
| O2 (apical) | 17 | 7.0 | 5.7 |
| O3 | 7.1 | 15 | 16 |
| O4 | 7.5 | 14 | 15 |

*For $Ca_3Mn_2O_7$ U=8.0 eV and J=1.0 eV (see supplementary document).

## Figure Captions

**Fig. 1**. (a) Low-temperature unit cell indicating unique O sites and (b) the $MnO_6$ polyhedral rotation about the long axis ($X_2^+$, black curved arrow) and the tilts ($X_3^-$, blue arrows) of the polyhedra normal to the long axis. (c) Partial density of states derived from DFT simulations showing the Mn, Ca and O site projected components. Note that the phonon DOS corresponding to the Ca sites vanishes for energies above ~ 350 $cm^{-1}$.

**Fig. 2.** (a) Raman spectra of a single crystal for photon beam normal to the long axis for data taken between 100 and 500 K. Note that the weak peak near ~146 $cm^{-1}$ vanishes as temperature increases (measured on cooling from 500 K). The vertical line indicates the abrupt change in structure near 370 K (b) The weak peak (arrow in (a) ) area vs. temperature is displayed showing that it disappears above ~370 K. (c) comparison of the peak position vs. temperature for the main Raman peak for powder and single crystal samples. Data are for cooling unless both warming and cooling results are shown.

**Fig. 3.** DSC curves for $Ca_3Mn_2O_7$ (cooling rate=20 K/min) from single crystal materials. Multiple scans reveal the high reproducibility of the transitions in both systems. The hysteresis in the complete loops (see offset transitions here at 405 K on warming and 367 K on cooling) reveal that the transition in $Ca_3Mn_2O_7$ is first order.

**Fig. 4**. XAFS structure function of $Ca_3Mn_2O_7$ measured between 320 K and 1100 K (in (a) and (b)) revealing that the Mn-Mn first and second nearest neighbor in plane correlations persist for the full temperature range. In (b) the inset shows that there is an abrupt reduction of the position of the Mn-O peak near 1060 K indicating the transition to the I4/mmm high symmetry structure.

**Fig. 5.** Local structure results derived from XAFS measurements. In the high-temperature region the (a) Mn-O and (b) Mn-Ca bond lengths reveal abrupt changes near 1060 K.

**Fig. 6.** Intermediate structure results derived from PDF measurements. (a) Temperature-dependent volume (cooling data) with b axis length shown as inset (warming and cooling). Clear transitions are seen near ~350 K on cooling and near ~400 on warming (b-axis). The volume data also reveal a transition near 105 K which is the magnetic ordering temperature (see also heat capacity data in Fig S1). (c) Position of the Ca2 and Mn ions vs. temperature revealing that only the Mn site position changes near 105 K and both Mn and Ca sites change near 370 K (cooling run). Motion of ions going from 300 K to 400 K. The arrows indicate exclusively a tilt rotation about an axis normal to the long axis exclusively.

**Fig. 7.** Crystal structure at 480 K derived from single crystal diffraction data. Note that the tilt angle vanishes. The planar oxygen atoms exist in two positions with 80 % probability indicating multiple domain structure but the rotation about the local z-axis is nonzero relative to the cubic structure.

**Fig. 8.** High-pressure Raman measurements for the incident beam normal to (a) and parallel to (b) the long axis. High-pressure Raman data reveal a phase transition between 1.0 and 2.3 GPa indicated by suppression of features A and D in panels (a) and (b). The feature D is expanded in the inset to panel (b). Panel (a) shows a second transition onsetting near 5.8 GPa as feature labeled B also associated with the splitting of a peak near 380 $cm^{-1}$ indicated as feature C.

**Fig. 1.** Liu *et al.*

**(a)**

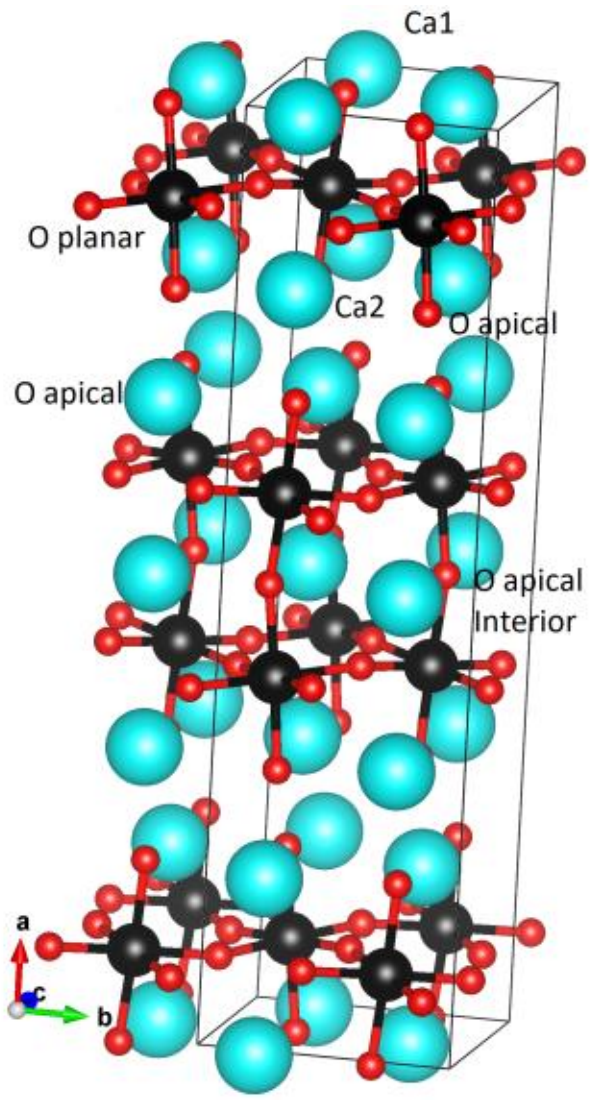

**(b)**

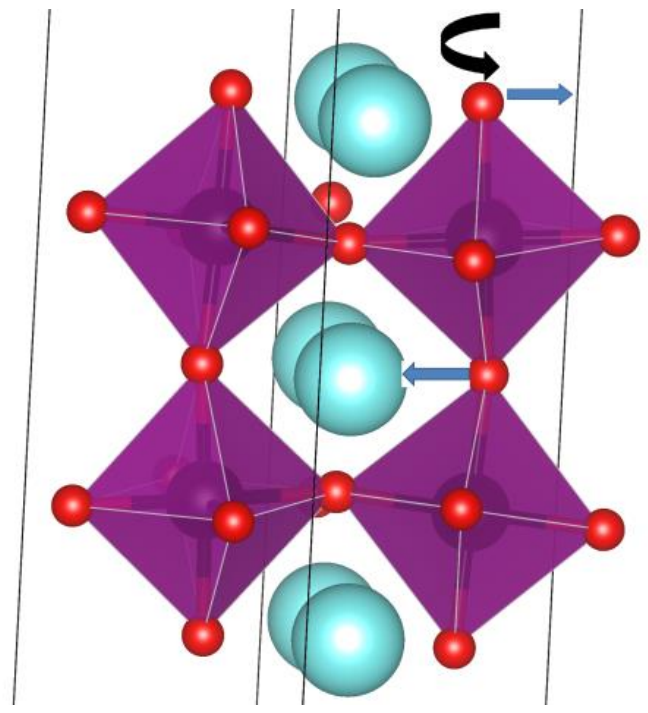

**(c)**

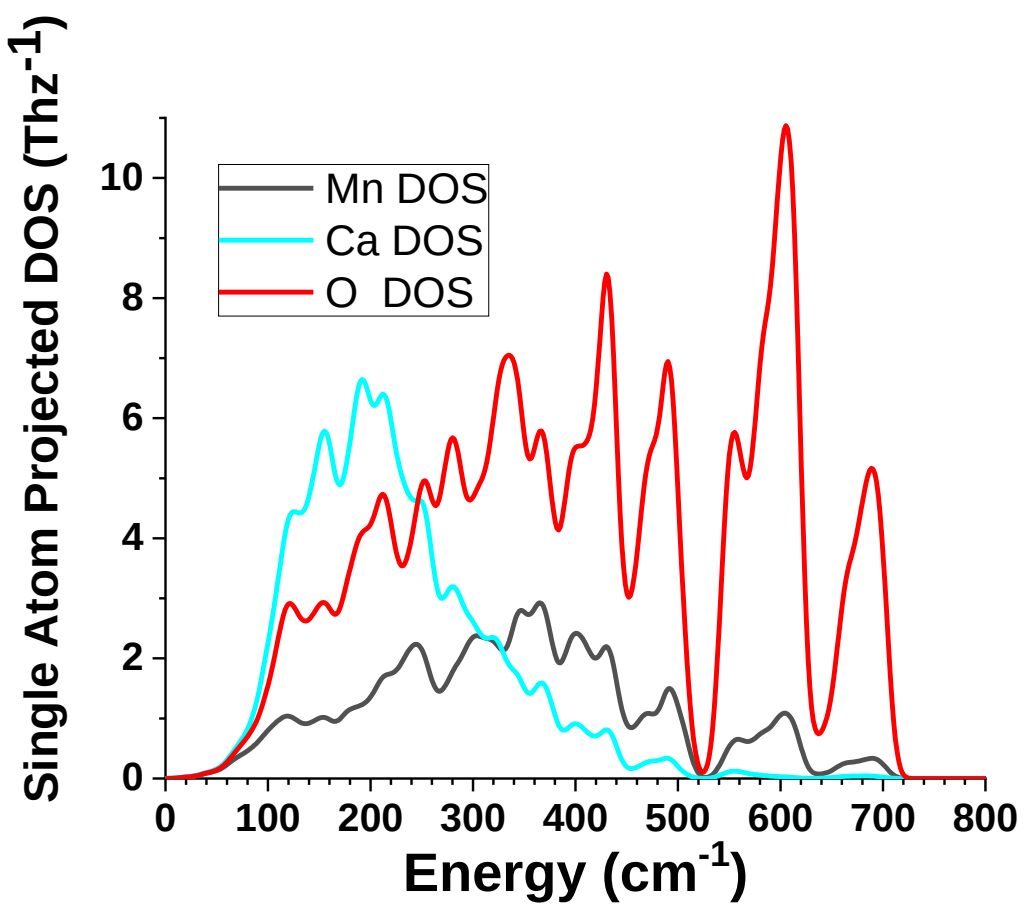

**Fig. 2** Liu *et al.*

**(a)**

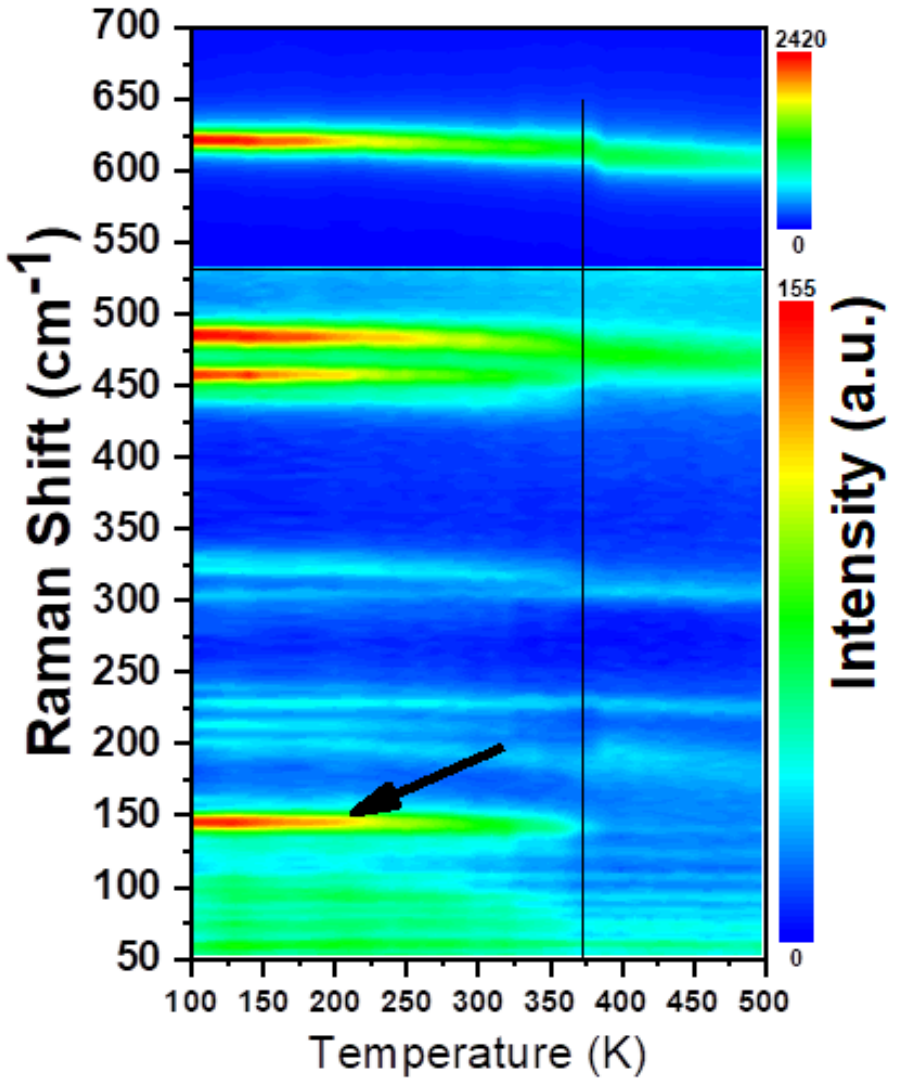

**(b)**

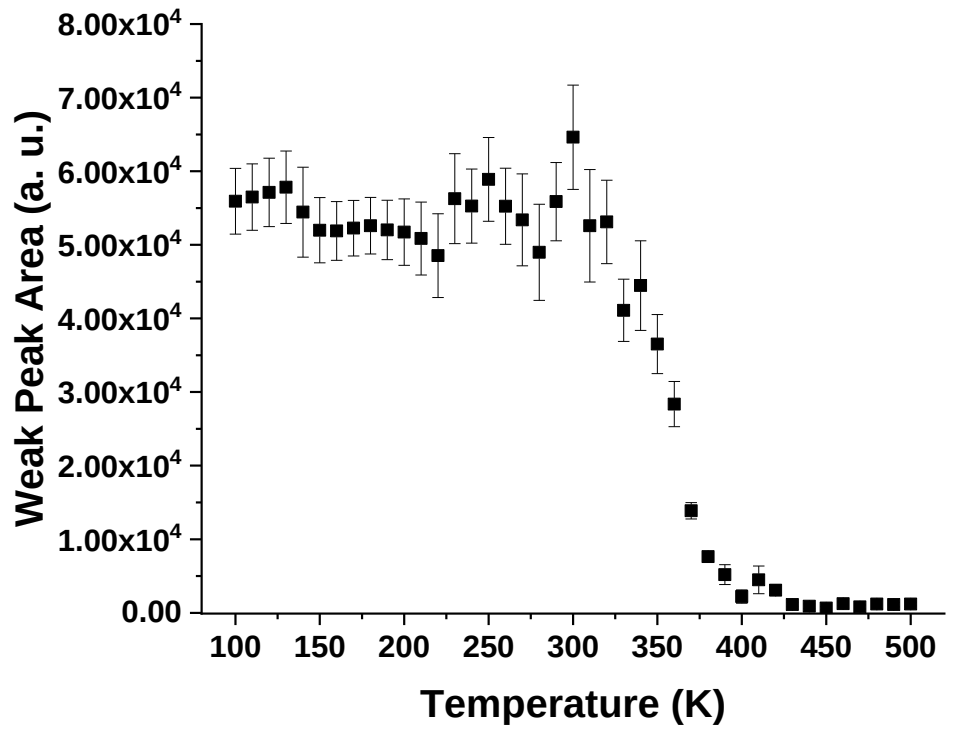

**(c)**

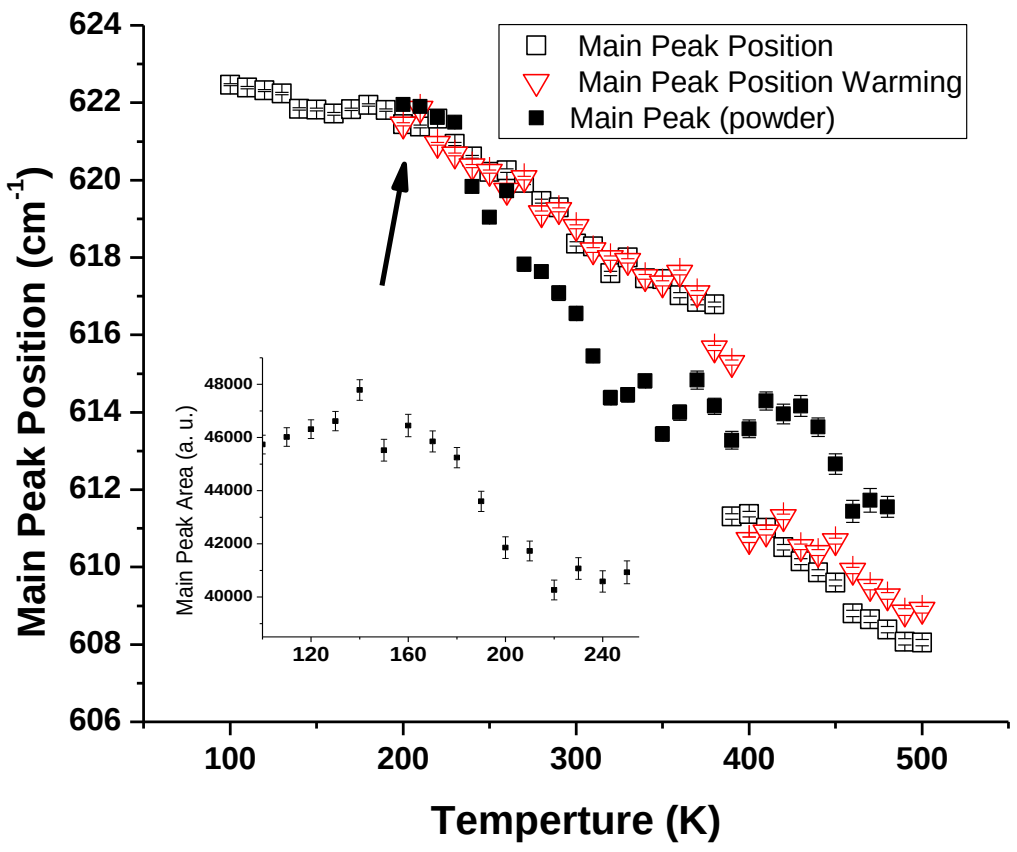

**Fig. 3.** Liu *et al*.

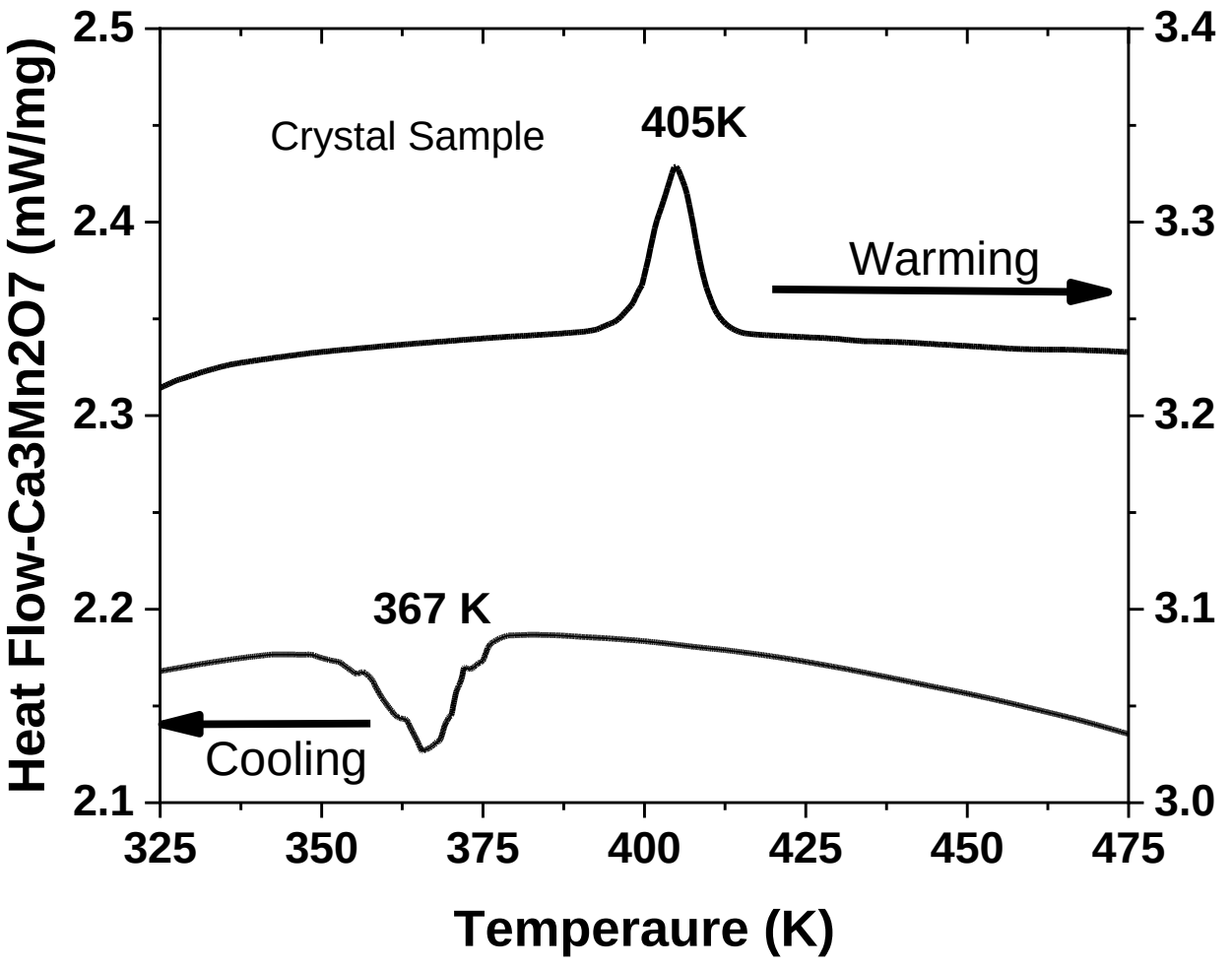

**Fig. 4.** Liu *et al.*

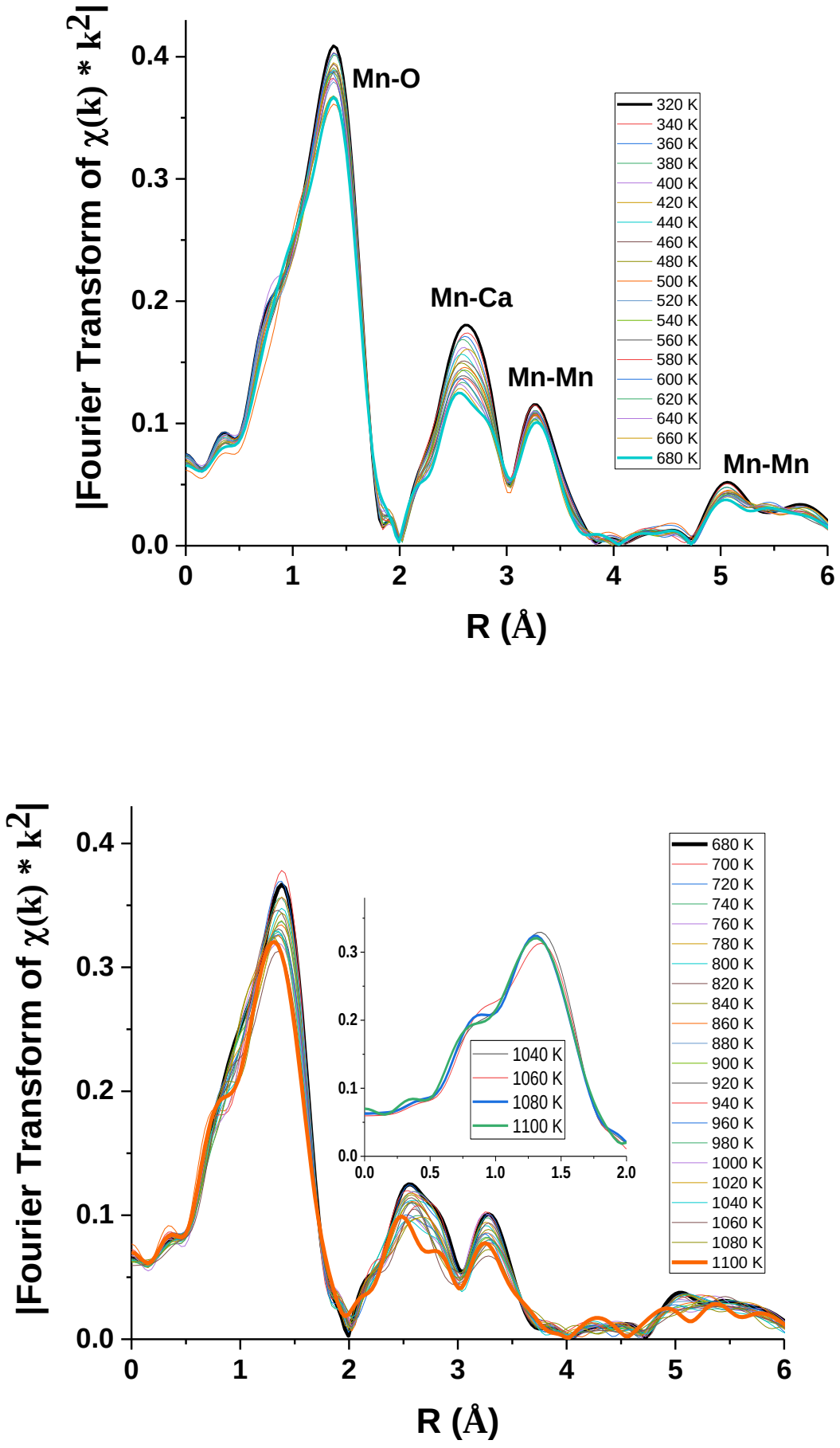

Fig 5. Liu *et al.*

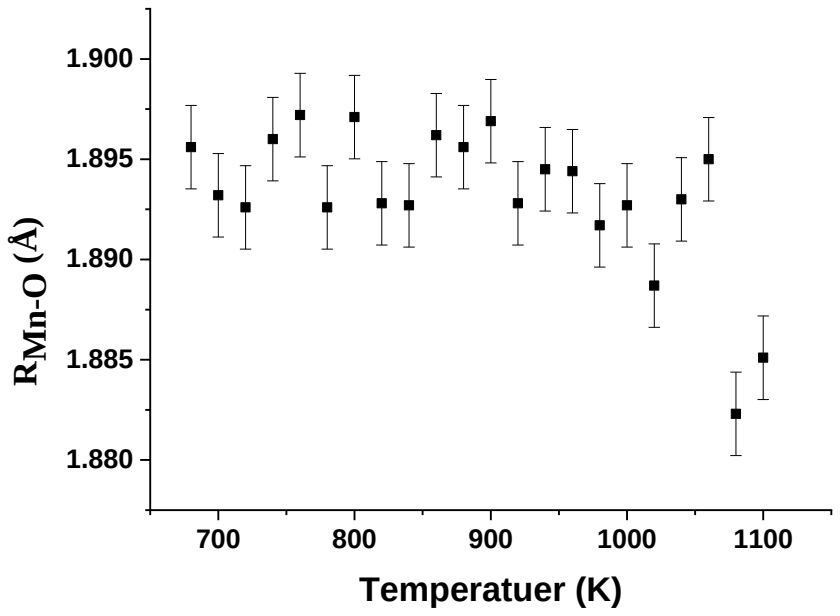

(a)

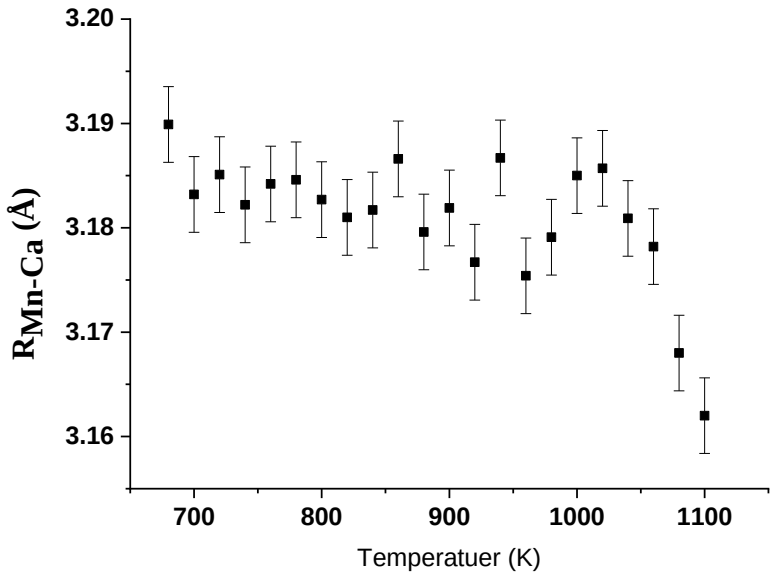

(b)

Fig 6. Liu *et al.*

**(a)**

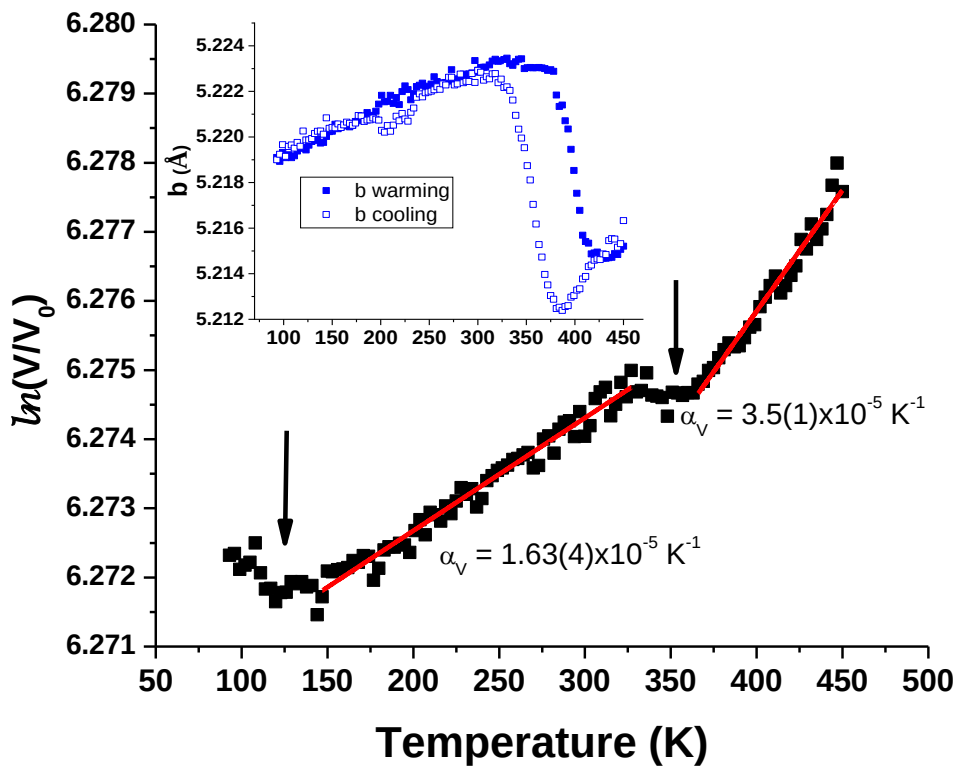

**(b)**

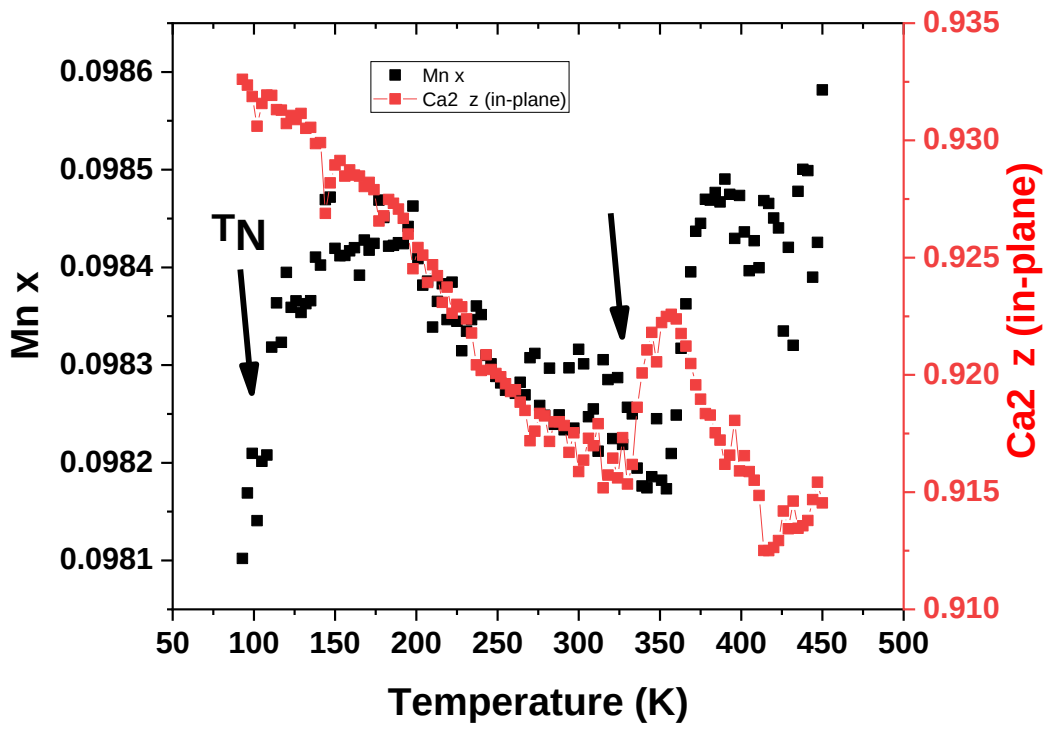

**(c)**

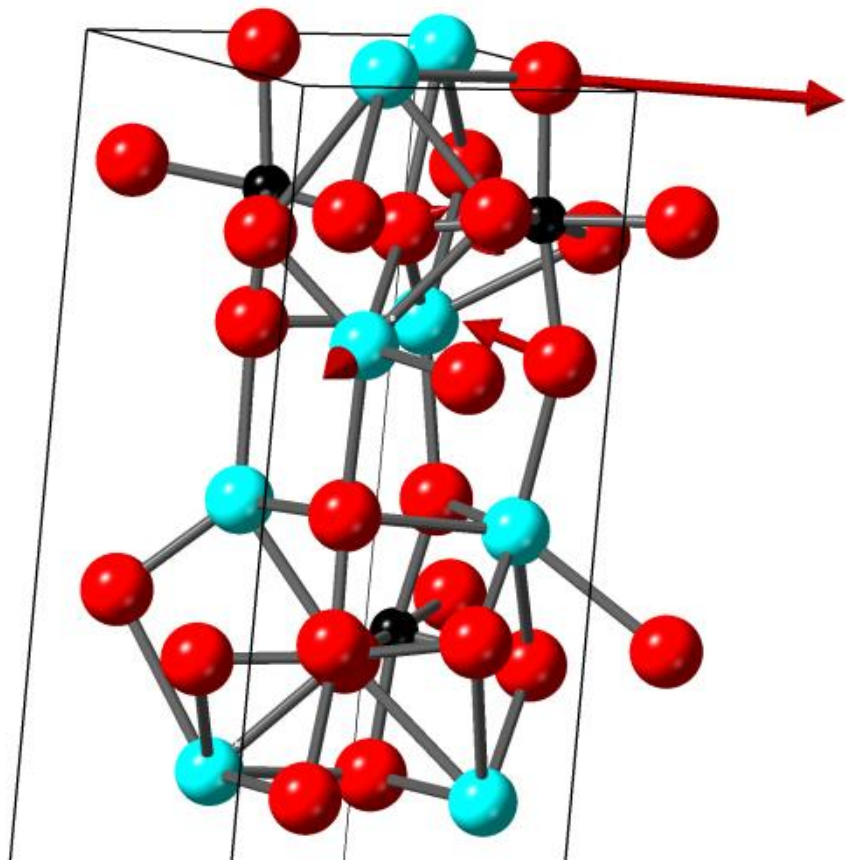

Fig 7. Liu *et al.*

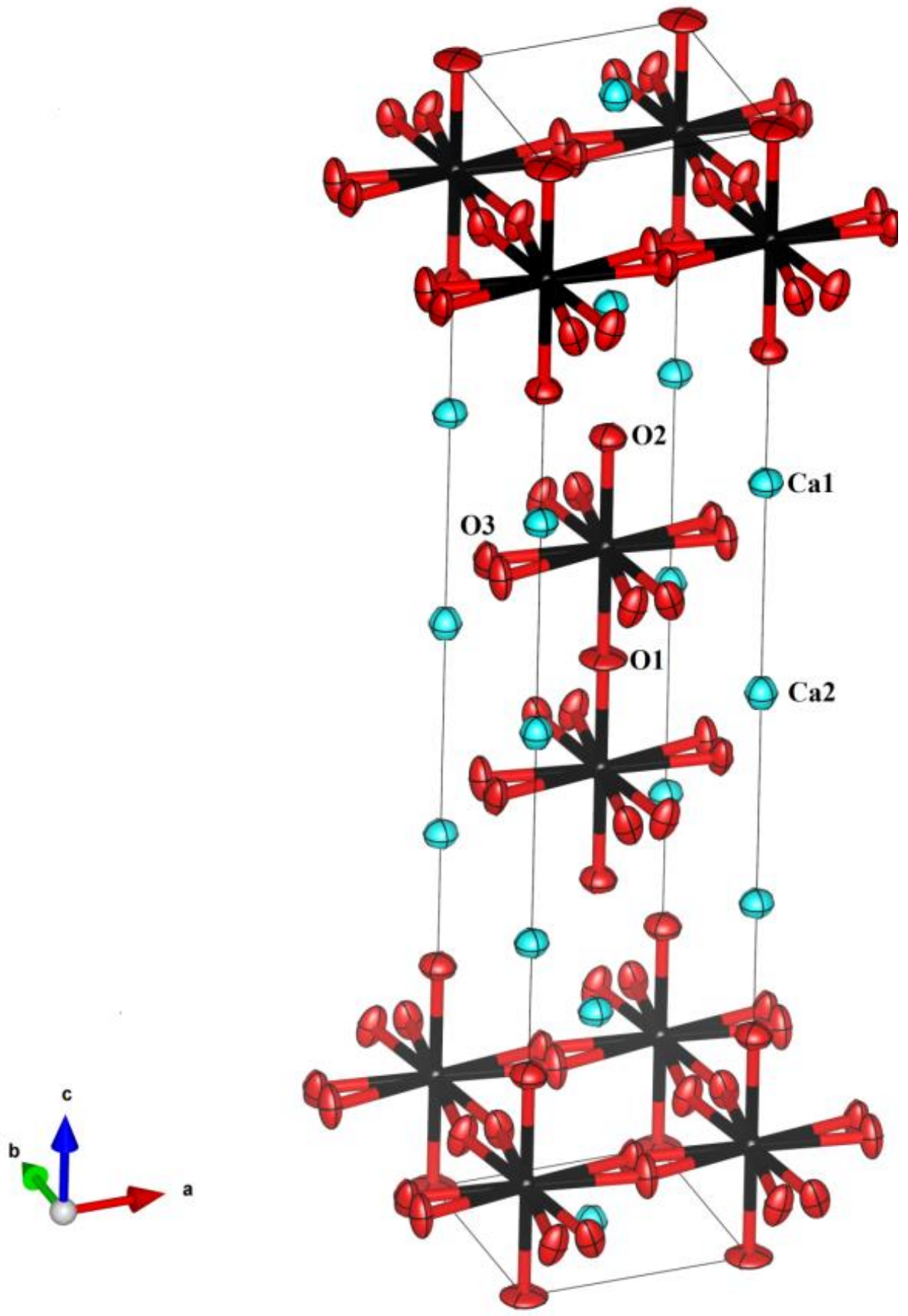

Fig. 8. Liu *et al.*

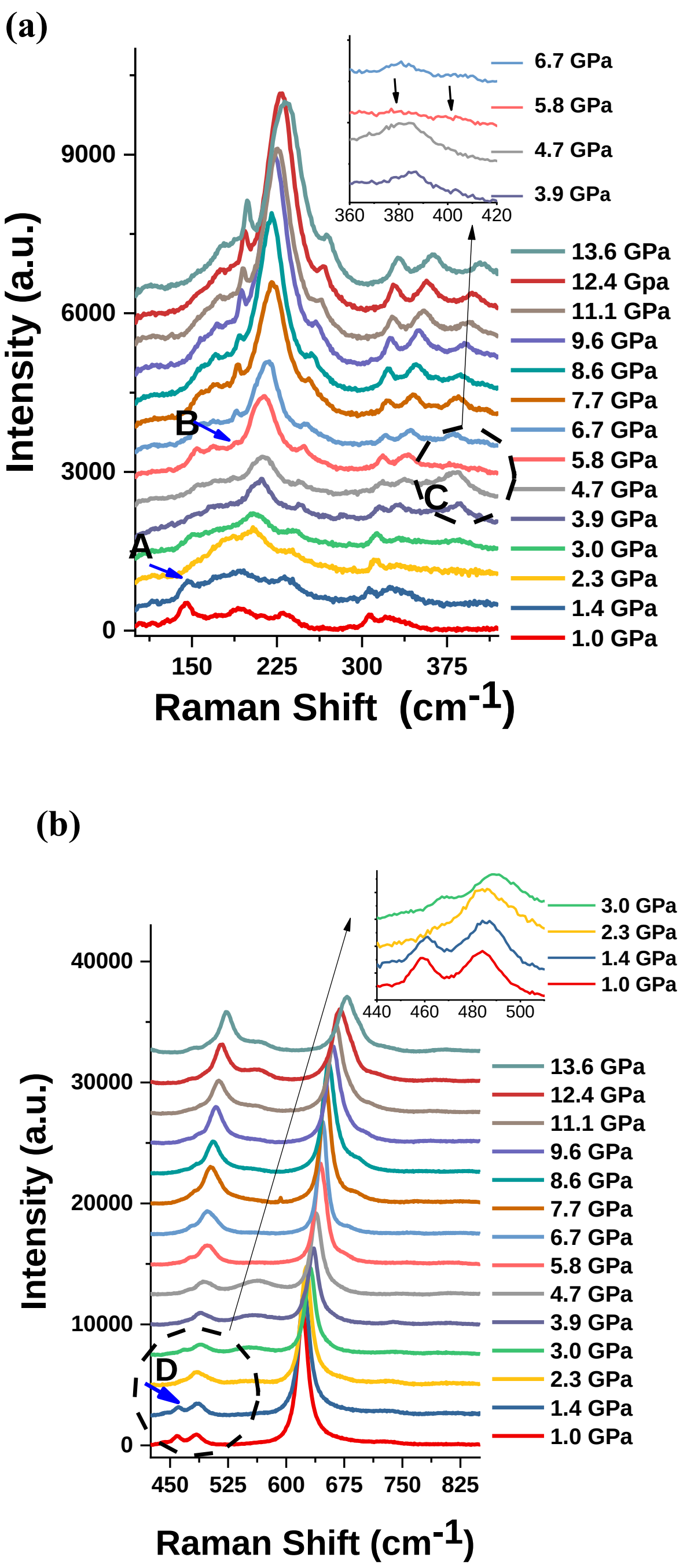

# Nature of the Structural Symmetries Associated with Hybrid Improper Ferroelectricity in $Ca_3X_2O_7$
# (Supplementary Document)

S. Liu[1], H. Zhang[1], S. Ghose[2,*], M. Balasubramanian[3], Zhenxian Liu[4], S. G. Wang[5], Y-S. Chen[5], B. Gao[6,7], J. Kim[6,7], S.-W. Cheong[6,7], and T. A. Tyson[1,7,*]
[1]Department of Physics, New Jersey Institute of Technology, Newark, NJ 07102
[2]National Synchrotron Light Source II, Brookhaven National Laboratory, Upton, NY 11973
[3]Advanced Photon Source, Argonne National Laboratory, Argonne, IL 60439
[4]Institute of Materials Science, Department of Civil and Environmental Engineering, The George Washington University, Washington, DC 20052
[5]Center for Advanced Radiation Sources, University of Chicago, Argonne, IL 60439
[6]Department of Physics and Astronomy, Rutgers University, Piscataway, NJ 08854
[7]Rutgers Center for Emergent Materials, Rutgers University, Piscataway, NJ 08854

*Corresponding Authors: T. A Tyson, e-mail: tyson@njit.edu
S. Ghose, e-mail: sghose@bnl.gov

Samples of $Ca_3Mn_2O_7$ and $Ca_{3-x}Sr_xTi_2O_7$ were prepared in single crystal form as given in Ref. [1]. For powder sample measurements, the crystal samples were crushed and sieved to obtain 500 mesh powders. Polycrystalline samples prepared by solid-state reaction were also examined for comparison, but the primary work is on materials derived from single crystal samples.

Specific heat measurements on single crystals were conducted on cooling from 390 K to 2 K and on warming from 60 K to 390 K using the relaxation method in a Quantum Design PPMS system. Temperature steps of 1 K were utilized, each temperature point was measured three times, and the average result is reported (Fig. S1). Approximately 10 minutes was required to measure each temperature data point. Hence the system was never in a quenched state. Differential scanning calorimetry measurements were conducted under flowing $N_2$ using a Perkin Elmer DSC 6000. Ambient pressure Raman Spectra were measured with an excitation wavelength of 532 nm in backscattering geometry using a Thermo Scientific DXR Raman Microscope. A 20x objective was used with 10 mW power setting. The sample was found to be stable under this laser power after test were done on a range of laser powers (3 to 10 mW). The original spectra were recovered after warming and cooling cycles. Both the Raman and DSC measurements were conducted at the NJIT York Center.

High-pressure Raman measurements were conducted at National Synchrotron Light Source (NSLS2) beamline 22-IR-1 National. Measurements were conducted in a symmetric diamond cell with (100) oriented diamonds. The culet size was 500 μm and tungsten gaskets were used. The pressure medium utilized was methanol:ethanol:water in a 14:3:1 ratio by volume. Pressure calibration was conducted using both ruby fluorescence mainline shifts [2] and the pressure-dependent changes in the diamond Raman edge (near ~1300 $cm^{-1}$) [3]. Pressure calibration measurements were made before and after each Raman spectrum was collected. In addition, calibration measurements as a function of position at multiple points in the cell at the highest pressure showed variation below 0.1 GPa indicating a high level of hydrostatic behavior of the pressure medium. The custom micro-Raman system at beamline 22-IR-1 consisted of a 532 nm solid-state laser (Spectral-Physics Excelsior, 150 mW), a Princeton Instruments liquid-nitrogen cooled PyloN CCD detector, a PI Acton SpectraPro SP-2556 Imaging

Spectrograph and a 50 X objective. Based on the ambient pressure laser power tests, a laser power was used which caused no change in sample spectra during the measurements. The accumulation time was 20 seconds and averaged 20 times. For all Raman measurements, no change in the spectra was observed over time at a given pressure.

Diffraction measurements on ~20 μm diameter crystals were conducted at the Advanced Photon Source (Argonne National Laboratory) beamline 15-ID-B with a wavelength of 0.41328 Å. The data were collected with a PILATUS 1M detector (maximum count rate = $10^7$ cps/pixel, counter depth =20 bit) between 300 K and 480 K (and at 104 K). Refinement of the data was done using the program Olex2 [4] after the reflections were corrected for absorption. Anomalous scattering corrections were induced for all atoms. Single crystal refinement results are presented in Tables S3 to S8.

To determine force constants and phonon DOS for $Ca_3Mn_2O_7$, spin density functional calculations in the projector augmented wave approach [5] were carried out utilizing the VASP code. Full structural optimization was conducted for both lattice parameters and atomic positions and the LDA+U approximation was implemented to obtain the relaxed structure. All calculations were conducted with ferromagnetically ordered Mn sites. The parameters U = 8.0 eV and J = 0.88 eV were used for the Mn on-site coulomb and exchange parameters (as done for $YMnO_3$ [6]). An energy cutoff of 500 eV was implemented. The ground state structure was optimized so that forces on each atom were below 1 x $10^{-5}$ eV/Å. A second set of simulations with U = 4.5 eV and J = 1.0 eV was also conducted (as done in Ref. [3] main text). For the 1x1x1 cell calculations a 2x4x4 gamma centered k-space grid was utilized and for the the 1x2x2 cell calculations a 2x2x2 gamma centered k-space grid was used.

Force constants were determined using density functional perturbation theory for a 1x1x1 cell with U = 8.0 eV and J = 0.88. In addition, force constants were calculated by the frozen phonon method with 1x2x2 cells for both U=4.5 eV with J =1.0 eV and for U=8.0 with J=0.88 eV. The lattice parameters derived for the U/J parameters sets were compared with single crystal data taken at 104 K (Table S8) and indicate that U = 8.0 eV and J = 0.88 eV gives a structure closest to experimental values (Table S9). The atom projected phonon DOS curves for all simulations are given in Fig. S22. We note that, qualitatively, the shaped of these integrated DOS spectra and the corresponded force constant are robust with respect to cell size and significant variations in J. The collected self-force constants (Table 1, Table S10 and S11), phonon modes (Table S1) and partial DOS (Fig. S22) are givin in this supplementary document. The force constants (Table 1), phonon modes (Fig. S1 and Table S1) and partial DOS (in Fig.1(c)) discussed in the main text are from the 1x2x2 cell for U=8.0 with J=0.88 eV. We also observe that the phonon DOS presented here for $Ca_3Mn_2O_7$ are quite similar to those of reference [7] for the $Ca_3Ti_2O_7$ system. This is not surprising since the mass of Mn and Ti are quite close ($(M_{Mn}/M_{Ti})^{1/2}$=1.07) and all other atoms are identical. The code Phonopy was utilized to determine the phonon density of states and phonon displacement mode from the force constants (see Fig 1, Fig. S1, Table I and Table S1) [8].

Mn and Ti K-edge XAFS spectra were collected at NSLS2 beamline 8-ID (ISS). Reduction of the x-ray absorption fine-structure (XAFS) data was performed using standard procedures [9]. Measurements were done in fluorescence mode with 500 mesh powders held in 1 mm quartz capillaries with 10 μm walls. Data were corrected for self-absorption. In the XAFS refinements, to treat the atomic distribution functions on equal footing, the Mn and Ti K-edge spectra were modeled in R-space by optimizing the integral of the product of the radial distribution functions and theoretical spectra with respect to the measured spectra. Specifically, the experimental spectrum is modeled by, $\chi(k) = \int \chi_{th}(k,r)\, 4\pi r^2 g(r) dr$ where $\chi_{th}$ is the theoretical spectrum and g(r) is the real space radial distribution function based on a sum of Gaussian functions (χ(k) is measured spectrum) [10] at each temperature (as in Ref. [11]). For each shell fit, the coordination number (N) was held at the crystallographic value but the position (R) and Gaussian width (σ) was fit to the data. For the Mn K-Edge, the k-range 2.50 < k < 12.1 $Å^{-1}$ and the R-range 0.63 < R < 3.80 Å were utilized. Coordination numbers for the atomic shells were fixed to the crystallographic values. For the Mn K-Edge $S_0^2$ (accounting for electron loss to multiple excitation channels) value was fixed at 0.81. The Gaussian widths and positions were fit for each component. Ca K-edge XAFS were collected at APS beamline 20

BM at Argonne National Laboratory. The measurements were done at 300 K, 350 K and 450 K. The k-range 2.39 < k < 10.4 $Å^{-1}$ and the R-range 1.02 < R < 3.40 Å were utilized in the fits.

The theoretical Ti K-edge x-ray near edge spectra (XANES) presented were computed, as in Ref. [12] using the RELXAS programs [13]. The potentials were computed for cubic perovskite structures, based on clusters containing the first 118 atoms surrounding the Ti site. A simple cluster based on $CaTiO_3$ was used to determine the changes in the XANES pre-edge structure on tilting and rotation of the $TiO_6$ polyhedra.

Pair distribution function experiments for all solid state reaction prepared powder samples were conducted at beamline the XPD-2 (28 ID) beamline at Brookhaven National Laboratory's National Synchrotron Light Source II using a wavelength $\lambda$ = 0.18372 Å (67.486 keV). For the samples derived from single crystal $Ca_3Mn_2O_7$, measurements between 90 K to 450 K were conducted at APS beamline 6 ID-D at Argonne National Laboratory ($\lambda$ = 0.12368 Å (100.25 keV)). A high-temperature measurement set on crystal derived $Ca_3Mn_2O_7$ powder between 300 K and 1100 K was conducted at NSLS2 beamline XPD-2 at 28 ID. All measurements utilized Perkin Elmer Area detectors with a sample to detector distance of ~200 mm. Exact detector to sample distances were derived by fits to Ni powder calibration standards. The range $Q_{mim}$ = 0.23 $Å^{-1}$ and $Q_{max}$ = 25.1 $Å^{-1}$ was used in data reduction. The methods utilized for analysis of the PDF data are described in detail in Refs. [1,14]. For the fits in R-space covered the range 1.3 < r < 20 Å was utilized. The time interval between temperature points was ~7 minutes. Combined with the small temperature steps, the approach kept the samples from being in a quenched state.

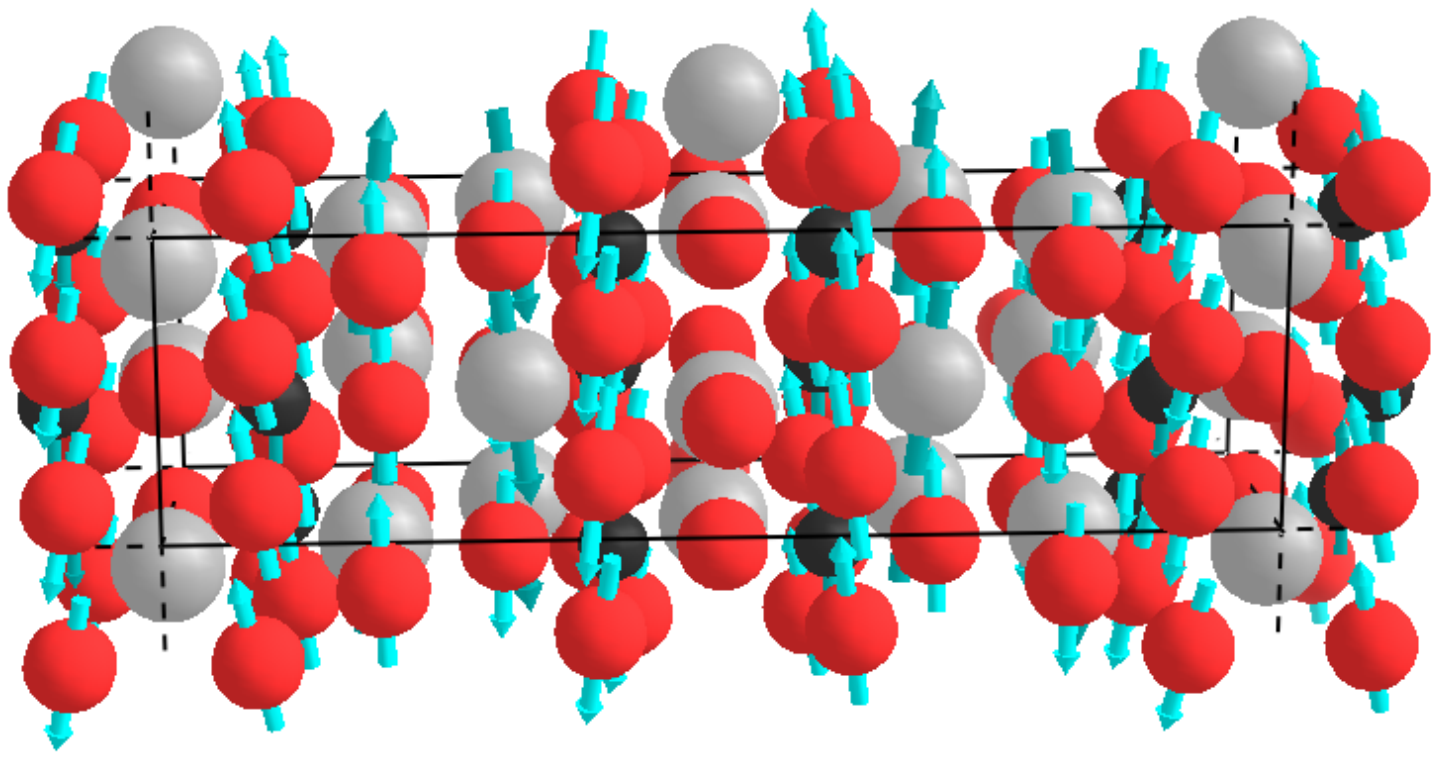

**(a)**

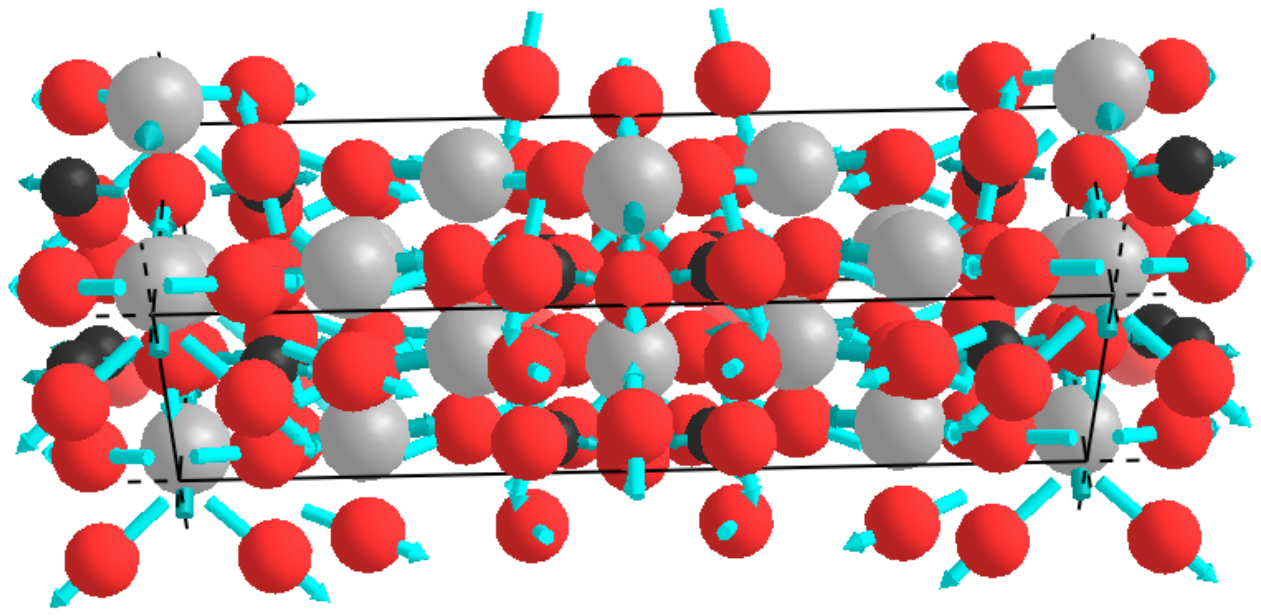

**(b)**

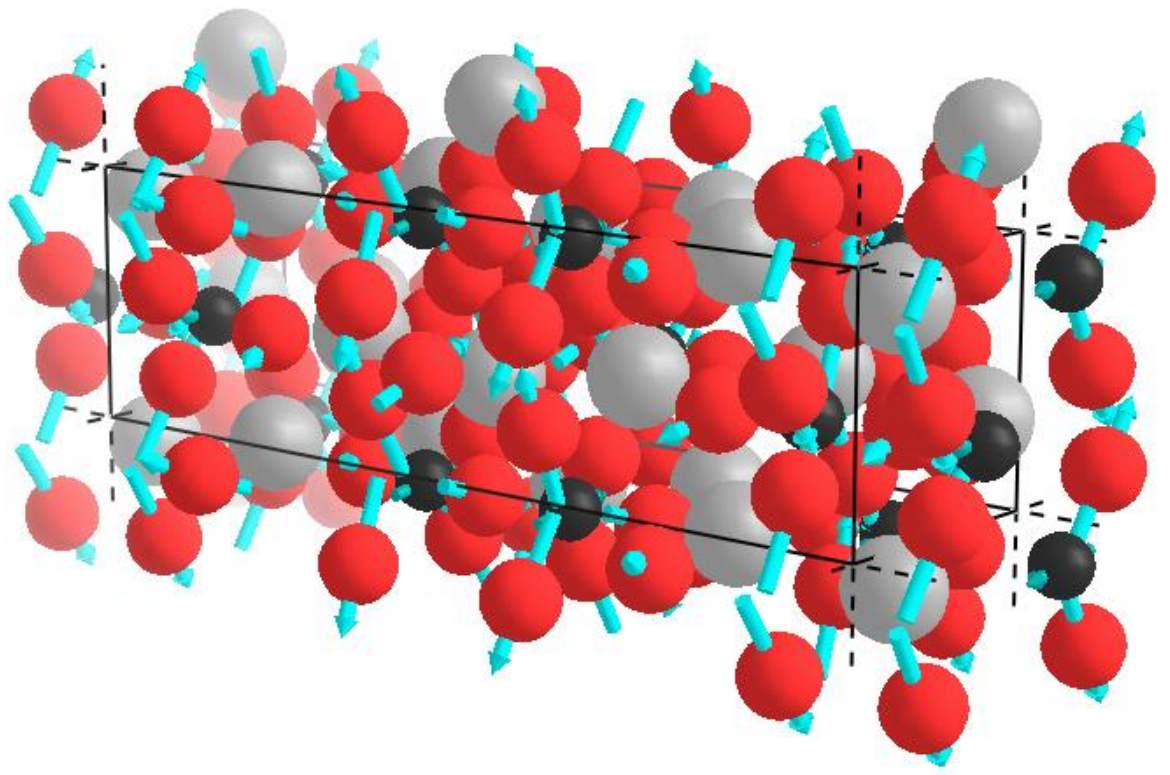

**(c)**

**Fig. S1**. Atomic motion for phonon modes from DFT simulations of $Ca_3Mn_2O_7$ for energies of (a) 128, (b) 141 and (c) 612 $cm^{-1}$. The black, grey and red symbols correspond to Mn, Ca and O atoms. Note that the low-frequency modes involve motion of the Ca and O ions only. The arrows indicated the atomic displacement directions for one half of the motion cycle (arrows change direction on the second half of cycle) with lengths indicating the corresponding amplitudes. The full set of phonon (gamma point) energies are given in Table S1.

# Table S1. Computed Phonon Energies for $Ca_3Mn_2O_7$ ($Cmc2_1$ Space Group)*

| Energy($cm^{-1}$) | Irreducible Rep. | Energy($cm^{-1}$) (cont.) | Irreducible Rep. |
|---|---|---|---|
| 125.3 | A2 | 338.9 | B2 |
| 127.9 | B1 | 344.8 | B1 |
| 141.2 | A1 | 350.3 | A1 |
| 158.9 | A1 | 370.7 | A2 |
| 160.6 | B2 | 372.8 | A1 |
| 169.2 | A2 | 385.1 | B2 |
| 172.7 | A1 | 403.1 | B1 |
| 178.0 | B1 | 414.9 | A2 |
| 191.4 | B2 | 422.6 | B2 |
| 193.8 | A2 | 435.4 | A1 |
| 197.5 | B1 | 437.4 | B1 |
| 211.5 | A2 | 437.6 | B2 |
| 225.2 | A1 | 446.7 | A2 |
| 230.0 | B2 | 452.4 | B2 |
| 233.1 | B1 | 456.5 | A1 |
| 237.0 | A2 | 466.3 | B1 |
| 237.4 | B1 | 486.2 | A2 |
| 239.4 | A1 | 493.3 | A1 |
| 241.4 | B2 | 493.6 | B2 |
| 244.9 | A1 | 540.5 | A2 |
| 245.8 | A2 | 546.4 | B1 |
| 254.2 | A1 | 555.1 | B2 |
| 262.0 | B2 | 575.8 | A2 |
| 274.8 | B1 | 590.8 | B1 |
| 275.6 | A2 | 595.9 | A1 |
| 277.4 | B2 | 605.0 | B1 |
| 280.1 | A1 | 605.7 | A2 |
| 290.8 | A2 | 606.0 | A1 |
| 296.5 | A1 | 612.3 | B2 |
| 297.5 | B1 | 615.9 | A1 |
| 312.9 | B2 | 644.7 | B1 |
| 319.8 | B1 | 684.3 | B2 |
| 322.9 | A1 | 697.1 | A2 |
| 324.6 | B2 | | |
| 333.7 | A2 | | |

*Phonons at gamma point based on 1x2x2 cell.

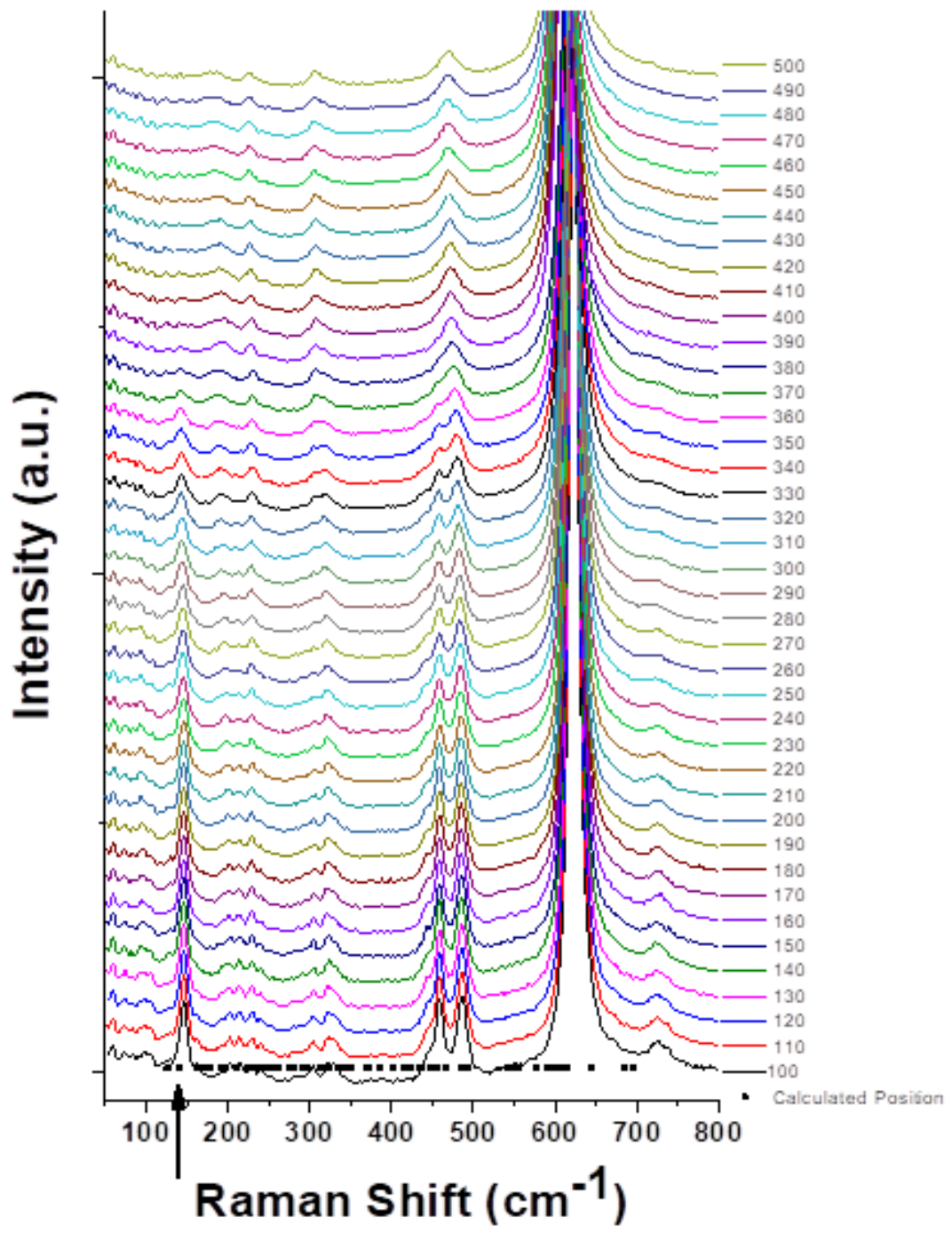

**(a)**

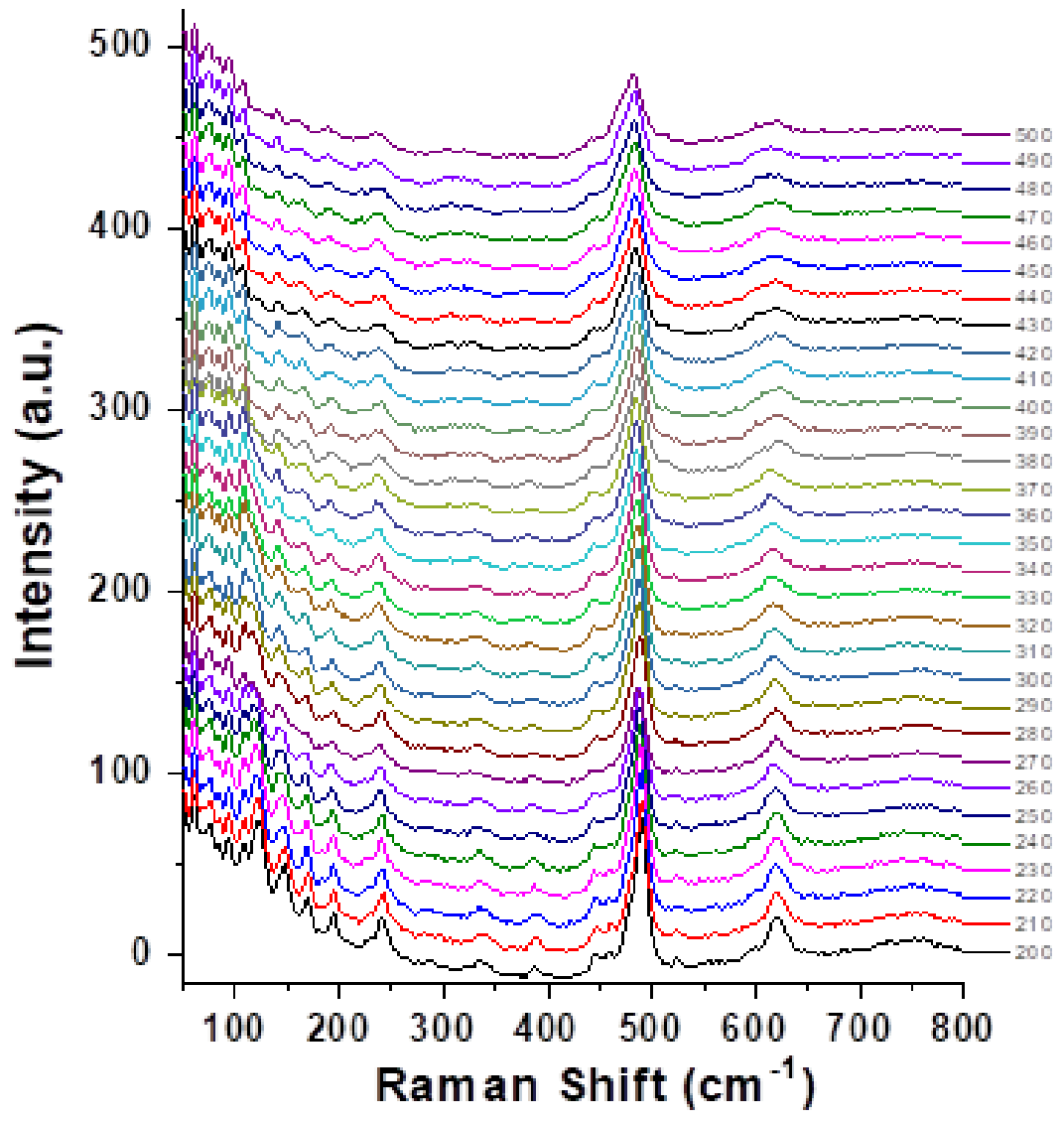

**(b)**

**Fig. S2**. Full Raman spectra for $Ca_3Mn_2O_7$ single crystal for beam normal to (a) and along (b) the long axis (in-plane). The full spectrum is given in (a). Note the reduction in amplitude of the feature near 130 $cm^{-1}$ indicated by the arrow. Data are from cooling curves.

\

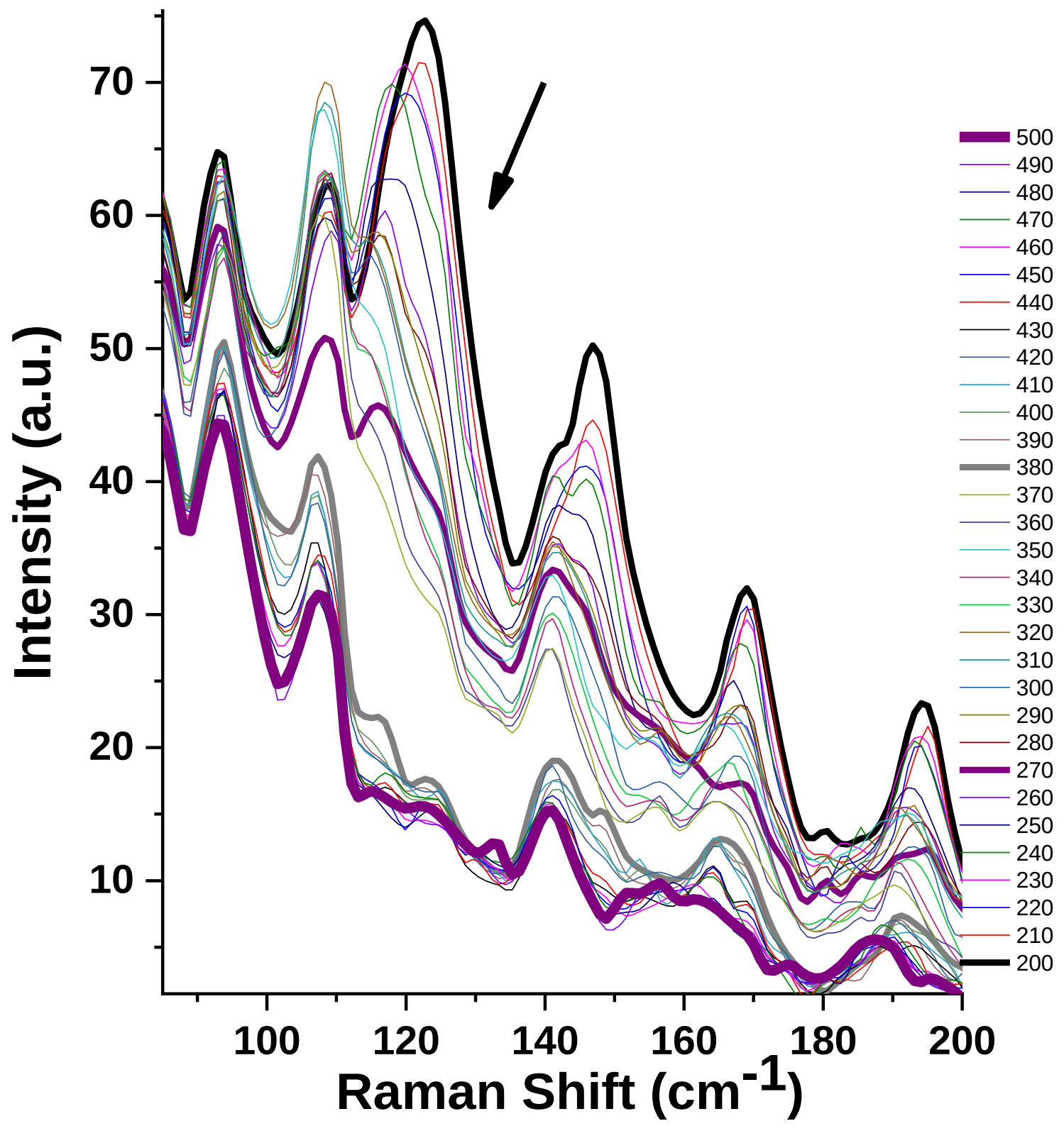

**Fig. S3**. Expanded low energy region Raman spectra for $Ca_3Mn_2O_7$ single crystal for beam along the long axis (in-plane). Note the reduction in amplitude of the feature near 120 $cm^{-1}$ indicated by the arrow.

**(a)**

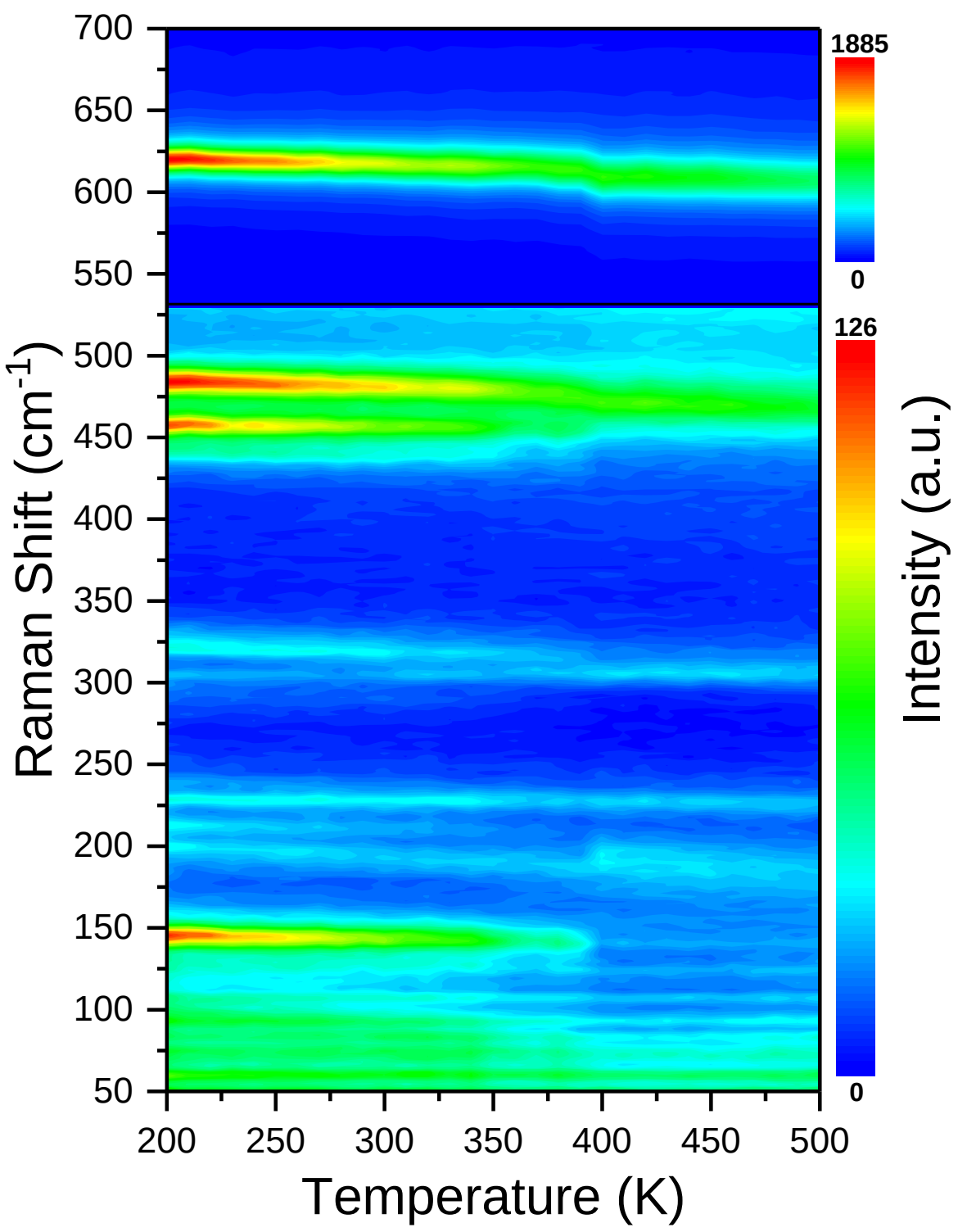

**(b)**

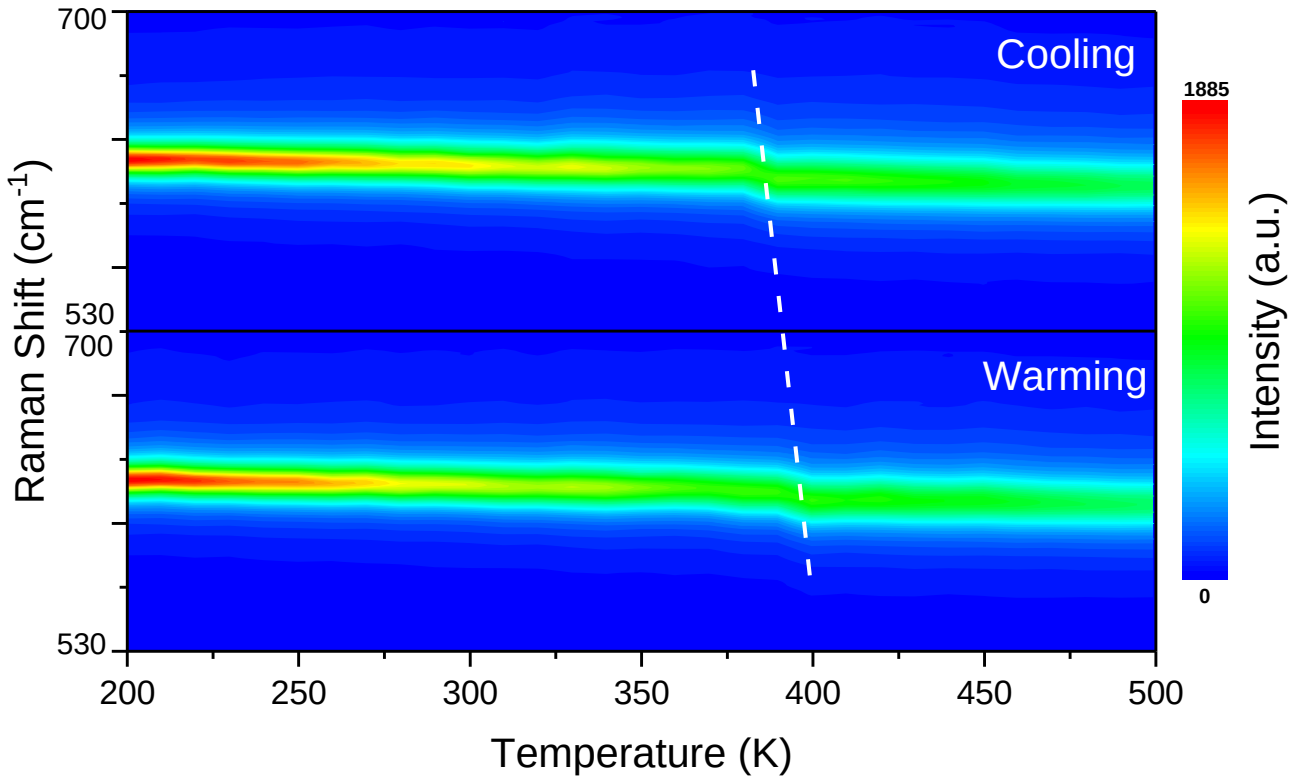

**Fig. S4**. (a) Surface plot for warming from 200 K to 500 K for incident beam normal to long axis. In (b), a surface plot is given for the intensity of the main peak vs. temperature for warming and cooling. As in Fig. 2, note the abrupt changes in the position of this peak near 370 K during cooling and near 400 K during warming.

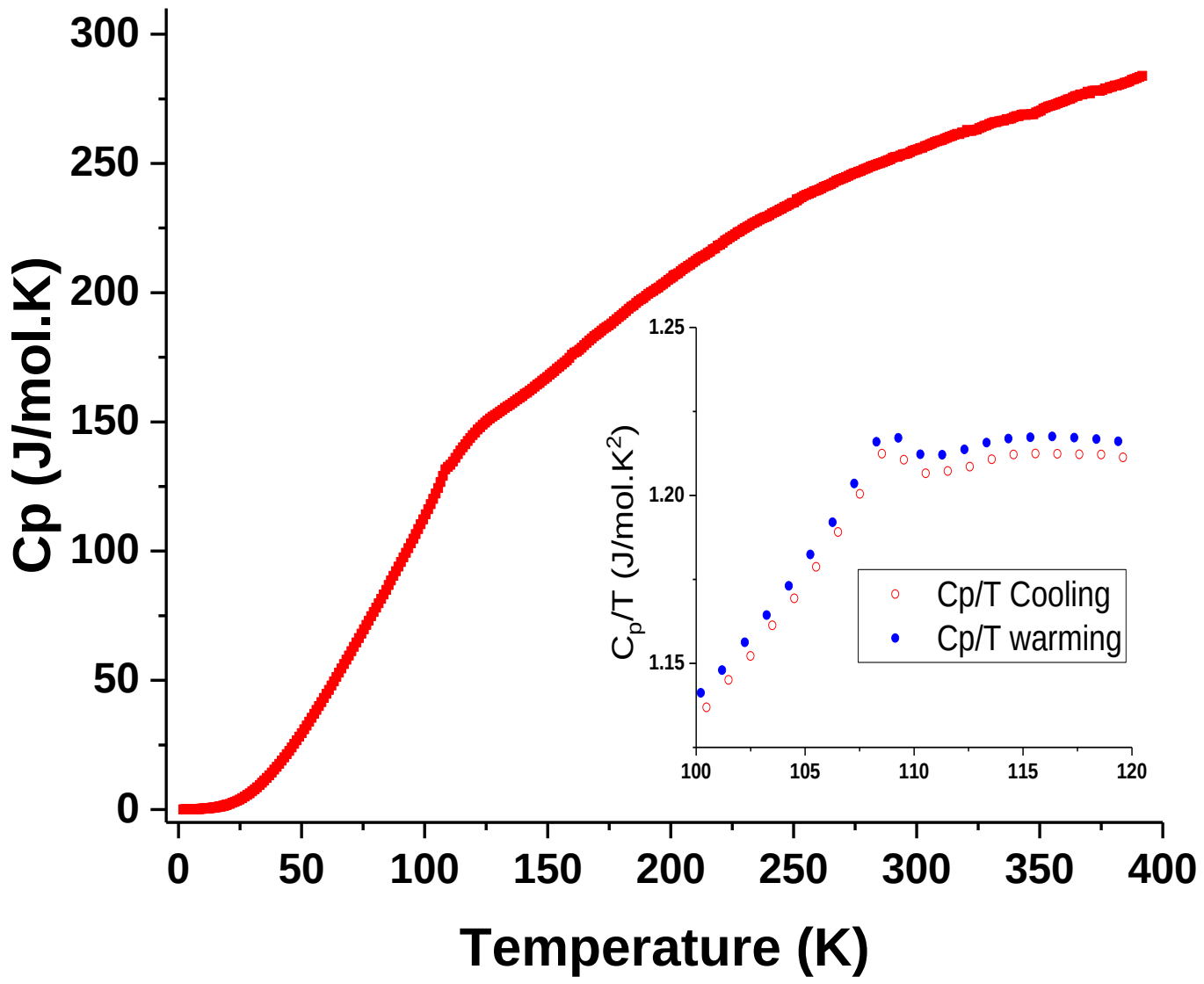

**Fig. S5**. The heat capacity of $Ca_3Mn_2O_7$ single crystal revealing the magnetic transition near ~108 K (see inset). Note that no additional clear features are seen up to the maximum temperature of 390 K.

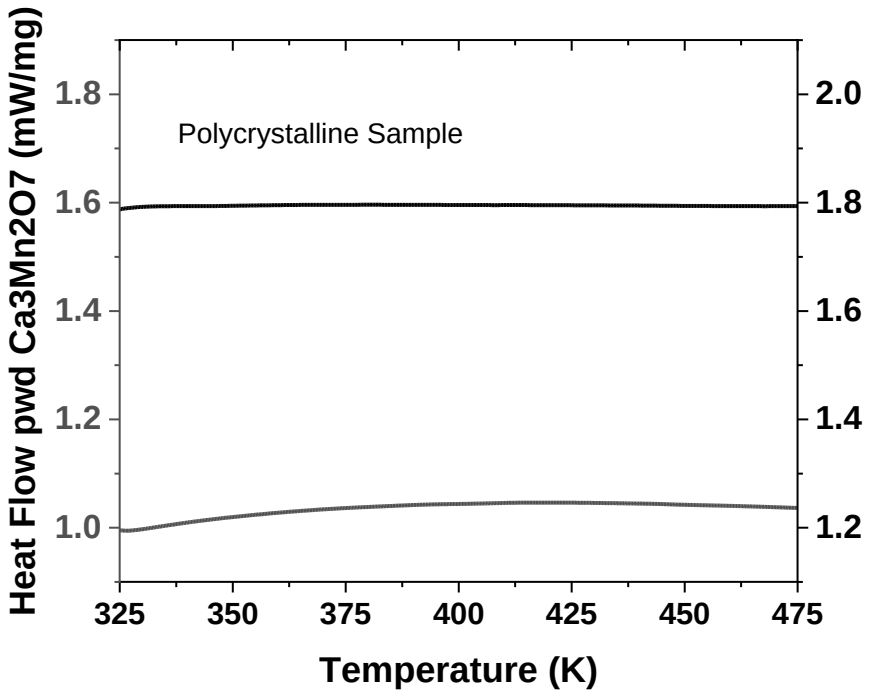

**(a)**

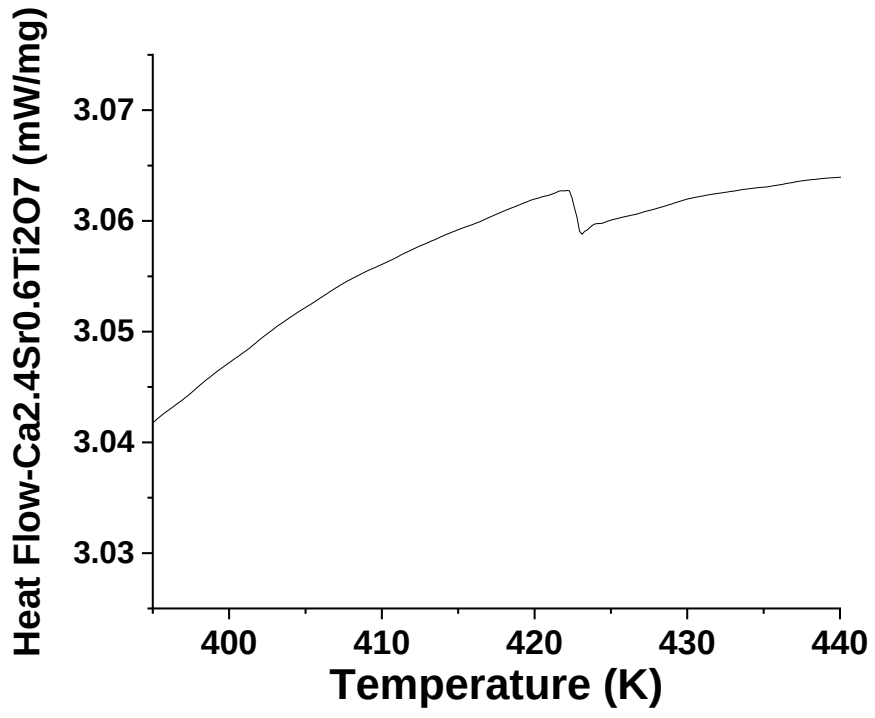

**(b)**

**Fig. S6**. (a) DSC curves for $Ca_3Mn_2O_7$ (cooling rate=20 K/min) from a polycrystalline sample (solid state synthesis) indicating no clear transition. Upper and lower curves have the same meaning as in (a). (b) Cooling curve DSC measurement for $Ca_{2.4}Sr_{0.6}Ti_2O_7$ crystals indicating a very weak feature with step change characteristic of a continuous or glass-like transition. Note that no background subtraction (of Al container) was done for samples.

(a)

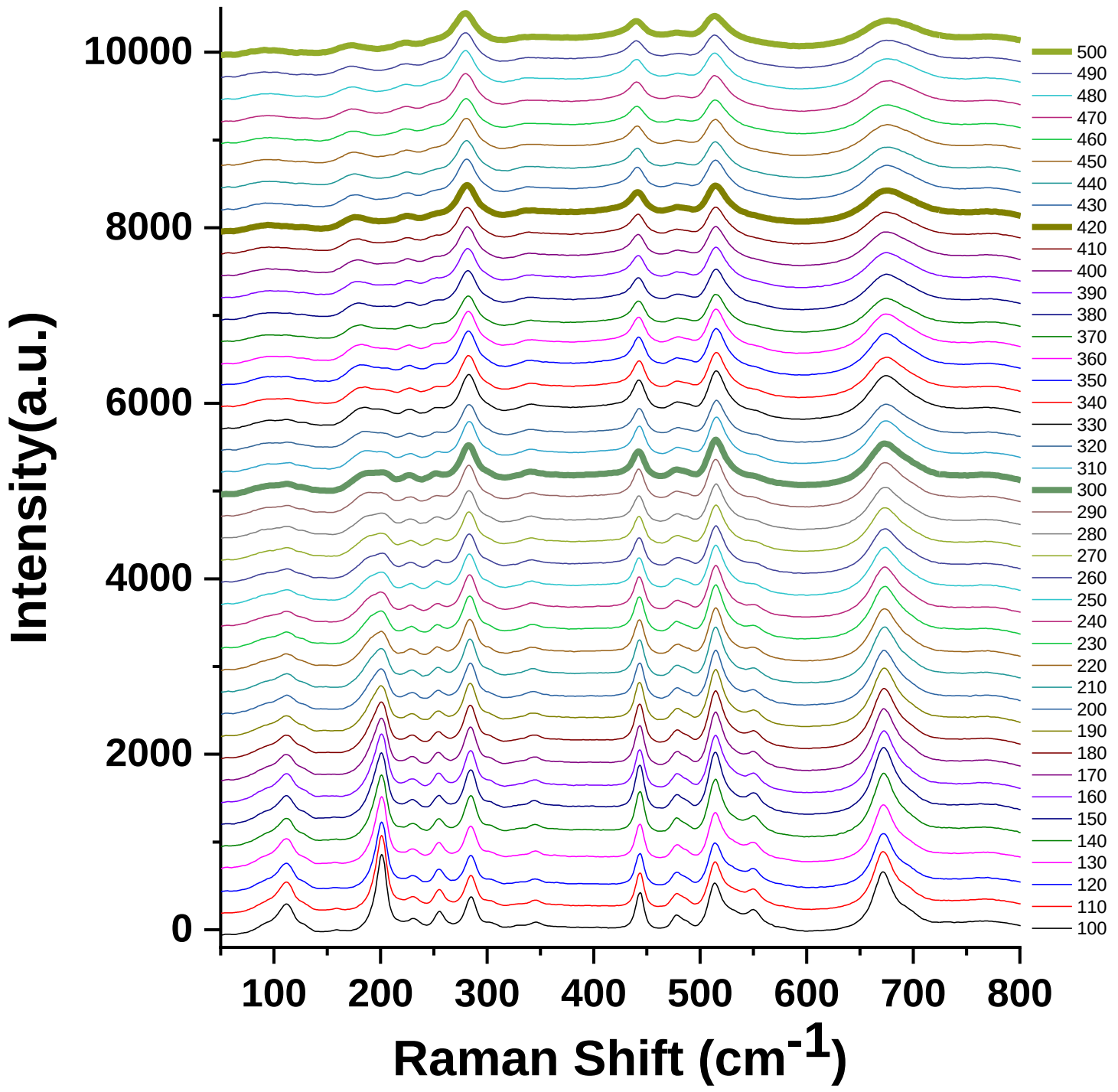

(b)

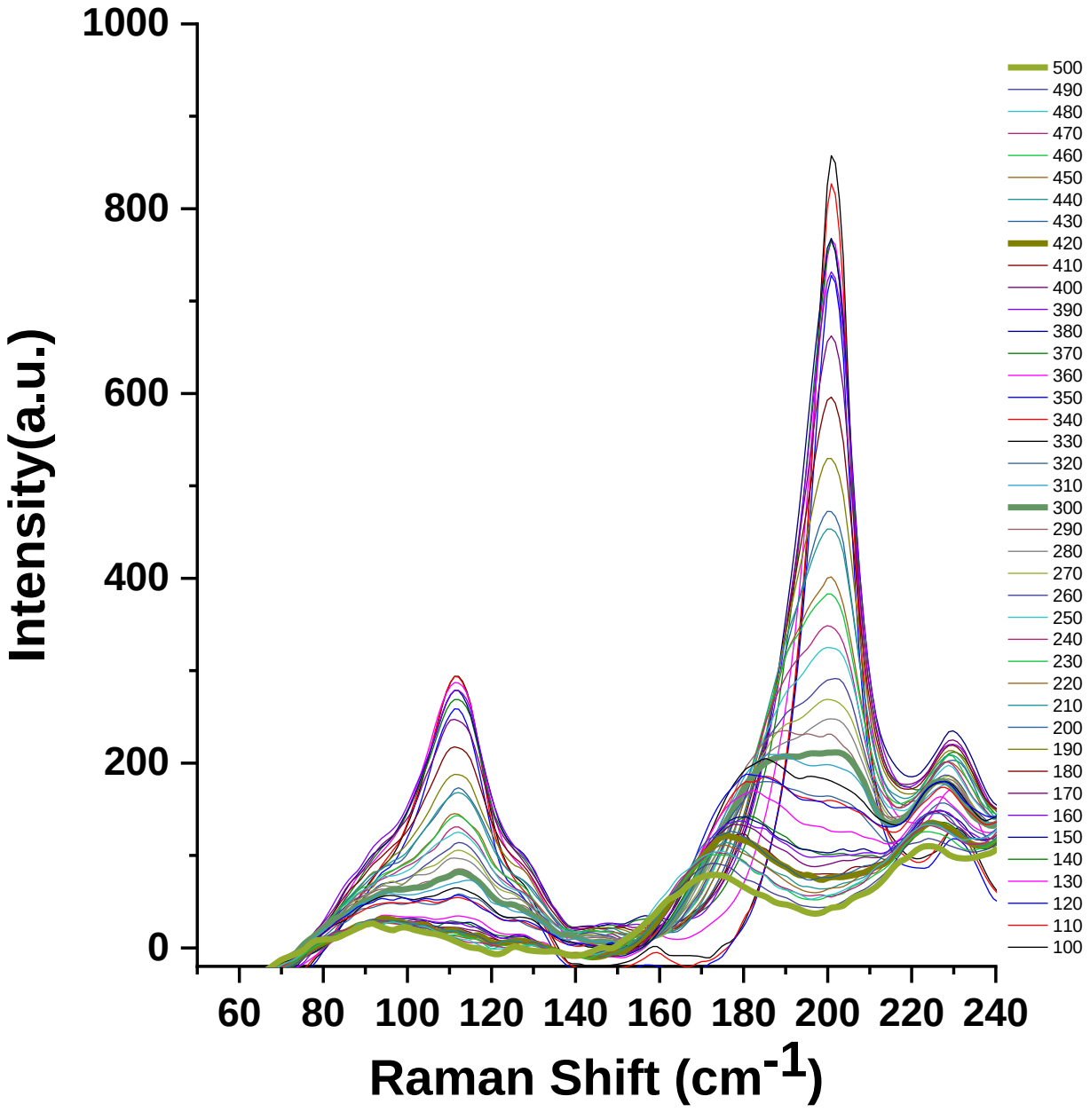

**Fig. S7**. Raman spectra for $Ca_{2.4}Sr_{0.6}Ti_2O_7$ single crystal (beam direction with respect to crystal not known). The full spectrum is given in (a). Note the reduction in amplitude of the features near 120 $cm^{-1}$ and 200 $cm^{-1}$ which are expanded in (b). Both peak are suppressed at ~ 420 K. This behavior is similar to that seen in $Ca_3Mn_2O_7$. Data are from cooling curves.

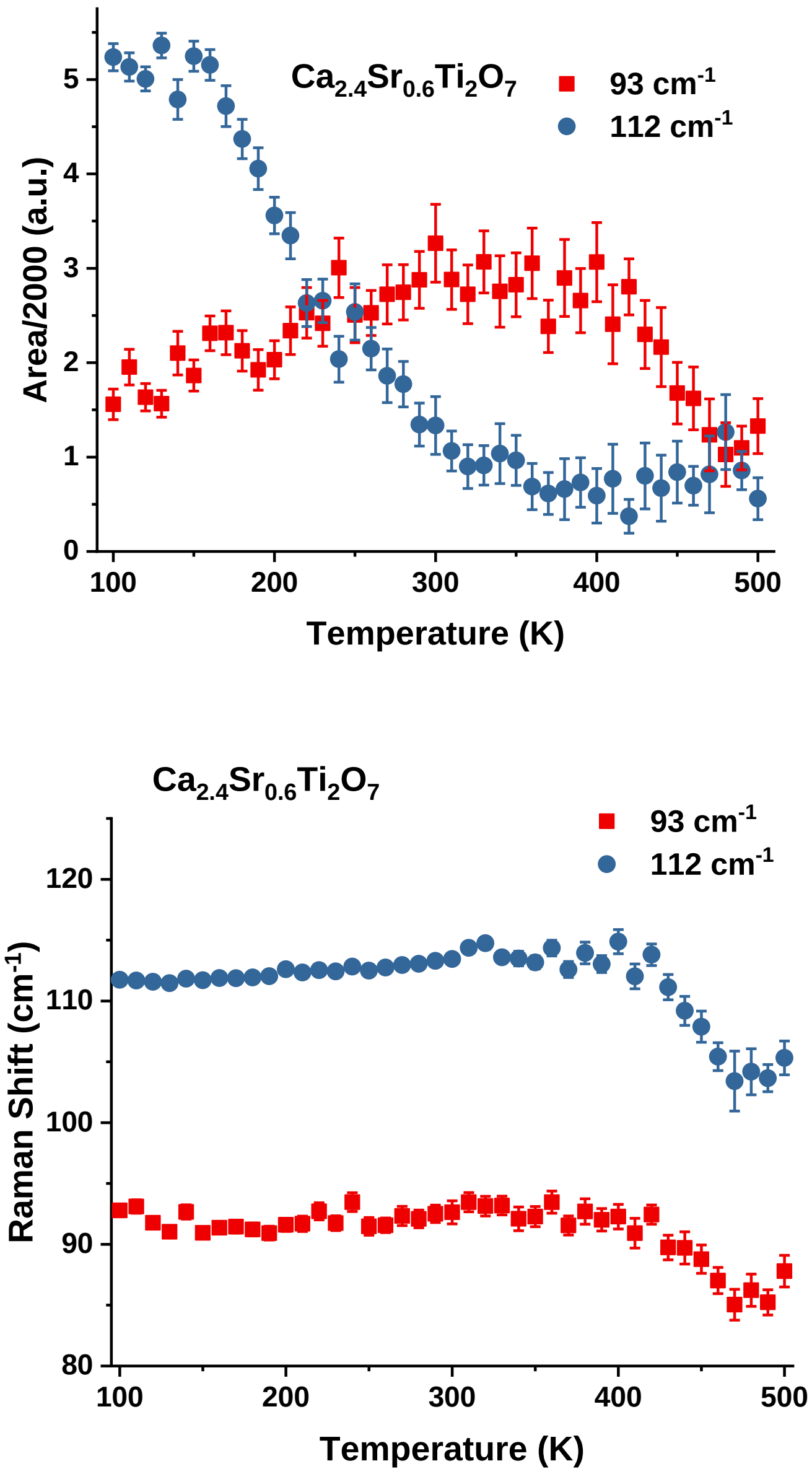

**(a)**

**(b)**

**Fig. S8**. Extracted parameters for 93 and 112 cm$^{-1}$ peaks for $Ca_{2.4}Sr_{0.6}Ti_2O_7$ single crystal. Note the decay of the 112 cm$^{-1}$ peaks amplitude by 300 K in (a), and the softening of the 93 and 112 cm$^{-1}$ peaks above 400 K. Data are from cooling curves.

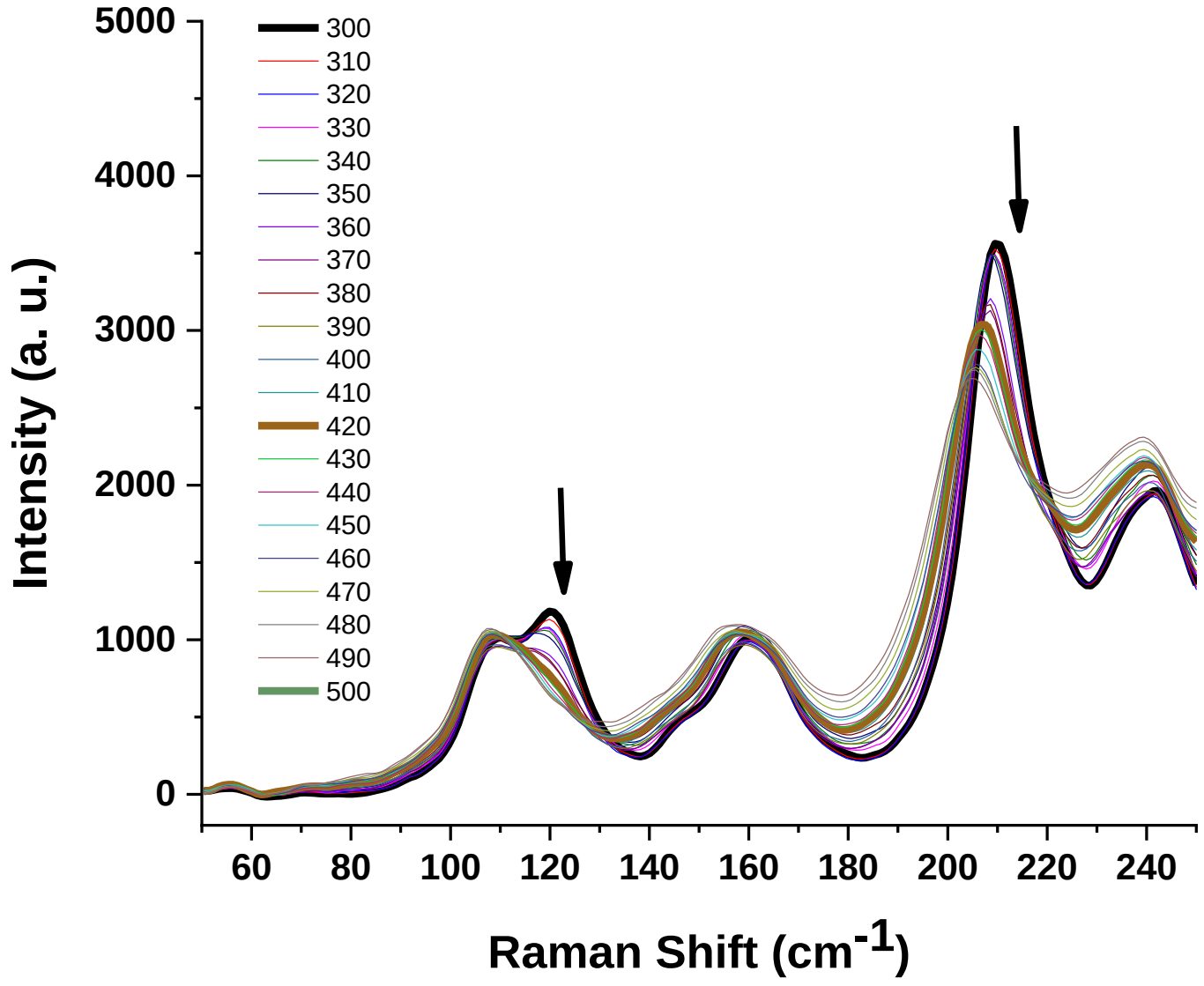

**Fig. S9**. Raman spectra for $Ca_3Ti_2O_7$ polycrystalline sample. Note the reduction in amplitude of features near 120 $cm^{-1}$ and 200 $cm^{-1}$ have the same behavior as the temperature increases up to ~420 K (vanish) as the $Ca_{2.4}Sr_{0.6}Ti_2O_7$ single crystal sample. Hence there is a common type of structural transition near ~ 400 K in the Ti and Mn-based systems. Data are from warming curves.

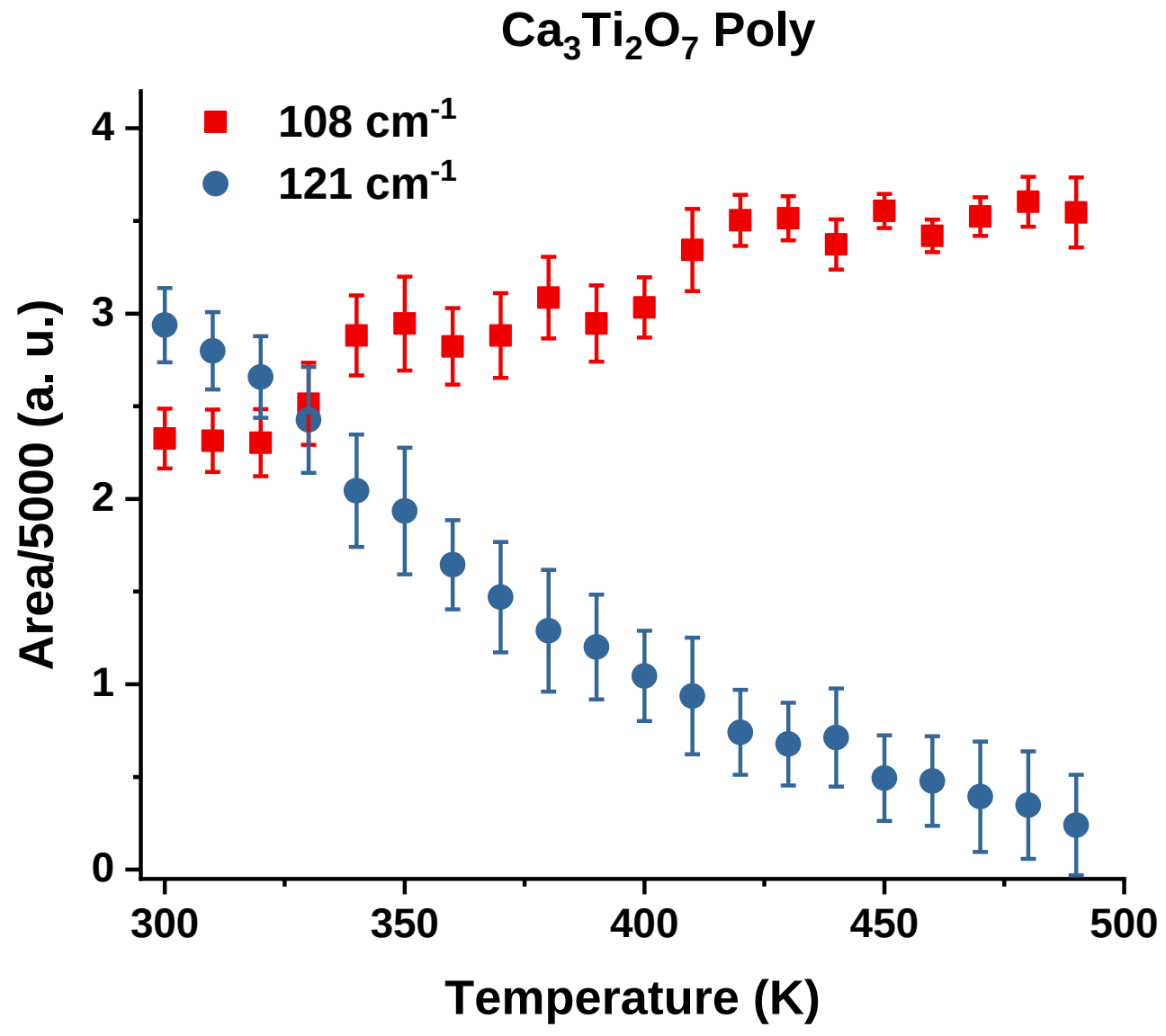

(a)

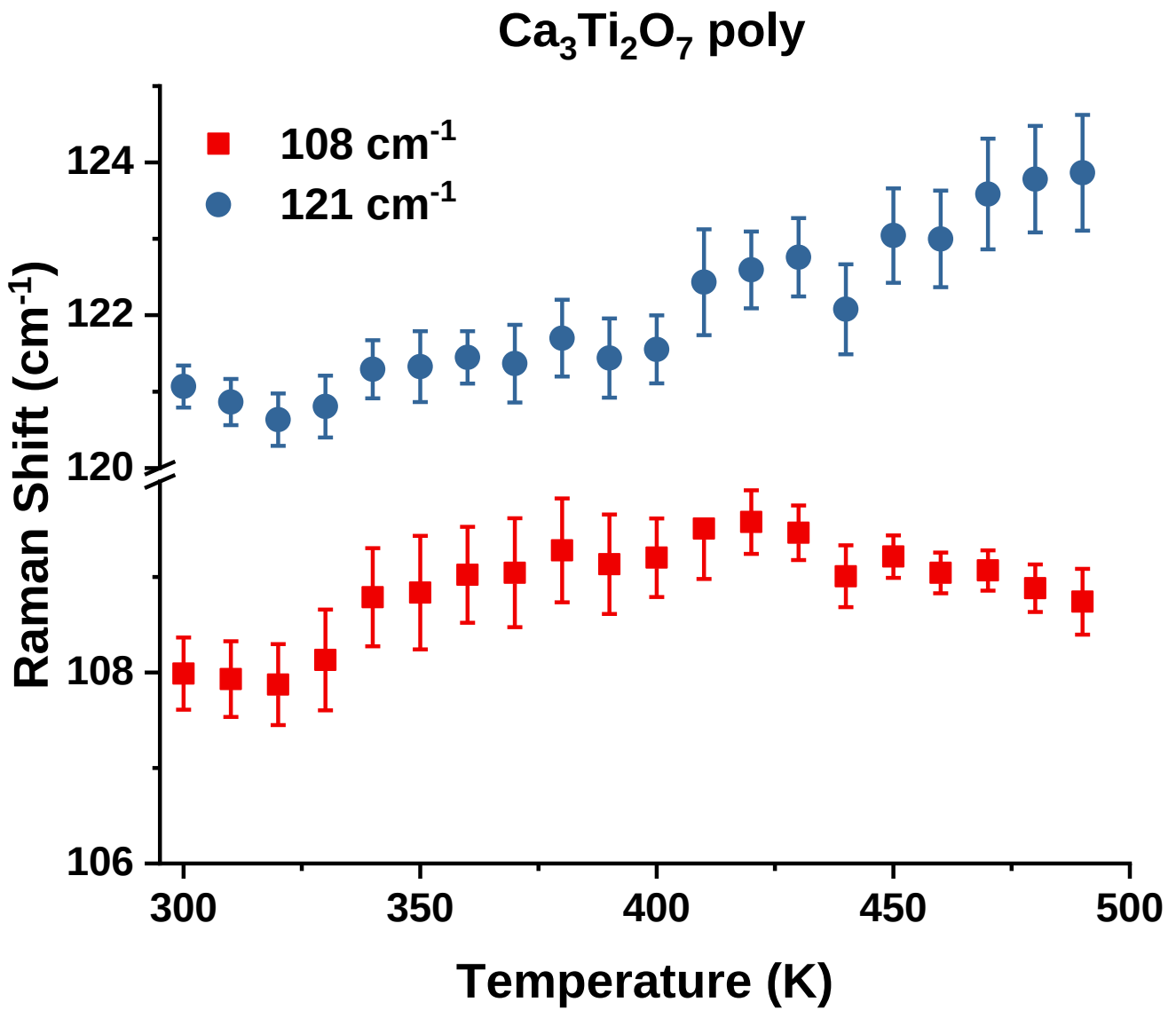

(b)

**Fig. S10**. Extracted parameters for 108 and 121 cm$^{-1}$ peaks for $Ca_3Ti_2O_7$ powder sample. Note the decay of the 121 cm$^{-1}$ peaks amplitude with temperature in (a). The phonon modes harden with increasing temperature. Data are from warming curves.

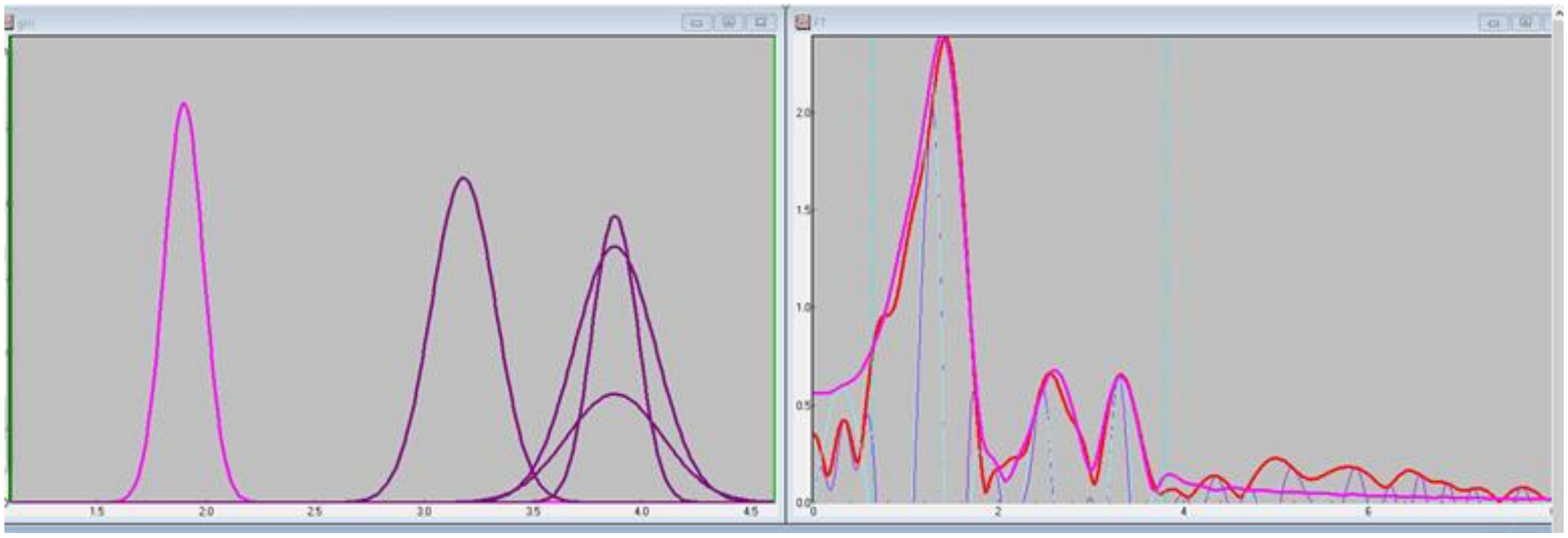

**Fig. S11**. (Left Panel) Fit to $Ca_3Mn_2O_7$ Mn K-edge data at 880 K showing the single scattering g(r) Gaussian peaks for the Mn-O shell and the Mn-Ca shell and both single and multiple scattering contributions for the nearest neighbor in plane Mn-Mn peak. The full R-space fit over the same three shells is also given (Right Panel). The peak near 4.0 Å (left panel) includes multiple scattering contributions.

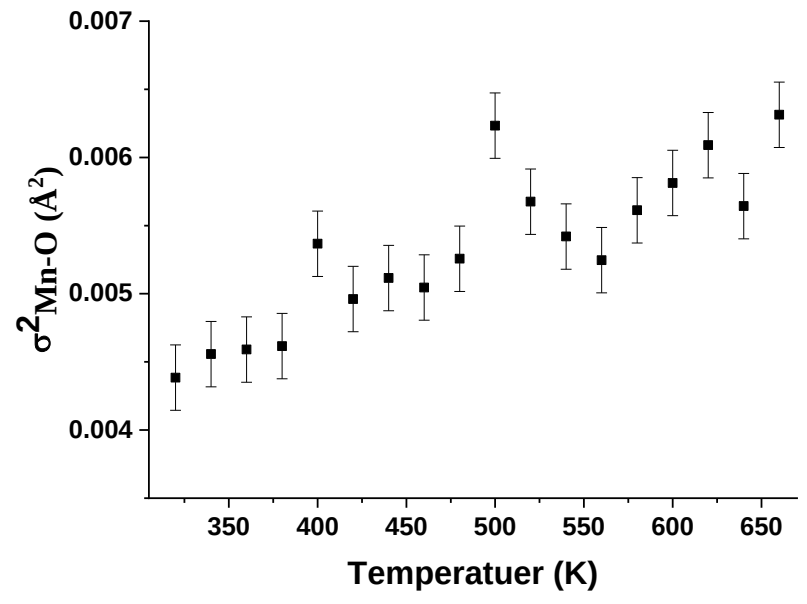

(a)

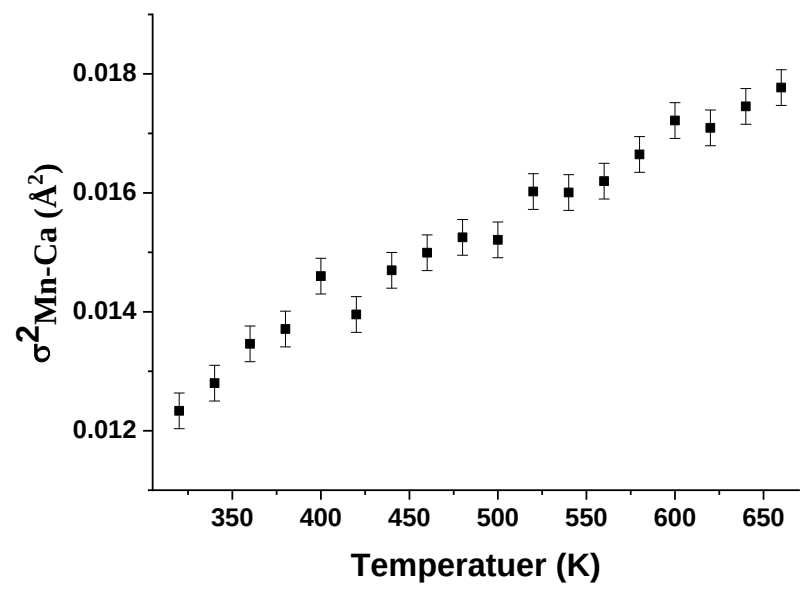

**Fig. S12**. XAFS structure function of $Ca_3Mn_2O_7$ measured between 320 K and 700 K (in (a) and (b)) revealing that the Mn-O and Mn-Ca Correlations. Note the change in the Mn-O correlation near 400 and 500K.

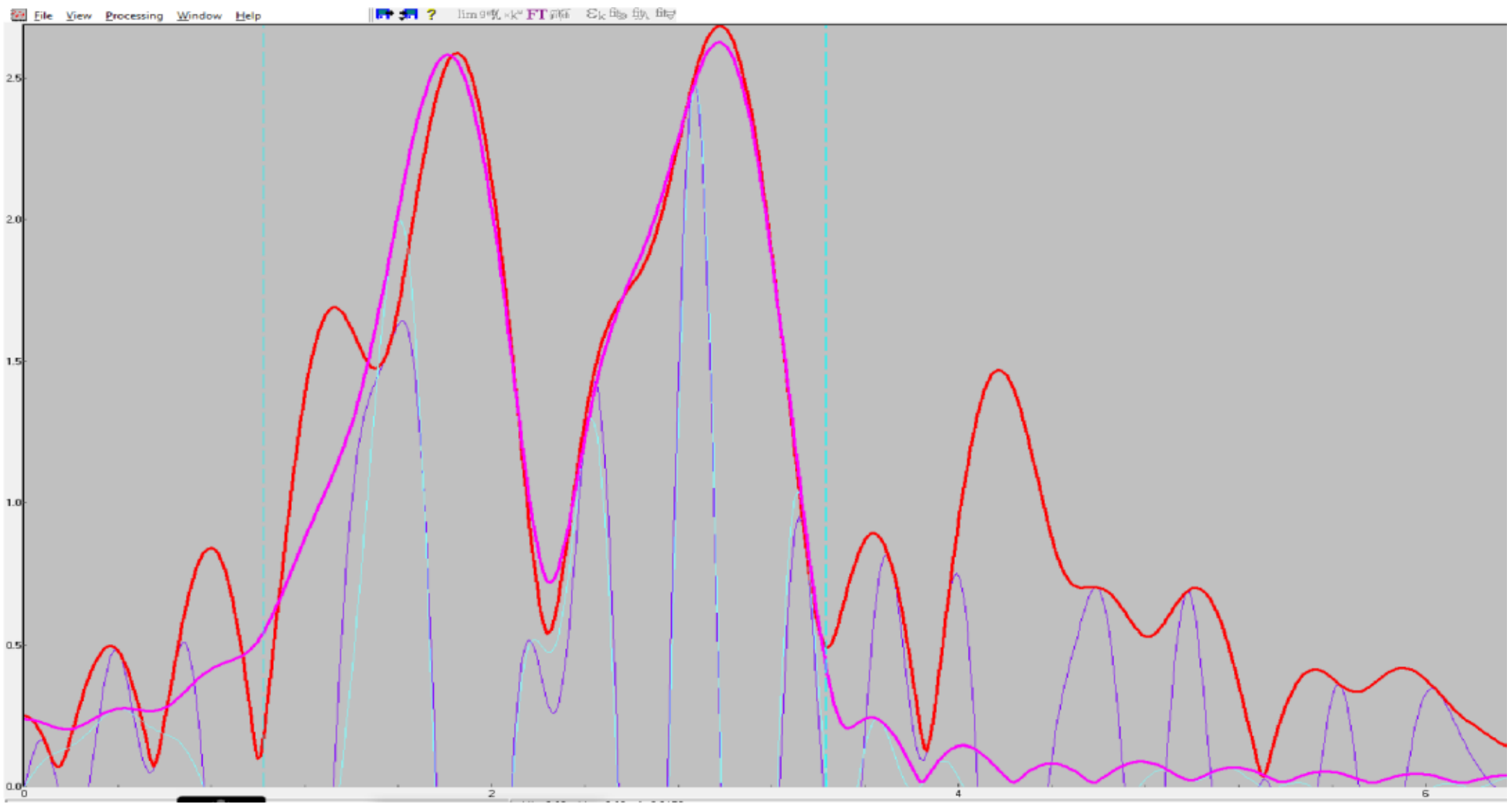

**Fig. S13.** Fit to Ca K-edge XAFS of $Ca_3Mn_2O_7$ in R-space for data (crystal derived powder) taken at 350 K. Extracted structural parameters are given in Table S2 at multiple temperatures (300, 350 and 450 K).

**Table S2. Ca K-Edge XAFS Derived Structural Parameters**

| Average Bond Distance (Å) | | Gaussian Width (Å) |
|---|---|---|
| 300 K | | |
| Ca-O | 2.30(1) | 0.007(1) |
| **Ca-Mn** | **3.13(1)** | **0.009(1)** |
| Ca-Ca | 3.62(1) | 0.003(1) |
| 350 K | | |
| Ca-O | 2.35(2) | 0.008(2) |
| **Ca-Mn** | **3.15(1)** | **0.007(1)** |
| Ca-Ca | 3.63(2) | 0.004(2) |
| 450 K | | |
| Ca-O | 2.34(2) | 0.011(2) |
| **Ca-Mn** | **3.17(1)** | **0.008(2)** |
| Ca-Ca | 3.64(2) | 0.006(2) |

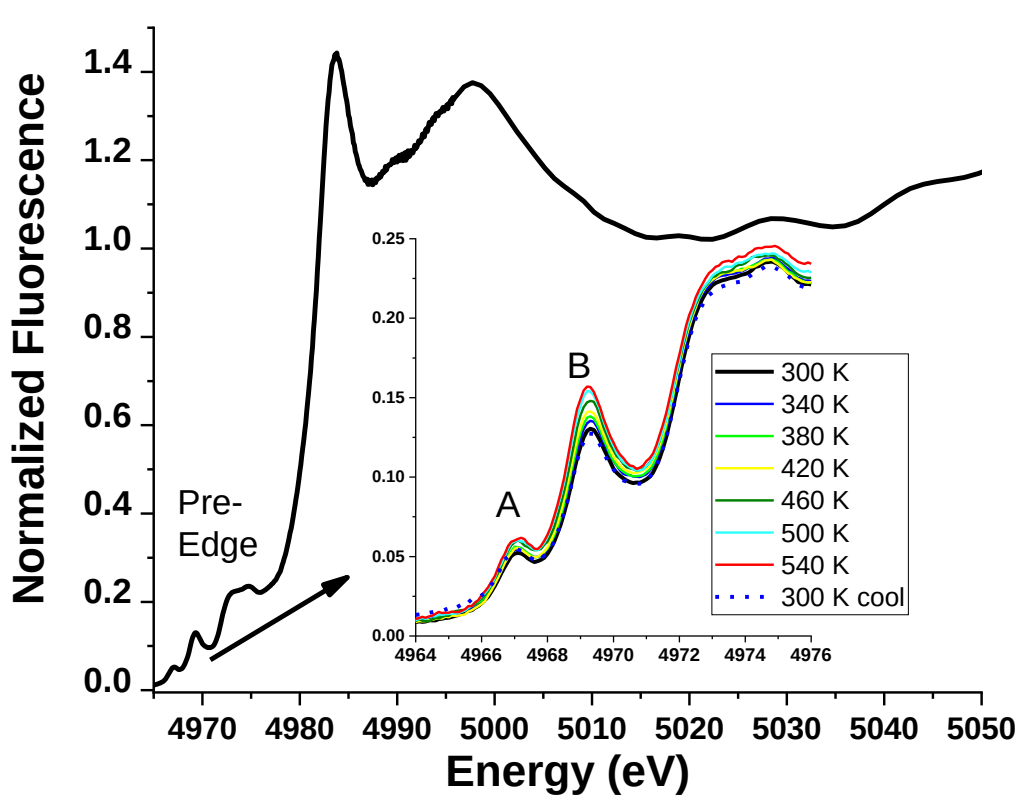

**(a)**

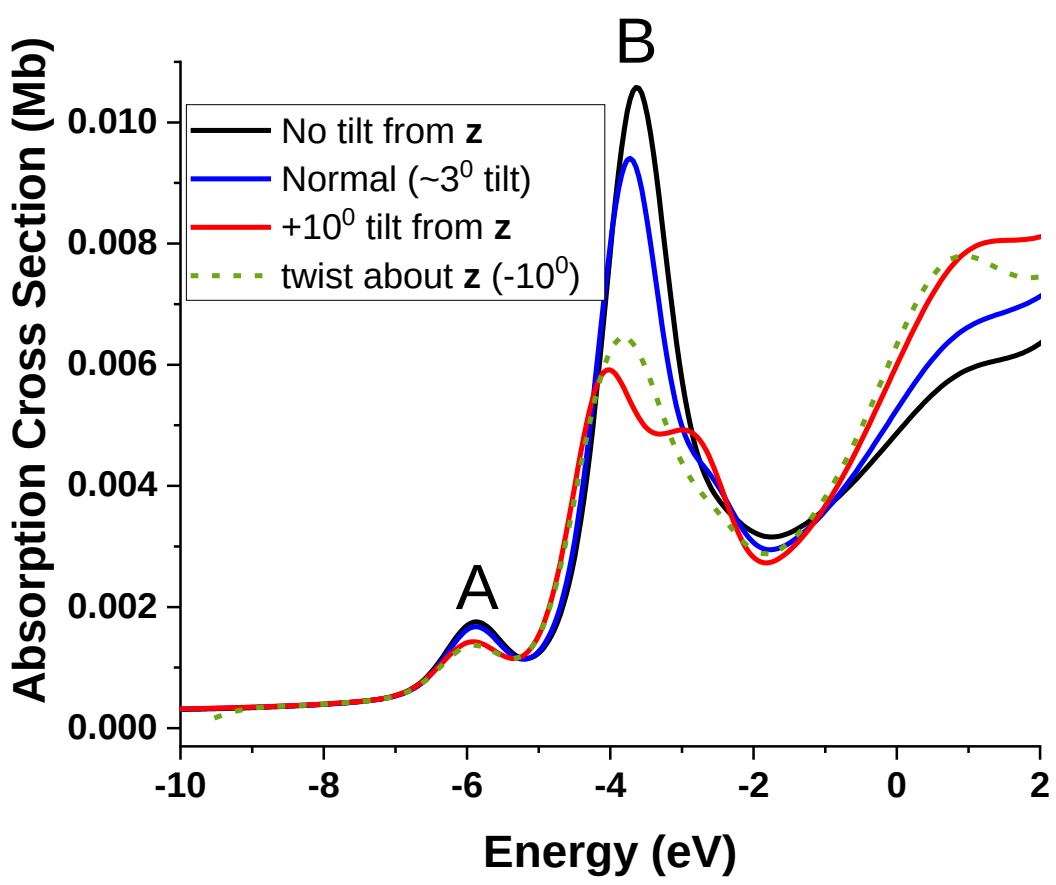

**(b)**

**Fig. S14**. (a) Ti K-edge XANES spectra of $Ca_{2.4}Sr_{0.6}Ti_2O_7$ between (crystal ground into powder sample) for temperatures between 300 and 540. In the inset, note that the pre-edge peak B intensity increase with increasing temperature. (b) Simulated spectra for a $CaTiO_3$ cluster for 118 atoms centered on Ti. Note that reduction of tiling relative to the long axis (called x here, Fig. 1(a)) increases the amplitude of peak B while reduction of twisting about the z-axis reduces the peak amplitude (dotted line). Hence the observed continuous enhancement of the B feature with temperature indicates a continuous reduction of the tilting of the $TiO_6$ polyhedra with increasing temperature in $Ca_{2.4}Sr_{0.6}Ti_2O_7$.

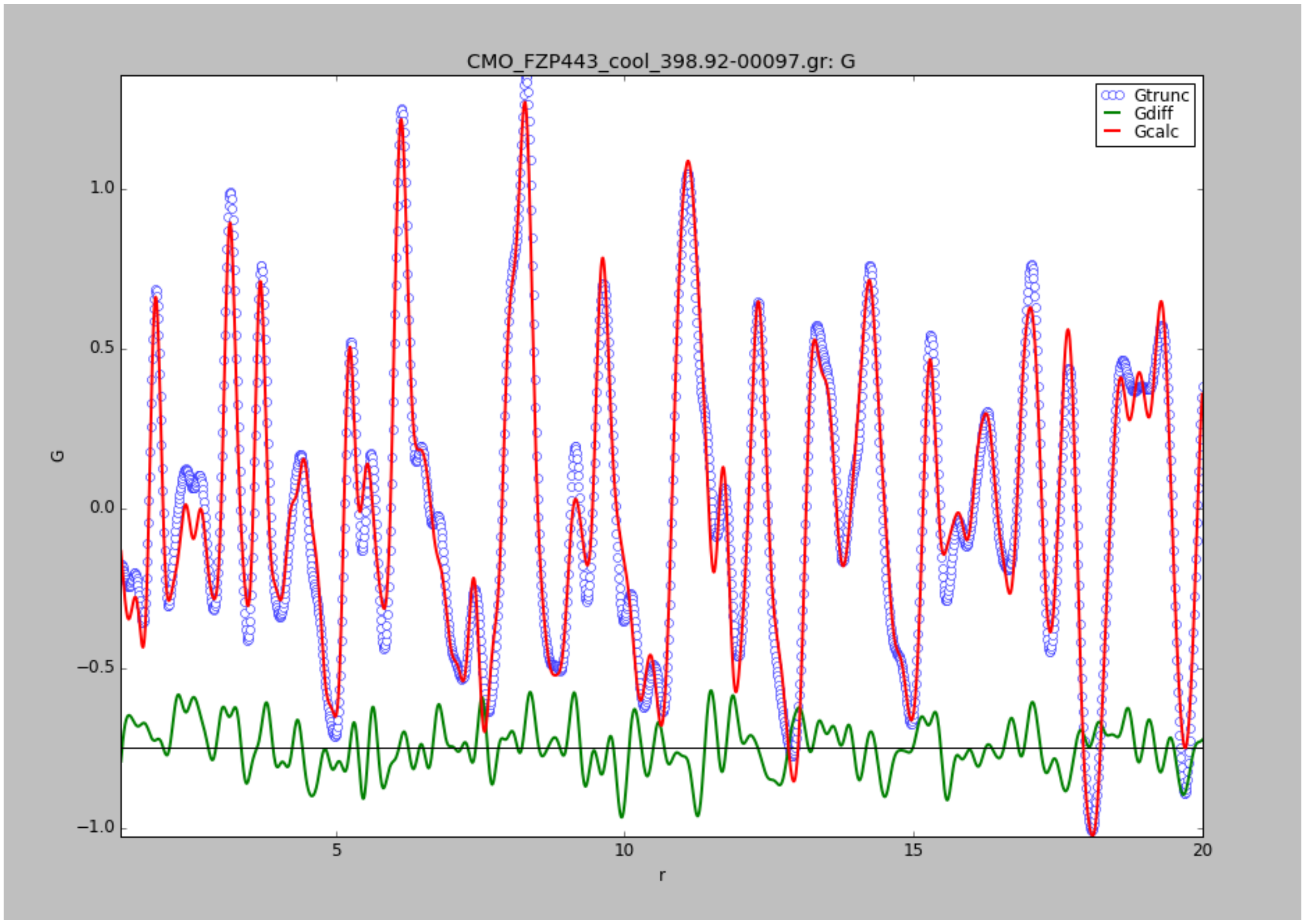

**Fig. S15**. R-Space PDF fit between 1.3 and 20 Å for the $Cmc2_1$ model for typical data taken at 399 K. The lower curve is the difference between the model (solid red line) and the data (open circles). Fits were conducted with PDFgui [15].

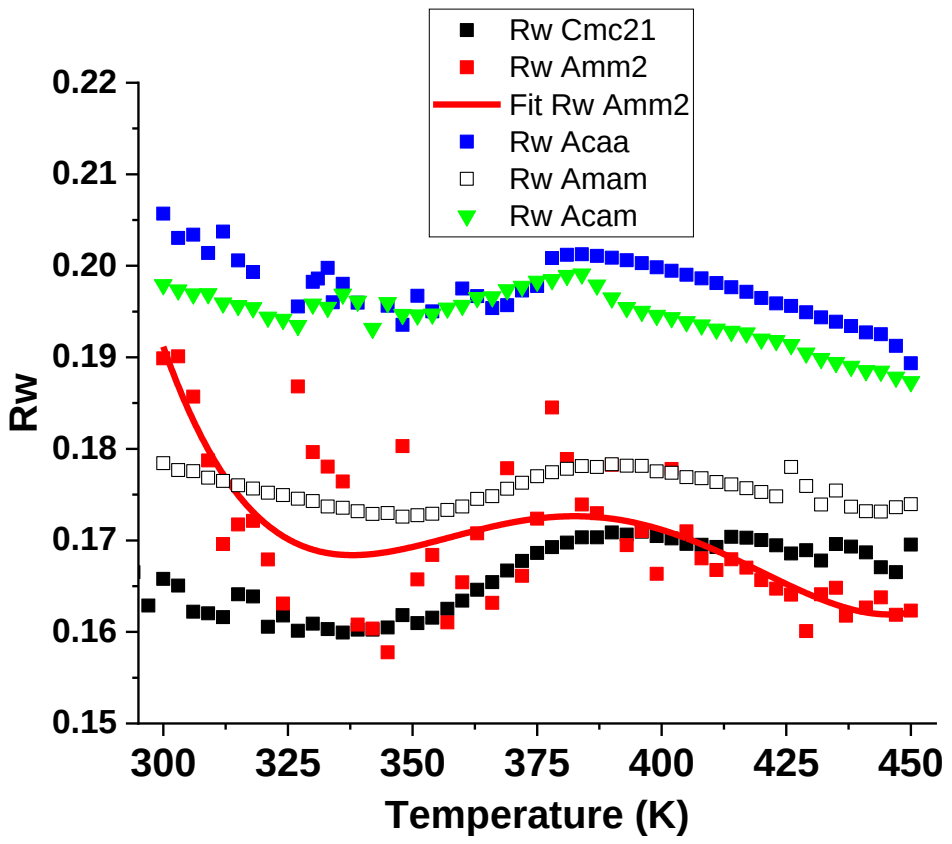

**(a)**

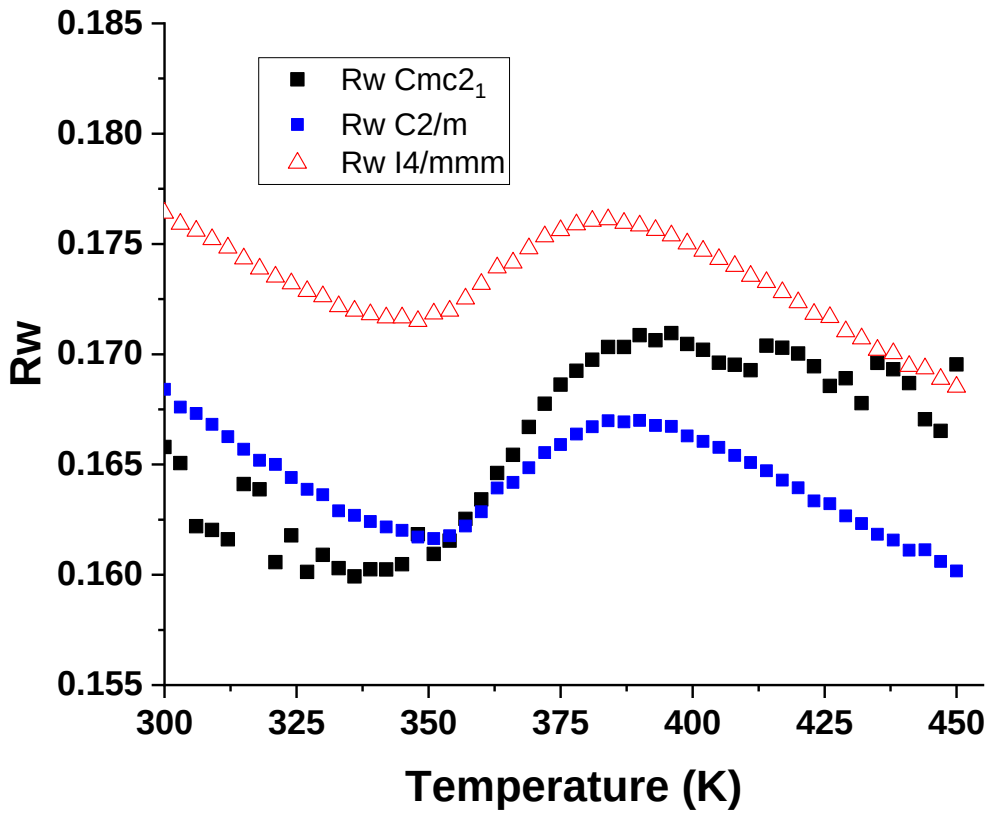

**(b)**

**Fig. S16**. (a) Rw fitting parameters for $Ca_3Mn_2O_7$ from PDF fits show that the Cmc2$_1$ space group gives the best model up to ~400 K while a possible space group for higher temperatures is the polar Amm2 space group. (b) Utilizing the space group (unit cell) results from single crystal measurements, it is seen that the best model above the first order transition is the C2/m cell with a volume ½ that of the low-temperature Cmc2$_1$ cell. Data are from cooling curves.

.

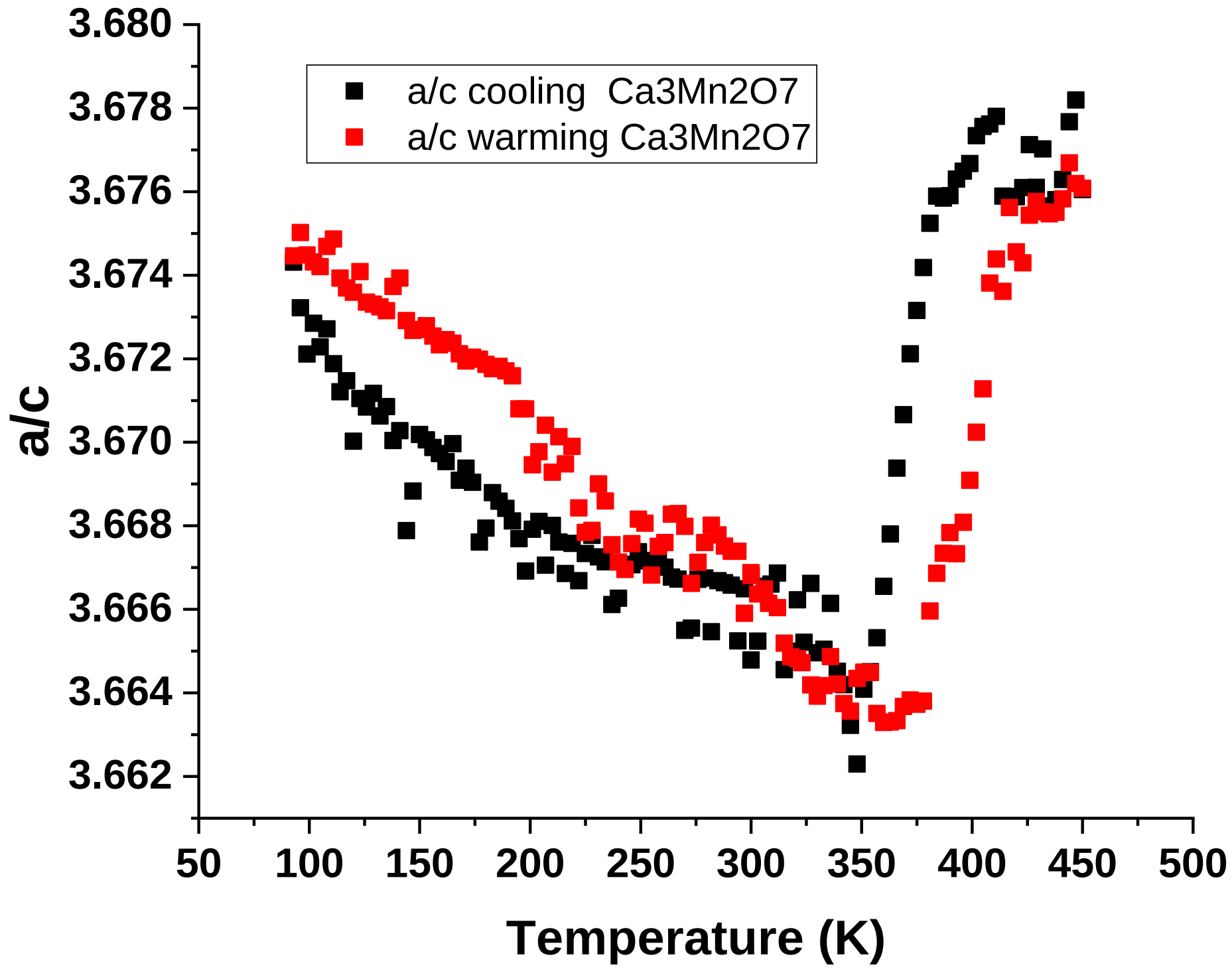

**Fig. S17**. Ratio of a to c lattice parameter $Ca_3Mn_2O_7$ (based on single crystal sample) during warming and cooling based on PDF fits.

(a)

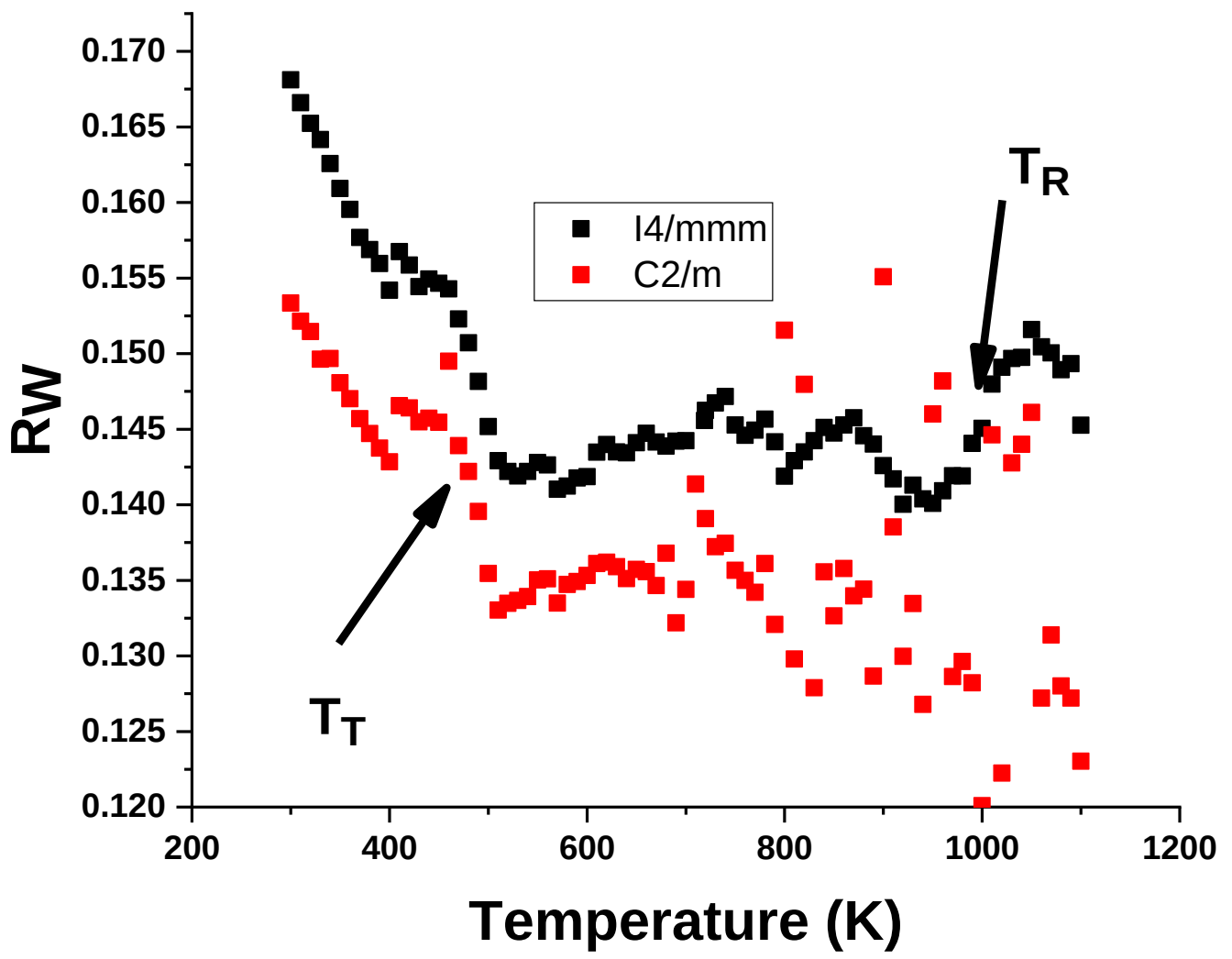

(b)

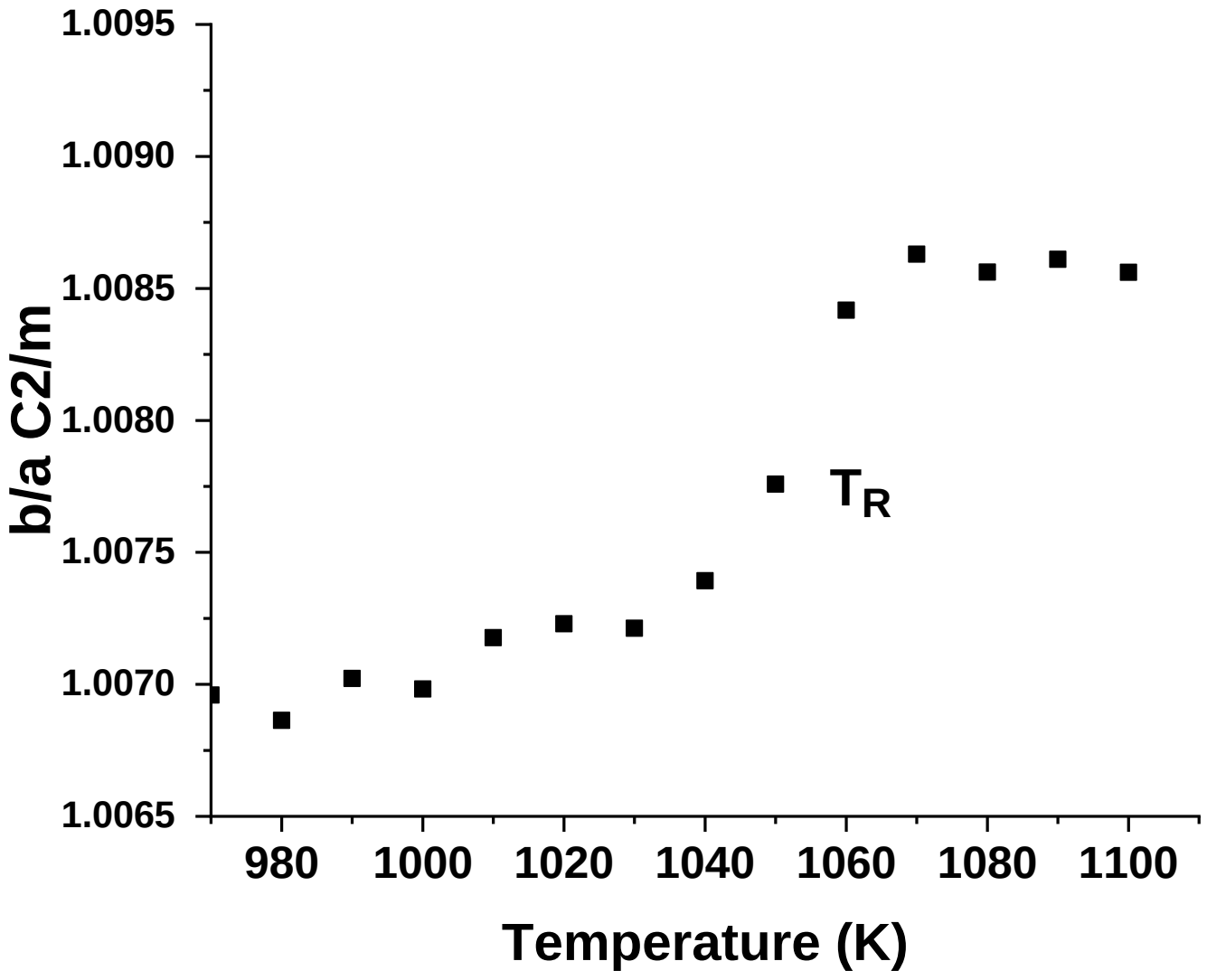

**Fig. S18.** (a) Fit of the parameter (for single crustal derived $Ca_3Mn_2O_7$ power) for short-range structure PDF data covering the same r-range as in Fig. S15 (1.3 to 20 Å). The local structure is consistent with the C2/m space group for the full temperature range. (a) b/a ratio (C2/m space group) near the high-temperature transition at ~1050 K. Measurements were made on warming.

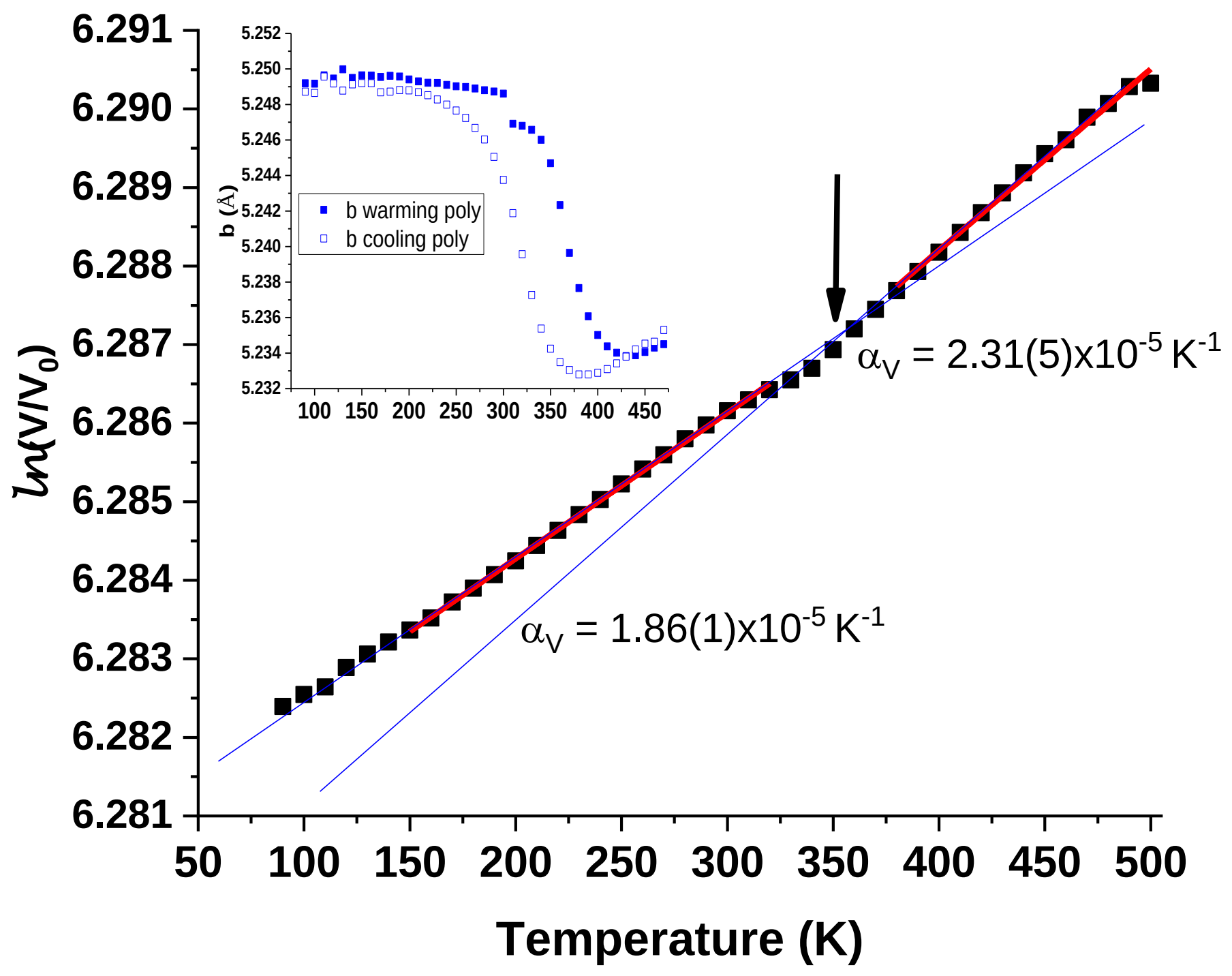

**Fig. S19**. Volume and b lattice parameter vs. temperature for polycrystalline $Ca_3Ti_2O_7$. Note broad range of separation for the b parameter (inset) for warming and cooling. The broad hysteresis is due to defects in powder prepared samples. Data in the main figure are from PDF cooling curves.

**Table S3. Structural Parameters from Single Crystal Solution for $Ca_3Mn_2O_7$ at 480 K (I4/mmm)**

| Atoms | x | y | z | Ueq (Å$^2$)×10$^3$ |
|---|---|---|---|---|
| Mn | 5000 | 5000 | 4018.7(6) | 6.6(4) |
| Ca1 | 5000 | 5000 | 1865.3(9) | 13.3(5) |
| Ca2 | 0 | 0 | 5000 | 13.7(6) |
| O1 (apical interior) | 5000 | 5000 | 5000 | 20.0(16) |
| O2 (apical) | 5000 | 5000 | 3040(3) | 14.3(10) |
| O3 (in-plane)* | 4188(12) | 0 | 4031(2) | 14.2(11) |

| Atom | $U_{11}$(Å$^2$)×10$^3$ | $U_{22}$(Å$^2$)×10$^3$ | $U_{33}$(Å$^2$)×10$^3$ | $U_{12}$(Å$^2$)×10$^3$ | $U_{13}$(Å$^2$)×10$^3$ | $U_{23}$(Å$^2$)×10$^3$ |
|---|---|---|---|---|---|---|
| Mn | 6.0(5) | 6.0(5) | 7.7(6) | 0 | 0 | 0 |
| Ca1 | 14.5(6) | 14.5(6) | 10.7(8) | 0 | 0 | 0 |
| Ca2 | 13.2(7) | 13.2(7) | 14.6(10 ) | 0 | 0 | 0 |
| O1 | 28(2) | 28(2) | 3(2) | 0 | 0 | 0 |
| O2 | 17.0(14) | 17.0(14) | 8.9(16) | 0 | 0 | 0 |
| O3 | 12(2) | 6.8(18) | 24(2) | 0 | -4.3(13) | 0 |

Space Group: I4/mmm (Z=2) Tetragonal Cell
a = 3.7054 (3) Å, c = 19.585 (3) Å, Dx = 4.225 g/cm$^3$
Measurement Temperature: 480 K
Crystal Dimensions: ~20 μm (diameter)
Wavelength: 0.41328 Å,
Absorption Coefficient: 1.783 mm$^{-1}$
F(000) = 333.4
2θ Range : 2.42° to 40.2°
$-6 \leq h \leq 6$, $-6 \leq k \leq 5$, and $-31 \leq l \leq 32$,
Number of Measured Reflections: 5325
Number of Independent Reflections: 241
Number of fitting parameters: 20
Restraints = 0
Max and Min Peak in Final Difference Map: 3.1/-1.0 e-/ Å$^3$
$R_1$ = 4.42 %, $wR_2$ = 18.8 %, Goodness of Fit = 1.63 ($I>=2\sigma(I)$)
$R_1$ = 4.56 %, $wR_2$ = 19.0 % (all data)

$$R_1 = \sum ||F_0| - |F_c|| / \sum |F_0|$$

$$wR_2 = \sum w(F_o^2 - F_c^2)^2) / \sum w(F_o^2)^2$$

* O3 atoms have 50% occupancy of two sites.

**Table S4. Bond Distances in $Ca_3Mn_2O_7$ at 480 K**

| Atoms | Distance (Å) |
|---|---|
| Mn-O3 | 1.8771(7) x 4 |
| Mn-O2 | 1.922(1) |
| Mn-O2 | 1.916(6) |
| | |
| Ca1-O2 | 2.301(6) |
| Ca1-O3 | 2.344(4) x 2 |
| Ca1-O2 | 2.6266(5) x 4 |
| Ca1-O2 | 2.779(5) x 2 |
| Ca2-O3 | 2.451(4) x 4 |
| Ca2-O2 | 2.6201(4) x 4 |
| Ca2-O3 | 2.870(4) x 4 |
| | |
| Mn-Ca1 | 3.1405(9) x 4 |
| Mn-Ca2 | 3.2494(7) x 4 |

$\langle R_{Mn-O} \rangle$ =1.89 Å, Standard Dev =0.02 Å

$\langle R_{Ca-O} \rangle$ = 2.61 Å, Standard Dev =0.18 Å

$\langle R_{Mn-Ca} \rangle$ = 3.19 Å, Standard Dev =0.05 Å

**Table S5. Structural Parameters from Single Crystal Solution for $Ca_3Mn_2O_7$ at 480 K (C2/m low symmetry solution)**

| Atoms | x | y | z | Ueq (Å$^2$)×10$^3$ |
|---|---|---|---|---|
| Mn | 5983.5(10) | 5000 | 8035.1(8) | 7.1(3) |
| Ca1 | 8135.4(19) | 5000 | 3731.0(12) | 14.1(3) |
| Ca2 | 5000 | 0 | 10000 | 14.4(4) |
| O1 (apical interior) | 5000 | 5000 | 10000 | 21.5(11) |
| O2 (apical) | 6960(6) | 5000 | 6081(4) | 14.0(7) |
| O3 | 3870(16) | 2103(17) | 8060(6) | 15.2(16) |
| O3P | 3052(18) | 2917(17) | 8064(7) | 14.6(16) |

| Atom | $U_{11}$(Å$^2$)×10$^3$ | $U_{22}$(Å$^2$)×10$^3$ | $U_{33}$(Å$^2$)×10$^3$ | $U_{12}$(Å$^2$)×10$^3$ | $U_{13}$(Å$^2$)×10$^3$ | $U_{23}$(Å$^2$)×10$^3$ |
|---|---|---|---|---|---|---|
| Mn | 6.6(4) | 6.7(4) | 8.2(4) | -0 | -2.2(3) | 0 |
| Ca1 | 15.1(5) | 15.5(6) | 11.7(5) | -0 | -3.6(4) | 0 |
| Ca2 | 13.3(6) | 14.5(7) | 15.4(7) | -0 | -3.6(5) | 0 |
| O1 | 27(3) | 31(3) | 5.9(18) | -0 | -3.5(19) | 0 |
| O2 | 12.4(14) | 20.5(17) | 7.8(12) | -0 | -0.1(10) | 0 |
| O3* | 11(3) | 13(3) | 21(2) | -1(2) | -3.5(19) | -4.3(19) |
| O3P | 12(3) | 10(3) | 25(3) | -3(2) | -10(2) | 2(2) |

Space Group: C2/m (Z=2) Monoclinic Cell
a = 5.2404(7) Å, 5.2401(7) Å, c = 10.1369(14)) Å, Dx = 4.225 g/cm$^3$
b= 75.024(3)°
Measurement Temperature: 480 K
Crystal Dimensions: ~20 μm (diameter)
Wavelength: 0.41328 Å,
Absorption Coefficient: 1.783 mm$^{-1}$
F(000) = 333.4
2θ Range : 2.42° to 40.2°
$-8 \leq h \leq 8$, $-8 \leq k \leq 7$, and $-16 \leq l \leq 16$
Number of Measured Reflections: 5325
Number of Independent Reflections: 687
Number of fitting parameters: 46
Restraints = 0
Max and Min Peak in Final Difference Map: 3.1/-1.7 e-/ Å$^3$
$R_1$ = 4.77 %, $wR_2$ = 15.5 %, Goodness of Fit = 1.53 (I>=2σ (I))
$R_1$ = 5.60 %, $wR_2$ = 19.7 % (all data)

$$R_1 = \sum ||F_0| - |F_c|| / \sum |F_0|$$

$$wR_2 = \sum w(F_o^2 - F_c^2)^2) / \sum w(F_o^2)^2$$

* O3 and O3P atoms have 50% occupancy of the two sites (derived from fit).

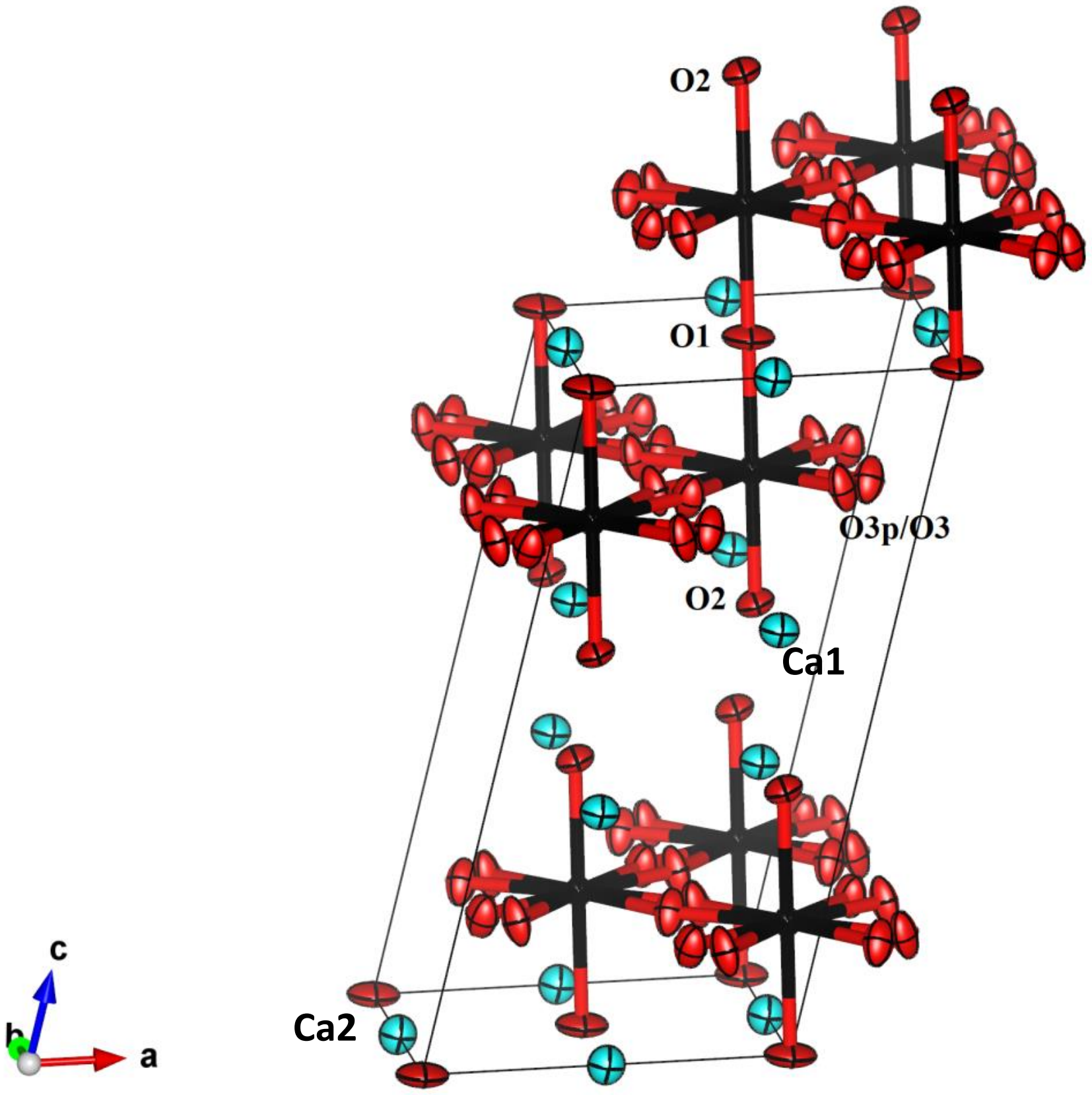

**Fig. S20**. Unit cell for the low symmetry solution (C2/m) to 480 K data for $Ca_3Mn_2O_7$. Note (as in I4/mmm solution) that there is a 50/50 split occupancy at the O3 site in-plane due to domain formation.

**Table S6. Structural Parameters from Single Crystal Solution of $Ca_3Mn_2O_7$ at 300 K**

| **Atoms** | **x** | **y** | **z** | Ueq (Å$^2$)×10$^3$ |
|---|---|---|---|---|
| Mn | -4018.5(5) | -7504.8(13) | -2684.3(19) | 5.8(2) |
| Ca1 | -3136.9(7) | -7453(2) | -7846(2) | 9.5(3) |
| Ca2 | 5000 | -2507(4) | -2487(3) | 10.9(4) |
| O1 | 5000 | -7948(11) | -2709(14) | 12.5(12) |
| O2 | -3040(2) | -7208(8) | -2627(10) | 9.5(8) |
| O3 | -4112(2) | -4623(6) | -4780(10) | 11.2(7) |
| O4 | -3943.2(19) | -9641(7) | -5515(11) | 12.0(7) |

| Atom | U11(Å$^2$)×10$^3$ | U22(Å$^2$)×10$^3$ | U33(Å$^2$)×10$^3$ | U12(Å$^2$)×10$^3$ | U13(Å$^2$)×10$^3$ | U23(Å$^2$)×10$^3$ |
|---|---|---|---|---|---|---|
| Mn | 6.2(4) | 7.0(4) | 4.1(4) | 0.05(15) | 1.6(3) | -0.19(16) |
| Ca1 | 8.5(5) | 12.2(5) | 7.7(6) | -0.5(3) | 0.3(5) | 1.2(3) |
| Ca2 | 10.6(7) | 12.3(6) | 9.7(9) | -0 | -0 | -2.1(5) |
| O1 | 4.8(18) | 16(3) | 17(3) | -0 | -0 | -2(3) |
| O2 | 7.2(13) | 9.9(15) | 11(2) | -0.9(10) | -2(2) | 0.1(15) |
| O3 | 15.1(15) | 8.1(16) | 10.3(15) | -1.1(12) | 0.1(13) | 0.9(13) |
| O4 | 15.4(14) | 8.9(15) | 11.6(17) | -0.6(11) | 0.7(14) | -3.0(16) |

Space Group: $Cmc2_1$ (Z=4) Orthorhombic Cell
a = 19.4387(12) Å, b = 5.2493(3) Å, c = 5.2433(3) Å, Dx = 4.247 g/cm$^3$
Measurement Temperature: 300 K
Crystal Dimensions: ~20 μm (diameter)
Wavelength: 0.41328 Å,
Absorption Coefficient: 1.792 mm$^{-1}$
F(000) = 666.8
2θ Range : 2.44° to 40.3°
$-32 \le h \le 32$, $-7 \le k \le 8$, and $-8 \le l \le 8$
Number of Measured Reflections: 10,565
Number of Independent Reflections: 1296
Number of fitting parameters: 58
Restraints = 1
Flack Parameter = 0.4(3)
Max and Min Peak in Final Difference Map: 3.2/-2.7 e-/ Å$^3$
$R_1$ = 5.23 %, $wR_2$ = 14.7 %, Goodness of Fit = 1.54 (I>=2σ (I))
$R_1$ = 7.15 %, $wR_2$ = 21.3 % (all data)

$$R_1 = \sum ||F_0| - |F_c|| / \sum |F_0|$$

$$wR_2 = \sum w(F_o^2 - F_c^2)^2) / \sum w(F_o^2)^2$$

**Table S7. Bond Distances in $Ca_3Mn_2O_7$ at 300 K**

| Atoms | Distance (Å) |
|---|---|
| Mn-O1 | 1.9221(12) |
| Mn-O2 | 1.908(4) |
| Mn-O3 | 1.897(5) |
| Mn-O3 | 1.879(4) |
| Mn-O4 | 1.887(5) |
| Mn-O4 | 1.866(5) |
| | |
| Ca1-O2 | 2.298(4) |
| Ca1-O2 | 2.457(4) |
| Ca1-O2 | 2.517(6) |
| Ca1-O2 | 2.811(4) |
| Ca1-O2 | 2.457(4) |
| Ca1-O2 | 2.457(4) |
| Ca1-O3 | 2.895(5) |
| Ca1-O3 | 2.410(4) |
| Ca1-O4 | 2.295(5) |
| Ca1-O4 | 2.597(5) |
| Ca2-O1 | 2.748(8) |
| Ca2-O1 | 2.517(8 |
| Ca2-O1 | 2.859(5) |
| Ca2-O1 | 2.396(5) |
| Ca2-O3 | 2.379(5) |
| Ca2-O3 | 2.696(5) |
| Ca2-O3 | 2.379(5) |
| Ca2-O3 | 2.696(5) |
| Ca2-O4 | 2.561(4) x 2 |
| | |
| Mn-Ca1 | 3.0617(15) |
| Mn-Ca1 | 3.1171(14) |
| Mn-Ca1 | 3.1543(13) |
| Mn-Ca1 | 3.2032(12) |
| Mn-Ca2 | 3.2457(18) |
| Mn-Ca2 | 3.2473(18) |
| Mn-Ca2 | 3.1596(19) |
| Mn-Ca2 | 3.3264(18) |

$\langle R_{Mn\text{-}O} \rangle$ =1.89 Å, Standard Dev =0.02 Å

$\langle R_{Ca\text{-}O} \rangle$ = 2.55 Å, Standard Dev =0.18 Å

$\langle R_{Mn\text{-}Ca} \rangle$ = 3.19 Å, Standard Dev =0.08 Å

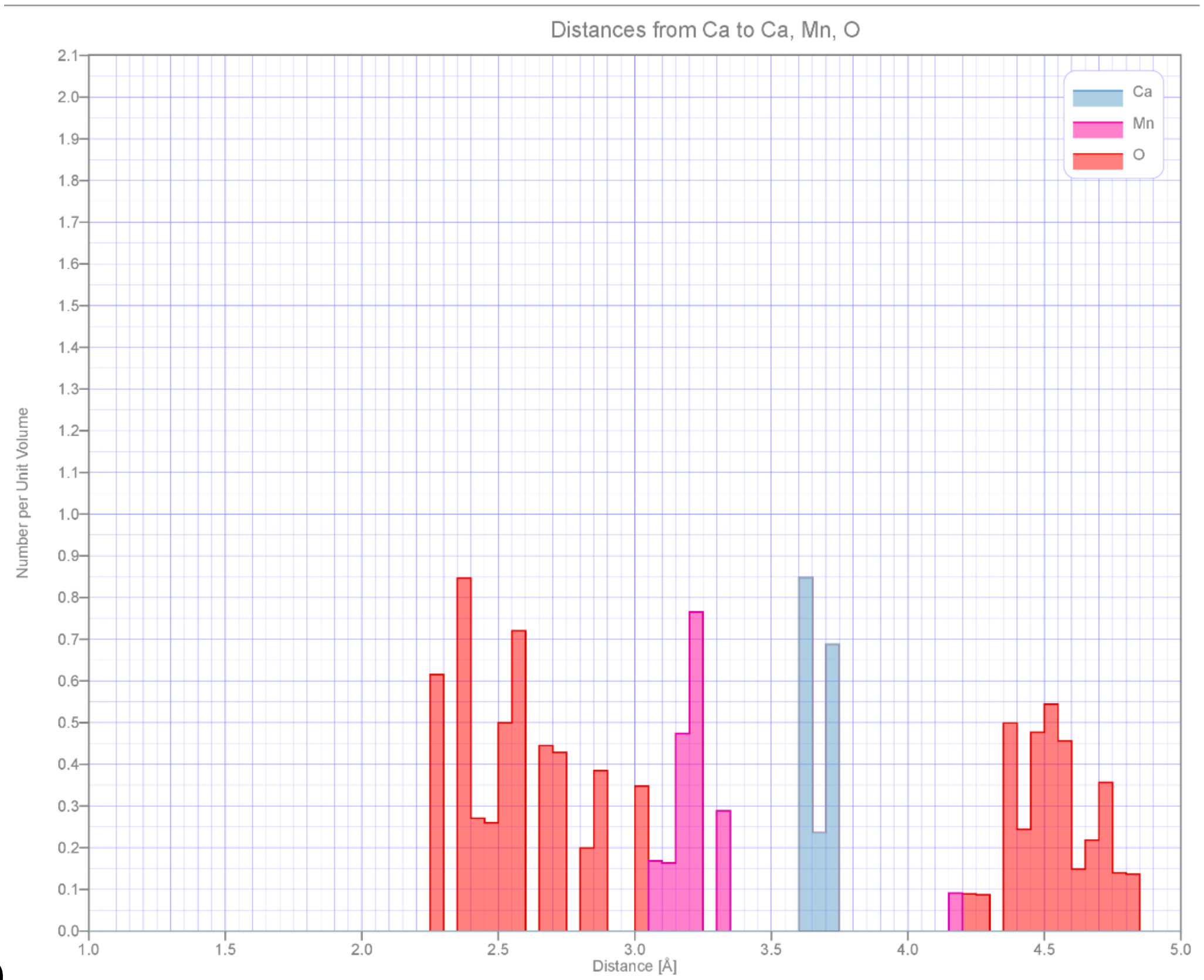

**(a)**

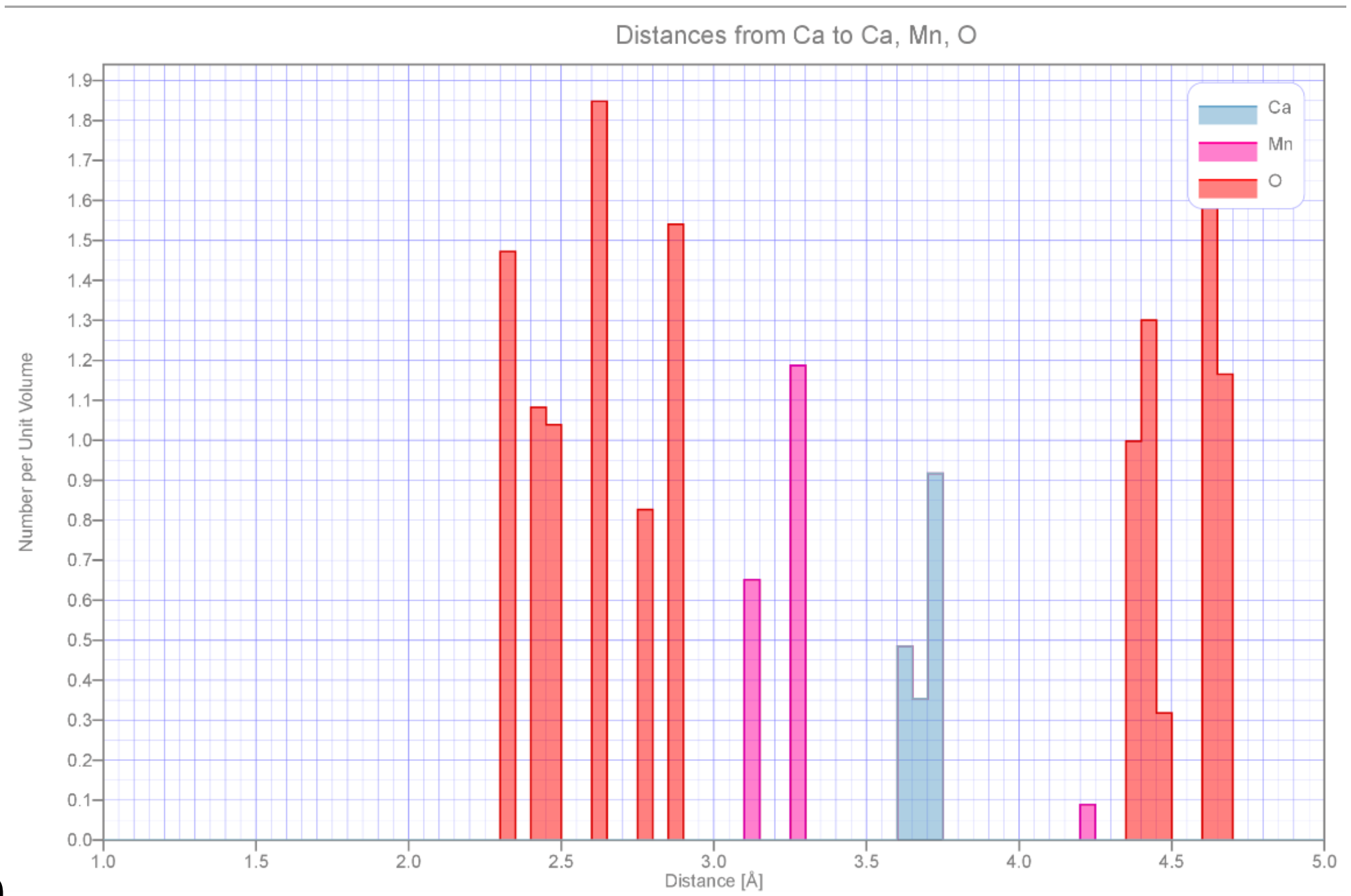

**(b)**

**Fig. S21.** Distribution of atoms about Ca at 300 K (a) and at 480 K (b) taken from the single crystal structure solution. Note the strong sharpening of the Ca-Mn distribution (pink blocks) at high temperature.

**Table S8. Structural Parameters from Single Crystal Solution of $Ca_3Mn_2O_7$ at 104 K**

| Atoms | x | y | z | Ueq (Å$^2$)×10$^3$ |
|---|---|---|---|---|
| Mn | 979.3(3) | 2506.1(11) | 4140.3(12) | 2.45(14) |
| Ca1 | 0 | 2489(3) | -1076.7(19) | 6.9(2) |
| Ca2 | 1865.7(4) | 2458.5(16) | 9327.2(14) | 3.15(16) |
| O1 | 0 | 2937(9) | 4136(10) | 9.9(9) |
| O2 | 1956.7(13) | 2165(7) | 4045(7) | 6.4(5) |
| O3 | 1061.9(15) | 5354(5) | 1953(7) | 7.0(5) |
| O4 | 881.7(16) | -377(6) | 6244(7) | 8.3(6) |

| Atom | U11(Å$^2$)×10$^3$ | U22(Å$^2$)×10$^3$ | U33(Å$^2$)×10$^3$ | U12(Å$^2$)×10$^3$ | U13(Å$^2$)×10$^3$ | U23(Å$^2$)×10$^3$ |
|---|---|---|---|---|---|---|
| Mn | 3.5(2) | 2.1(2) | 1.8(2) | -0.06(12) | -1.2(2) | -0.40(15) |
| Ca1 | 4.9(4) | 9.7(5) | 6.3(6) | -0 | -0 | 0.6(4) |
| Ca2 | 4.8(3) | 0.3(3) | 4.3(4) | 0.5(2) | 0.7(3) | 0.9(3) |
| O1 | 1.9(12) | 15(2) | 13(2) | -0 | -0 | 8(2) |
| O2 | 4.2(9) | 6.2(12) | 8.8(15) | 2.3(8) | 0.2(14) | -3.2(13) |
| O3 | 9.1(10) | 3.8(13) | 8.1(13) | -1.8(8) | 3.3(11) | 3.5(12) |
| O4 | 8.7(10) | 8.6(16) | 7.4(13) | 4.2(10) | -2.0(9) | 1.2(11) |

Space Group: $Cmc2_1$ (Z=4) Orthorhombic Cell
a = 19.3692(8) Å, b = 5.2339(2) Å, c = 5.2301(2) Å, Dx = 4.285 g/cm$^3$
Measurement Temperature: 104 K
Crystal Dimensions: ~20 μm (diameter)
Wavelength: 0.41328 Å,
Absorption Coefficient: 1.635 mm$^{-1}$
F(000) = 666.8
2θ Range : 2.44° to 40.2°
$-32 \leq h \leq 32$, $-9 \leq k \leq 9$, and $-9 \leq l \leq 8$
Number of Measured Reflections: 5,103
Number of Independent Reflections: 1227
Number of fitting parameters: 60
Restraints = 1
Flack Parameter = 0.3(5)
Max and Min Peak in Final Difference Map: 2.3/-1.5 e-/ Å$^3$
$R_1$ = 4.19 %, $wR_2$ = 10.6 %, Goodness of Fit = 1.02 (I>=2σ (I))
$R_1$ = 4.46 %, $wR_2$ = 10.7 % (all data)

$$R_1 = \sum ||F_0| - |F_c|| / \sum |F_0|$$

$$wR_2 = \sum w(F_o^2 - F_c^2)^2) / \sum w(F_o^2)^2$$

**Table S9. DFT Structural Results for $Ca_3Mn_2O_7$**

| Lattice Parameters | a (Å) | b (Å) | c (Å) |
|---|---|---|---|
| U=8.0 eV/J=0.88 eV | | | |
| a,b,c | 19.0906 | 5.2185 | 5.2587 |
| % error | -1.4% | -0.29% | 0.55% |
| | | | |
| U=4.5 eV/J=1.0 eV | | | |
| a,b,c | 19.0629 | 5.1792 | 5.2012 |
| % error | -1.6% | -1.0% | -0.55% |
| | | | |
| Single Crystal Data | 19.3692(8) | 5.2339(2) Å | 5.2301(2) |
| Space Group Found | $Cmc2_1$ | | |
| 104 K results (see Table S8 in this supplementary document) | | | |

**Table S10. $Ca_3Mn_2O_7$** (U=4.5 eV and J=1.0 eV)

| Ion Site | $k_x$ (long axis) (eV/Å$^2$) | $k_y$ (eV/Å$^2$) | $k_z$ (eV/Å$^2$) |
|---|---|---|---|
| **$Ca_3Mn_2O_7$** (1x2x2) cell | | | |
| Mn | 26 | 28 | 28 |
| Ca1 | 8.6 | 8.3 | 7.9 |
| Ca2 | 12 | 7.9 | 7.2 |
| O1 (apical interior) | 22 | 9.0 | 6.1 |
| O2 (apical) | 18 | 7.0 | 5.9 |
| O3 | 7.8 | 16 | 16 |
| O4 | 7.6 | 16 | 17 |

**Table S11. $Ca_3Mn_2O_7$ and $Ca_3Ti_2O_7$ Self-Force Constants***

| Ion Site | $k_x$ (long axis) (eV/Å$^2$) | $k_y$ (eV/Å$^2$) | $k_z$ (eV/Å$^2$) |
|---|---|---|---|
| **$Ca_3Mn_2O_7$** (1x2x2) cell | | | |
| Mn | 26 | 28 | 28 |
| Ca1 | 8.2 | 8.2 | 8.0 |
| Ca2 | 11 | 7.9 | 7.2 |
| O1 (apical interior) | 21 | 9.0 | 5.9 |
| O2 (apical) | 17 | 7.0 | 5.7 |
| O3 | 7.1 | 15 | 16 |
| O4 | 7.5 | 14 | 15 |
| **$Ca_3Mn_2O_7$** (1x1x1) cell | | | |
| Mn | 26 | 26 | 26 |
| Ca1 | 8.6 | 7.9 | 7.7 |
| Ca2 | 11 | 7.5 | 6.8 |
| O1 (apical interior) | 21 | 8.8 | 5.7 |
| O2 (apical) | 17 | 6.6 | 5.4 |
| O3 | 7.2 | 14 | 15 |
| O4 | 7.6 | 14 | 14 |
| **$Ca_3Ti_2O_7$** (1x1x1) cell | | | |
| Ti | 24 | 19 | 19 |
| Ca1 | 8.0 | 6.8 | 6.3 |
| Ca2 | 11 | 7.8 | 6.7 |
| O1 (apical interior) | 21 | 8.1 | 6.0 |
| O2 (apical) | 18 | 6.8 | 5.2 |
| O3 | 7.6 | 13 | 15 |
| O4 | 8.1 | 13 | 13 |

*For $Ca_3Mn_2O_7$ U=8.0 eV and J=1.0 eV.

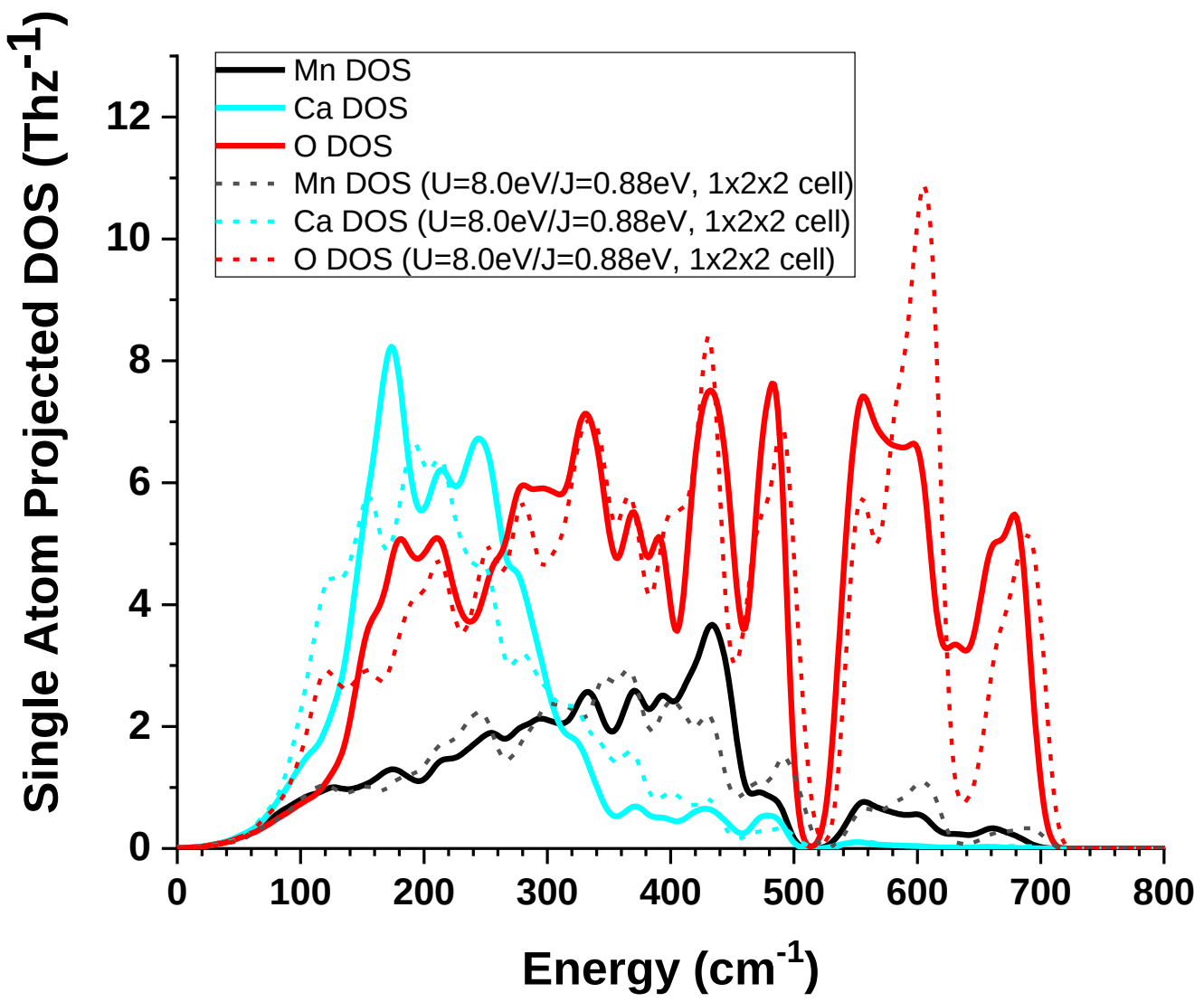

(a)

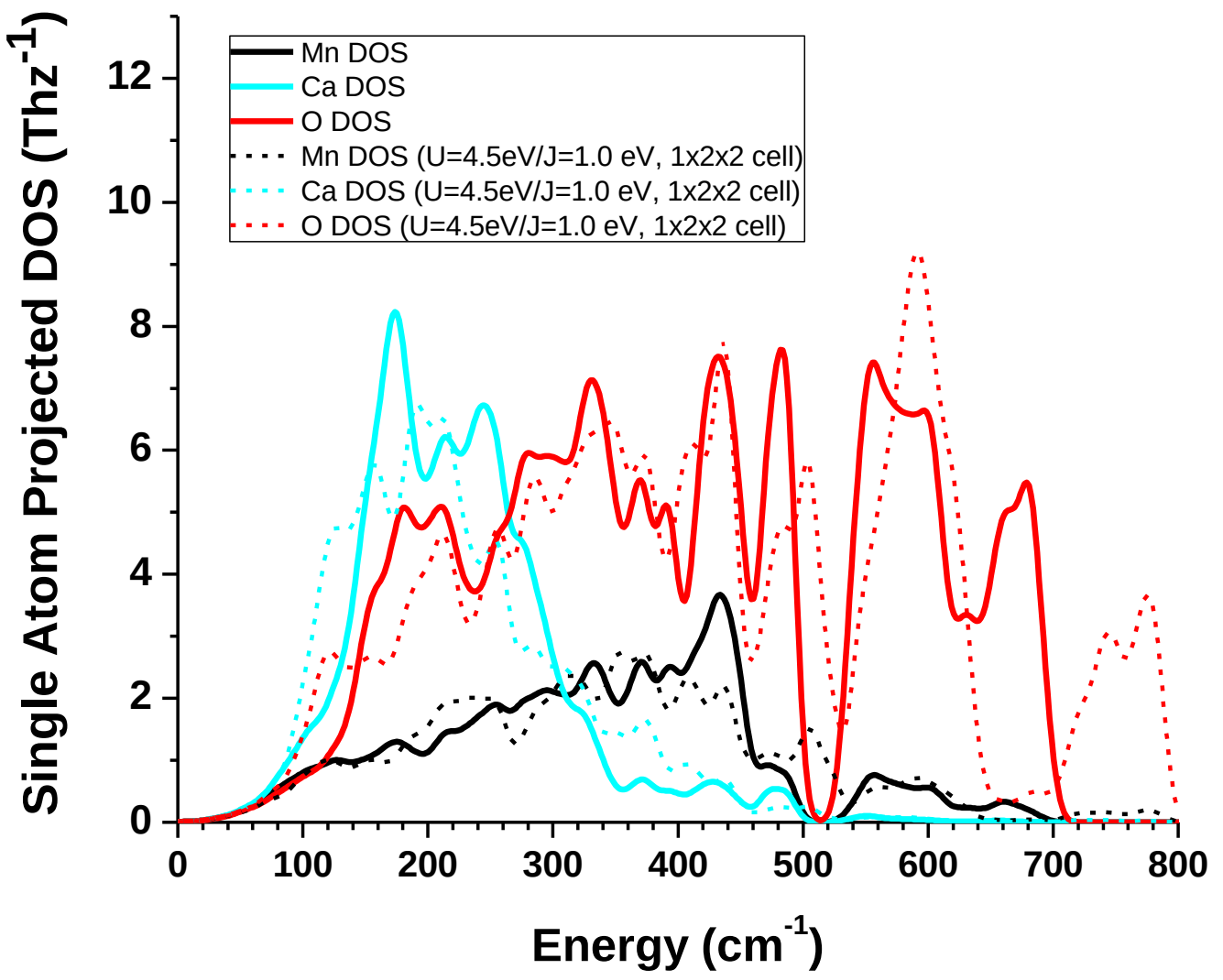

(b)

**Fig. S22.** Site projected phonon density of states for $Ca_3Mn_2O_7$. The thick solid lines in both panels show the DOS for a 1x1x1 unit cell computed with U = 8.0 eV and J = 0.88 eV. The dashed lines in panel (a) correspond to the phonon DOS for a 1x2x2 unit cell with the same U and J parameters as the 1x1x1 cell while the dashed line in panel (b) give the phonon DOS for a 1x2x2 unit cell with U = 4.5 eV and J = 1.0 eV. Note that the motion of O and Mn ions dominate the modes above 500 $cm^{-1}$.